\documentclass[superscriptaddress,aps,prd,showpacs,preprint]{revtex4-1}
\usepackage{lingmacros}
\usepackage{tree-dvips}
\usepackage{graphicx}
\usepackage{amsmath}
\usepackage{amssymb}
\usepackage{latexsym}
\usepackage{color}
\usepackage{fullpage}
\usepackage{commath}
\usepackage{cancel}
\usepackage{ulem}

\usepackage{dcolumn}
\usepackage[latin1]{inputenc}
\usepackage{graphicx, psfrag}
\usepackage{float}
\usepackage{tensor}
\usepackage{amsfonts}
\usepackage{hyperref}
\usepackage{subfigure}
\usepackage{pgfplots}
\usepackage{epstopdf}
\usepackage{booktabs}
\usepackage[11pt]{moresize}

\newcommand{\fixme}[1]{{\color{red}{#1}}}

\begin{document}

\title{Emergent Phase, Thermodynamic Geometry and Criticality of Charged Black Holes from R\'enyi Statistics}

\author{Ekapong Hirunsirisawat} \email{ekapong.hir@kmutt.ac.th}
\affiliation{Theoretical and Computational Physics (TCP); Theoretical and Computational Science Center (TaCS), Faculty of Science, King Mongkut's University of Technology Thonburi (KMUTT), Pracha Uthit Road, Bangkok, 10140, Thailand}
\affiliation{Learning Institute, King Mongkut's University of Technology Thonburi (KMUTT), Pracha Uthit Road, Bangkok, 10140, Thailand}

\author{Ratchaphat Nakarachinda} \email{tahpahctar\_net@hotmail.com}
\affiliation{The Institute for Fundamental Study, Naresuan University, Phitsanulok, 65000, Thailand}

\author{Chatchai Promsiri} \email{chatchaipromsiri@gmail.com} 
\affiliation{Department of Physics, Faculty of Science, King Mongkut's University of Technology Thonburi (KMUTT), Pracha Uthit Road, Bangkok, 10140, Thailand}
\affiliation{Theoretical and Computational Physics (TCP); Theoretical and Computational Science Center (TaCS), Faculty of Science, King Mongkut's University of Technology Thonburi (KMUTT), Pracha Uthit Road, Bangkok, 10140, Thailand}

\begin{abstract}
Recently, a novel emergent phase can occur from thermodynamic consideration of the asymptotically flat Reissner-Nordstr\"om black hole (RN-AF) using R\'enyi statistics. We present an analysis of the thermodynamical and mechanical stabilities of the RN-AF in both the Gibbs-Boltzmann (GB) and the alternative R\'enyi statistics when charge $q$ and electrostatic potential $\phi$ are treated as pressure and volume, respectively. Interestingly, the emergent phase of the RN-AF can be both thermodynamically and mechanically stable in some range of parameters in the framework of R\'enyi thermodynamics.  With the construction of the Maxwell equal area law in $q-\phi$ plane, the coexistence line between the near-extremal black hole phase and the emergent phase can be found in some values of charge which can be associated as the vapor pressure at which the liquid and gas phases coexist. In the aspect of thermodynamic geometry, the microscopic interaction between the black hole microstructures can be repulsive in the R\'enyi description. This implies that a novel correlation between the microstates of a self-gravitating system could be emerged via the nonextensive nature of long-range interaction systems. Finally, we also investigate the critical phenomena of the RN-AF in R\'enyi statistics compared to that of the van der Waals (vdW) fluid and find that the critical exponents of the relevant physical quantities of both systems are identical. This implies that both systems are in the same universality class of the phase transition.

\end{abstract}

\maketitle{}

\section{Introduction and Motivations}

According to the Gibbs-Boltzmann (GB) statistical approach, Bekenstein argued that the entropy of black hole is proportional to its horizon area~\cite{Bekenstein1973}. Later on, by taking into account the quantum fluctuation around the black hole's horizon, Hawking had shown that black holes emit thermal radiation at the Hawking temperature~\cite{Hawking1975}. Moreover, both quantities, namely, the Bekenstein-Hawking entropy $S_\text{BH}$ and Hawking temperature $T_\text{H}$ are related through the first law of black hole thermodynamics~\cite{Bardeen1973}. However, black holes are fascinating compact objects where the spacetime surrounded them is severely curved in such a way that an observer, being outside their event horizons, cannot observe anything inside the black hole event horizons. Therefore, the statistical origin of the black hole entropy is hidden from the rest of the universe, which leads to some important open questions on counting black hole microstates and calculating its corresponding entropy from the first principle~\cite{Strominger:1996sh, Callan:1996dv, Horowitz:1996fn, Emparan:2006it, Dabholkar:2001if, Giveon:2012kp, Brustein:2021cza, Giveon:2021gsc}.

One of the most peculiarities about black holes is that their entropy obeys the area law and thus it is seemingly nonextensive~\cite{Tsallis2013, Tsallis2020}. Consequently, when two black holes of entropies $S_\text{A}$ and $S_\text{B}$ merge adiabatically, the resulting entropy of the black hole system $S_\text{AB}$ can be expressed in a nonadditive form~\cite{Davies}.  To investigate the effects from nonextensitivity, it has been proposed that $S_\text{AB}$ can be written in the form of the Abe's generalized nonadditve composition rule~\cite{Abe}
\begin{equation} 
	H\left(S_\text{AB}\right)=H\left(S_\text{A}\right)+H\left(S_\text{B}\right)+\lambda H\left(S_\text{A}\right) H\left(S_\text{B}\right), \label{Abe}
\end{equation} 
where $H(S)$ is a differentiable function of $S$ and $\lambda$ is a nonextensivity parameter. From statistical mechanics, one can derive the thermodynamic quantities of macroscopic system from the microscopic description  via the standard statistical methods. In the GB approach, it is usually assumed that the interaction between the microstructures of the system is much smaller than the size of the system. In this way, the GB entropy formula might not be appropriate for strong-gravitating systems because the long-range interaction cannot be ignored. This is a microscopic description to the origin of nonextensive nature for black hole thermodynamical systems.

However, there is an obscure notion of thermal equilibrium and empirical temperature in thermodynamics of nonextensive entropy $H(S)$ with nonadditive composition rule \eqref{Abe}. Fortunately, this problem can be solved by transforming $H(S)$ into another entropic functional form \cite{Biro}
\begin{eqnarray}
	L(S) = \frac{1}{\lambda}\ln \left[ 1+\lambda H(S) \right], \label{Biro}
\end{eqnarray}
which makes the composition rule \eqref{Abe} turning out to be additive
\begin{eqnarray}
	L(S_\text{AB}) = L(S_\text{A}) + L(S_\text{B}).
\end{eqnarray}
In the simplest case, we identify the differentiable function $H(S)$ to be the Tsallis entropy $H(S) = S_\text{T}$ \cite{Tsallis1988}. In this case, the logarithmic functional form of entropy in Eq.~\eqref{Biro} is indeed the well-known R\'enyi entropy \cite{Renyi1959}
\begin{eqnarray}
	L(S_\text{T})=\frac{1}{\lambda} \ln (1+\lambda S_\text{T}) \equiv S_\text{R}.
\end{eqnarray}
Since the R\'enyi entropy is additive despite the presence of nonextensive nature, this tends to a well-defined thermal equilibrium state for nonextensive systems and then the empirical temperature function can be uniquely determined from the standard relation 
\begin{eqnarray}
	\frac{1}{T_\text{R}} = \frac{\partial S_\text{R}}{\partial E},
\end{eqnarray}
where $T_\text{R}$ is the corresponding so-called R\'enyi temperature and $E$ is the energy of the system.

Interestingly, by considering $S_\text{BH}$ as $S_\text{T}$ \cite{Tsallis2013}, one can obtain the R\'enyi entropy associated to a black hole. The results show that the black holes can be thermodynamically stable with its surrounding environment in some range of parameter space in the cases of asymptotically flat spacetime \cite{Biro:2013cra, Czinner2016, Czinner2017, Promsiri2020, Promsiri:2021hhv, Alonso-Serrano:2020hpb} and asymptotically dS spacetime \cite{Tannukij2020, Nakarachinda2021, Samart:2020klx}. In addition, the stability of a black string is also investigated using the R\'enyi statistics. \cite{Sriling:2021lpr}. Recently, as suggested in Refs. \cite{Promsiri2020, Promsiri:2021hhv}, the nonextensivity parameter $\lambda$ can be identified as a thermodynamic pressure $P = \frac{3\lambda}{32}$ and its conjugate variable as the thermodynamic volume $V = \frac{4}{3}\pi r_h^3$, where $r_h$ is the horizon radius. This framework is called the R\'enyi extended phase space approach. In this formalism, it was first indicated that, in a canonical ensemble, the Reissner-Nordstr\"om black hole in asymptotically flat spacetime (RN-AF) allows a first-order phase transition which is reminiscent of the vdW liquid-gas phase transition and a Hawking-Page phase transition of charged AdS black holes \cite{Chamblin1999, Chamblin19992nd, Peca:1998cs, Burikham:2014gwa, Kubiznak:2012wp, Gunasekaran:2012dq, Majhi:2016txt} (for review, see \cite{Kubiznak:2016qmn, Altamirano:2014tva, Mann2015} and references therein).

In a conventional thermodynamic system, the well-known \textit{Maxwell equal area law} or \textit{Maxwell construction} is a qualitative explanation for the first-order phase transition between two different locally stable phases without passing the unstable phase. The vdW fluid undergone in the isothermal compression or expansion is an example of the Maxwell equal area law. On the $P-V$ plane, it has been found that there possibly exist three phases along each isothermal curve. Two of them are the liquid and gas phases with positive compressibility. Another one is the phase with negative compressibility. Hence, the phase transition of the fluid via the isothermal compression or expansion should pass the mechanically unstable phase. The argument states that the fluid will evolve under the first-order phase transition with a certain pressure instead of passing through the unstable phase. From the fact that the Gibbs free energy does not change \cite{David2016}, the certain pressure is evaluated in the way that the upper and lower areas between the isothermal curve and the line of constant pressure are the same. Interestingly, this equal area law is held in the charged black holes (see e.g. \cite{Spallucci:2013osa, Ma:2016aat, Zhou:2019xai}). In our case, the aforementioned equal area law should be obtained by considering the isothermal curve on the $q-\phi$ plane. It will be seen that the charged black hole with R\'enyi description undergoes the transition between two stable black hole phases in a similar way to what occurs in the vdW fluid.

One may argue that the correction due to nonextensive effect should be also seen in the microscopic structure of black hole systems.
In the phase transition, the thermodynamic systems are undergoing some reorganization processes of its microscopic structure. As mentioned above, the charged black hole cannot be only a thermal object, but it also interestingly undergoes a phase transition as the vdW liquid-gas system does. Therefore, it is believed that a black hole should have microscopic degrees of freedom even if its constituents are still unknown. There is a proposal on possible ways to study the microstructure or the interaction between \textit{black hole molecules} by using the geometrical approach of the thermodynamic phase space. The thermodynamic geometry method has been developed by Weinhold \cite{Weinhold1975} and Ruppeiner \cite{Ruppeiner1995}. It is an interesting approach to investigate some microscopic aspects of thermodynamic system from their macroscopic quantities. Ruppeiner \cite{Ruppeiner1995} proposed that an entropy plays a role of the thermodynamic potential rather than the internal energy \cite{Weinhold1975}, and then derived a Riemannian thermodynamic line element, which has physical meaning as the distance between two neighbouring thermal fluctuating states. The scalar curvature $R$ corresponding to the Ruppeiner metric can be computed in the same way as the Ricci scalar in the general relativity. By applying this formalism to various thermodynamic systems \cite{Ruppeiner1979, Brody1995, Brody2009, Ruppeiner1981, Janyszek1989, Oshima1999}, it has been argued that the sign of $R$ can be used to identify the type of interaction between the microstructure of a system. The positiveness and negativeness in $R$ are, respectively, related to the repulsive and attractive interactions, whereas $R=0$ corresponds to the systems with no interaction between its constituents, for a review see \cite{Ruppeiner2010} and the references therein.

There is a connection between the microscopic fluctuation of matters and macroscopic response function, such as the heat capacity and compressibility. At the second-order phase transition, these response functions go to infinity which means the correlated fluctuation being over large distance. Typically, such correlation occurs over a characteristic distance $\xi$, called the correlation length. Ruppeiner has shown that $R$ is proportional to the correlation volume of the system as~\cite{Ruppeiner1995}
\begin{eqnarray}
	R \sim \xi^d, \label{correlation}
\end{eqnarray}
where $d$ is the physical dimension of the system. This is a further evidence about the connection between $R$ and microscopic correlation of the system. According to the connections between $R$ and some microscopic aspects, the thermodynamic geometry approach might be suitable for probing some information about microstructure of a black hole. The interesting studies on thermodynamic geometry of various black holes in different theories of gravity can be seen in Refs. \cite{Aman:2003ug, Aman:2005xk, Shen:2005nu, Niu:2011tb, Wang:2019cax, Wei:2012ui, Wei:2015iwa, Wei:2019yvs, Wei:2021krr, Yerra:2020oph, Wu:2020fij, PhysRevD.101.046005, PhysRevD.104.104049}.

In the present work, we study further in a critical phenomena and microscopic origin of the R\'enyi entropy associated to the black hole. The rest of this paper is organized as follows. In section \ref{RN-GB}, we introduce an analogy $(q, \phi) \leftrightarrow (P, V)$ between RN-AF and vdW liquid-gas system from the first law of thermodynamics. In this way, we shall revisit thermal stability criteria based on the signs of the associated response functions such as the heat capacity and compressibility. In section \ref{RN-Renyi}, the thermal phase structure of RN-AF and its stability are studied within the R\'enyi statistics. In section \ref{Maxwell}, we present two different expressions of the Maxwell equal area law in the $q-\phi$ plane when the systems are near and far away from the critical point. At this point, the coexistence region and phase diagram are also studied. In section \ref{TherGeo}, we first briefly review the thermodynamic geometry of RN-AF with the GB statistics. The phase structure and types of interaction between microstructure of black hole within R\'enyi statistics are then studied in the thermodynamic geometry framework. In section \ref{criticality}, we investigate the thermal behavior of the RN-AF via the R\'enyi statistics near the critical point. All critical exponents have been calculated. We summarize and discuss our results in section \ref{conclusion} .

%%%%%%%%%%%%%%%%%%%%%%%%%%%%%%%%%%%%%%%%%%%%%%%%%%%%%%%%%%%%%%%%%%%%%%%%%%%%%%%%%%%%%%%%%%%%%%%%%%%%%%%%%%%%%%%%%%%%%%%%%%%%%%%%%%%%%%%%%%%%%%%%%%%%%%%%%%%%%%%%%%%%%%%%%%%%%%%%%%%%%%%%%%%%%%%%%%%%%%%%%%%%%%%%%%%%%%%%%%%%%%%%%%%%%%%%%%%%%%%%%%%%%%%%%%%%%%%%%%%%%%%%%%%%%%%%%%%%%%%%%%%%%%%%%%%%%%%

\section{Thermodynamics of Reissner-Nordstr\"om black holes from the Gibbs-Boltzmann statistics}\label{RN-GB}

In the asymptotically flat spacetime, the Reisser-Nordstr\"om solution to the Einstein field equation with the U(1) electric charge can be written in the form
\begin{eqnarray}
	ds^2 &=& -f(r)dt^2+\frac{dr^2}{f(r)}+r^2(d\theta^2+\sin^2\theta d\phi^2), \nonumber \\
	f(r) &=& 1-\frac{2M}{r}+\frac{q^2}{r^2}.\label{f of r}
\end{eqnarray}
By solving the horizon equation, $f(r)=0$, there exist the inner and outer horizon with the radii $r_-$ and $r_+$, following the relation
\begin{eqnarray}
	r_\pm=M \pm \sqrt{M^2-q^2}. \label{horizon}
\end{eqnarray}
The black hole is extremal when the inner and outer horizons are identical at $M=q$. For $M<q$, the black hole cannot exist because the event horizons in Eq.~\eqref{horizon} are complex numbers, and, hence, the cosmic censorship hypothesis will be violated \cite{Penrose:1969pc}. The mass and electric charge can be, respectively, expressed in terms of the horizon radii as follows:
\begin{eqnarray}
	M&=&\frac{r_+ + r_-}{2}, \\
	q&=&\sqrt{r_+r_-}. \label{charge}
\end{eqnarray}
In the standard approach, the Bekenstein-Hawking entropy of the black hole is obtained from the area law \cite{Bekenstein1973}
\begin{eqnarray}
	S_{\text{BH}}=\frac{\mathcal{A}}{4} = \pi r_+^2. \label{area}
\end{eqnarray}
where $\mathcal{A}$ is the surface area of the outer horizon radius. The first law of black hole thermodynamics reads \cite{Bardeen1973}
\begin{eqnarray}
	dM = T_\text{H}dS_\text{BH}+\phi dq, \label{1st}
\end{eqnarray}
which implies that the Hawking temperature $T_\text{H}$ and electrostatic potential $\phi$ on the outer event horizon are, respectively, given by
\begin{eqnarray}
	T_{\text{H}}&=&\left( \frac{\partial M}{\partial S_\text{BH}} \right)_q = \frac{r_+ - r_-}{4\pi r_+^2}, \label{Hawk temp}\\
	\phi &=& \left( \frac{\partial M}{\partial q} \right)_{S_\text{BH}}=\frac{q}{r_+}. \label{potential}
\end{eqnarray}
Note that Eq.~\eqref{Hawk temp} manifests the fact that that the Hawking temperature is vanishing at the extremal limit, where the outer and inner horizon radii are degenerate.

By considering the first and second derivatives of the temperature $T_\text{H}=T_\text{H}(r_+,q)$ with respect to $r_+$, this temperature has a maximum value given by $T_\text{max} = (6\sqrt{3}\pi q)^{-1}$ at the outer horizon radius being equal to  
\begin{eqnarray}
	r_0=\sqrt{3}q.\label{r0}
\end{eqnarray}
\begin{figure*}[!ht]
	\begin{tabular}{c c}
		\includegraphics[scale=0.5]{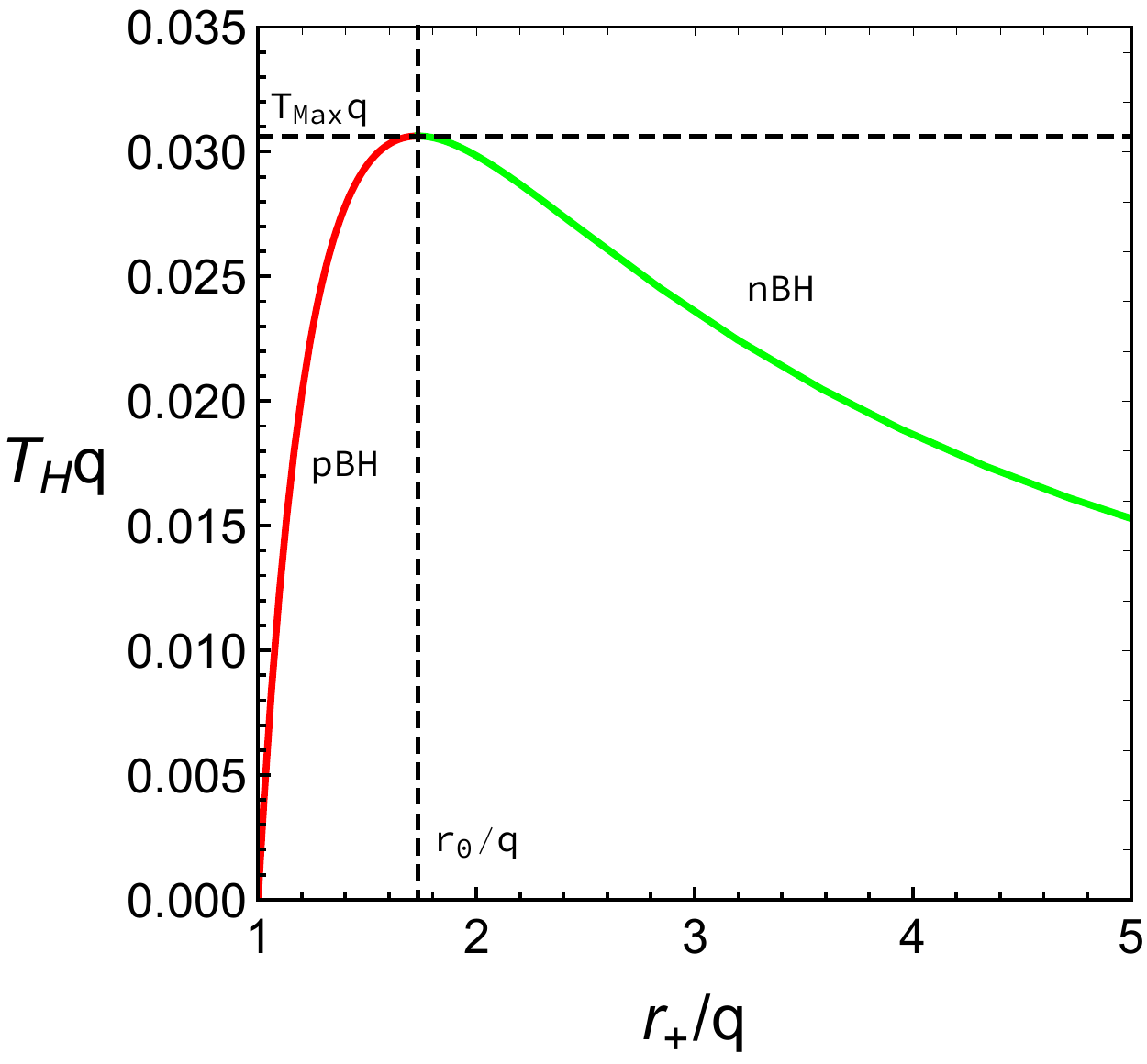} \hspace{1cm} 
		\includegraphics[scale=0.52]{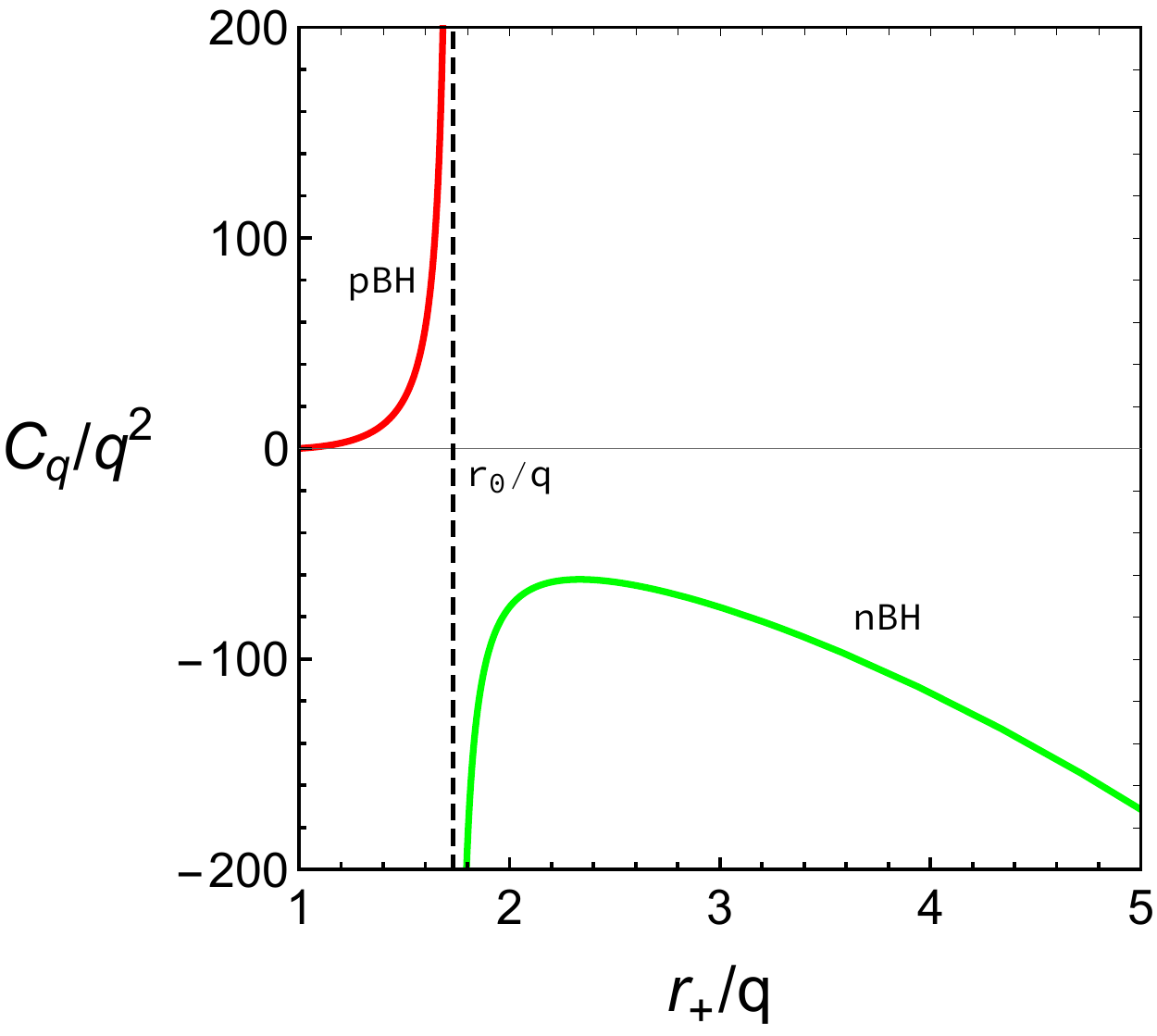}
	\end{tabular}
	\caption{The dimensionless Hawking temperature $T_\text{H}q$ (left) and heat capacity $C_q/q^2$ (right) of the RN-AF via the GB statistics with respect to the dimensionless outer horizon radius $r_+/q$.}\label{fig:T and Cq GB}
\end{figure*}
Therefore, the profile of this temperature can be divided into two different behaviours which are increasing and decreasing function in $r_+$ for $r_+<r_0$ and $r_+>r_0$, respectively. This is shown in the left panel in Fig.~\ref{fig:T and Cq GB}. As discussed previously, a black hole horizon exists when $r_+\geq q$. Using Eq.~\eqref{potential}, this implies that the electric potential is in the range $0<\phi\leq1$, where its upper limit corresponds to the extremal black hole.

The heat capacity for fixed charge is computed as
\begin{eqnarray}
	C_q 
	= T_\text{H}\left( \frac{\partial S_\text{BH}}{\partial T_\text{H}} \right)_q 
	= \frac{2\pi r^2_+ (r_+ - r_-)}{3r_- - r_+}.
	\label{C GB}
\end{eqnarray}
As a result, there are two possible black hole configurations, namely, the small black hole branch with positive heat capacity, and the large black hole branch with negative heat capacity in which we will henceforth denote as the pBH, and nBH, respectively. One also calls that the pBH (nBH) is locally stable (unstable). The nBH-pBH phase transition occurs when the outer event horizon equals $r_+=r_0$. At the radius $r_0$, the temperature reaches its maximum and the heat capacity diverges. The profiles of the temperature and heat capacity for the non-zero charged black hole with respect to the outer horizon are illustrated in Fig.~\ref{fig:T and Cq GB}. In this figure, the red (green) curves represent the quantities for pBH (nBH) branch. Since the entropy is indeed a monotonic function of $r_+$ as obviously seen in Eq.~\eqref{area}, the sign of the heat capacity can be obtained from that of the slope of  temperature profile graph (the left panel of Fig.~\ref{fig:T and Cq GB}). Note that eliminating $r_{-}$ in Eqs.~\eqref{Hawk temp} and \eqref{C GB} by using Eq.~\eqref{charge} allows us to express the outer horizon in the scale of $q$, \textit{i.e.} $r_{+}/q$, as seen in Fig.~\ref{fig:T and Cq GB}.This can be done due to the fact that the phase structure of this system is not changed for different values of non-zero charge $q$. However, it is very important to emphasize that the scaling by $q$ is not appropriate for black hole thermodynamics from the R\'enyi statistics because the phase structure depends on $q$ in that circumstance, as will be discussed in the next section.

It has been observed that the charge $q$ and electrostatic potential $\phi$ of the charged black holes can be identified as the pressure $P$ and volume $V$ of the vdW fluid, respectively \cite{Chamblin1999, Chamblin19992nd, Peca:1998cs}. 
From the first law of black hole thermodynamics, we have
\begin{eqnarray}
	dM
	&=&T_\text{H}dS_\text{BH}+\phi dq \label{first law} \\ 
	&=&T_\text{H}dS_\text{BH}-qd\phi+d(q\phi) , \nonumber
\end{eqnarray}
then we can write
\begin{eqnarray}
	d(M-q\phi)
	&=&T_\text{H}dS_\text{BH}-qd\phi . \label{du in S phi}
\end{eqnarray}
By comparing this relation to the conventional first law, $dU=TdS-PdV$, with identifying $q d\phi$ as $PdV$, one can define the appropriate internal energy of the RN-AF as
\begin{eqnarray}
	U=M-q\phi, \label{energy}
\end{eqnarray}
where the contribution of the electric potential energy to the mass of black hole is excluded \cite{Shen:2005nu}. In this analogy, the black hole mass $M$ is indeed a function of $S_\text{BH}$ and $q$, $M=M(S_\text{BH}, q)$. Therefore, the mass should be interpreted as the enthalpy rather than the internal energy.

In order to investigate the phase structure of the thermodynamic system, one needs to consider the Gibbs free energy which is defined as $\mathcal{G} = U-TS+PV$. As mentioned previously, the internal energy of the black hole should be taken in the form as expressed in Eq.~\eqref{energy}. The charge and its conjugate variable, \textit{i.e.}, the electrostatic potential are then interpreted as a pressure and volume, respectively. We, therefore, define an appropriate Gibbs free energy for RN-AF in the following form
\begin{eqnarray}
	\mathcal{G}
	=U -T_\text{H}S_\text{BH} + q\phi
	=M-T_\text{H}S_\text{BH}
	=\frac{1}{4}(r_++3r_-).
\end{eqnarray}
\begin{figure*}[!ht]
	\begin{tabular}{c c}
		\includegraphics[scale=0.5]{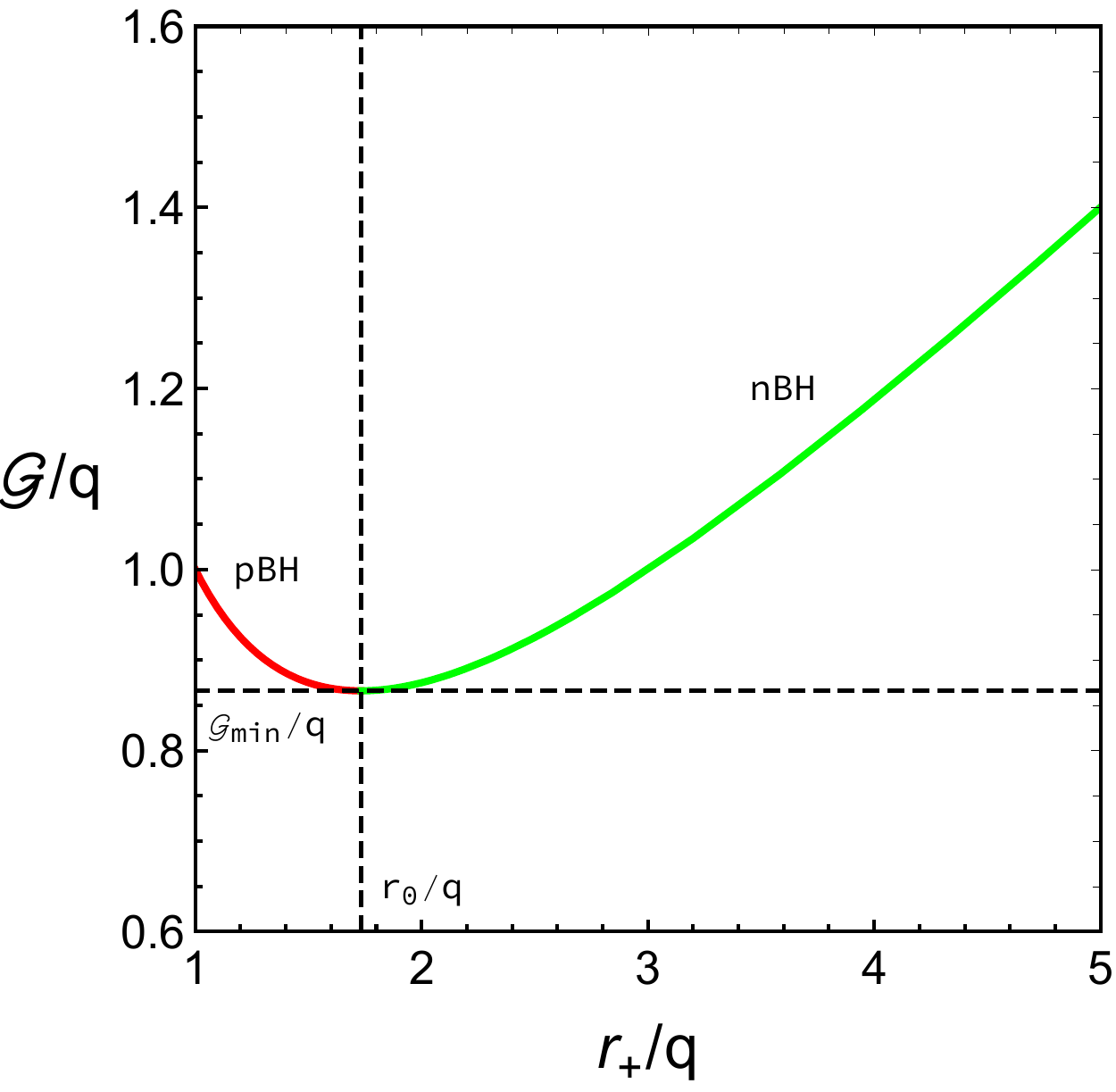}\hspace{1cm}
		\includegraphics[scale=0.52]{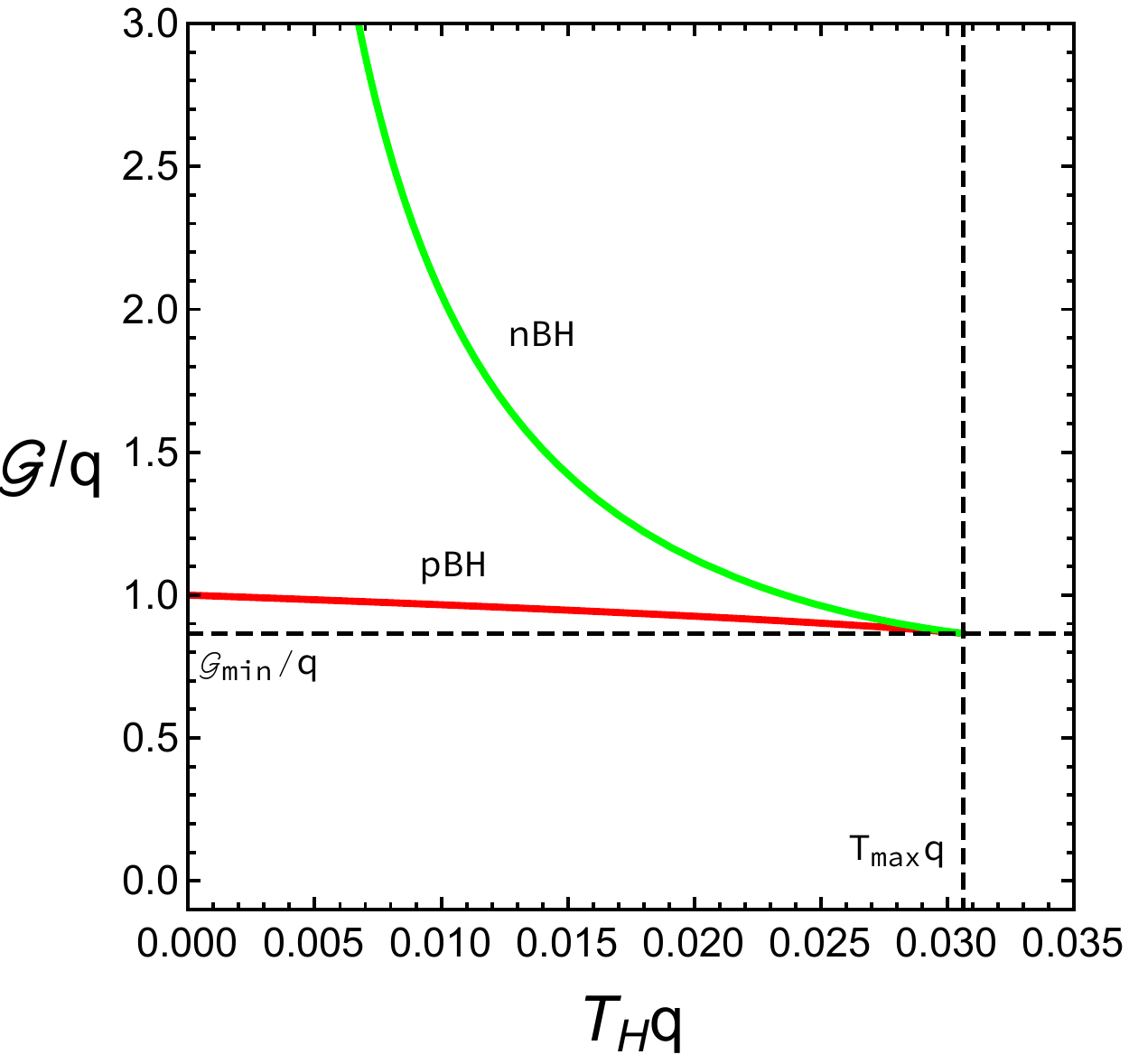}
	\end{tabular}
	\caption{The profile of the dimensionless Gibbs free energy $\mathcal{G}/q$ of the RN-AF via GB statistics with respect to the dimensionless outer horizon radius $r_+/q$ (left) and the Hawking temperature $T_\text{H}/q$ (right).}\label{fig:Gibbs GB}
\end{figure*}

The infinitesimal change of $\mathcal{G}=\mathcal{G}(T_\text{H}, q)$ can be expressed as follows:
\begin{eqnarray}
	d \mathcal{G}
	&=& dM - T_\text{H}dS_\text{BH} - S_\text{BH}dT_\text{H} \nonumber \\
	&=& ( T_\text{H}dS_\text{BH}+\phi dq ) - T_\text{H}dS_\text{BH} - S_\text{BH}dT_\text{H} \nonumber \\
	&=& - S_\text{BH}dT_\text{H} + \phi dq, 
\end{eqnarray}
where we have used \eqref{first law} for $dM$, this yields
\begin{eqnarray}
	\left(\frac{\partial\mathcal{G}}{\partial T_\text{H}}\right)_q=-S_\text{BH},\hspace{1cm}
	\left( \frac{\partial \mathcal{G}}{\partial q} \right)_{T_\text{H}}=\phi.
\end{eqnarray}
It is seen that the Gibbs free energy is always positive for both pBH and nBH branches, and reaches its minimum value $\mathcal{G}_\text{min}=\sqrt{3}q/2$ at the same point where the nBH-pBH phase transition occurs ($r_+=r_0$). The profile of the free energy is illustrated in Fig.~\ref{fig:Gibbs GB}. Note that, for the black hole with a certain value of free energy $\mathcal{G}$, there are two possible black hole phases which are small and large ones. The small (large) branch is locally stable (unstable). As seen in a right panel in Fig.~\ref{fig:Gibbs GB}, the nBH-pBH phase transition occurs at the cusp where the second derivative of $\mathcal{G}$ proportional to the heat capacity as $\Big(\frac{\partial^2\mathcal{G}}{\partial T_\text{H}^2}\Big)_q=-\frac{C_q}{T_\text{H}}$ is discontinuous. Hence, the nBH-pBH transition is actually the second-order type of phase transition. 

\begin{figure*}[!ht]
	\begin{tabular}{c c}
		\includegraphics[scale=0.5]{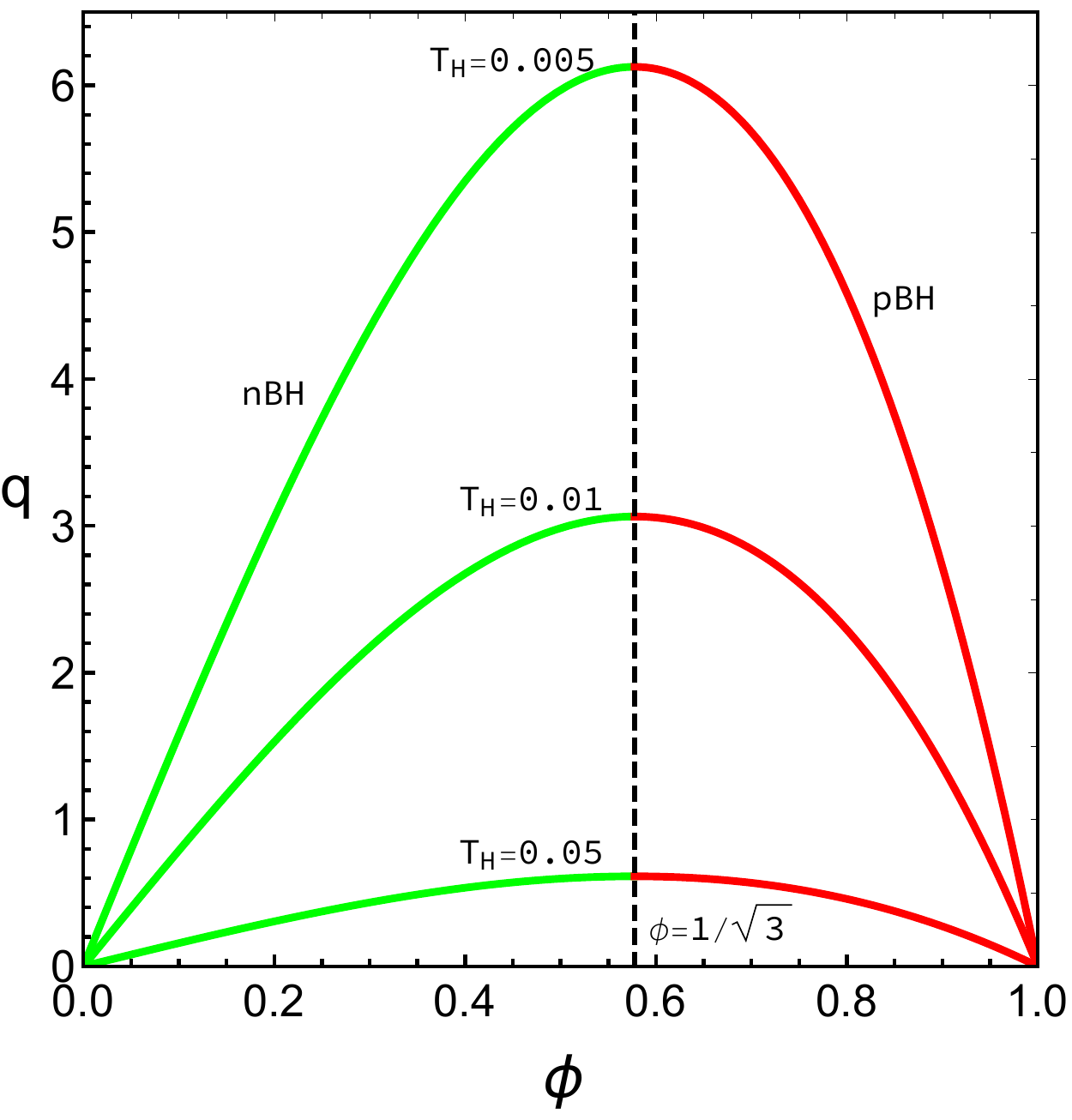}\hspace{1cm}
		\includegraphics[scale=0.525]{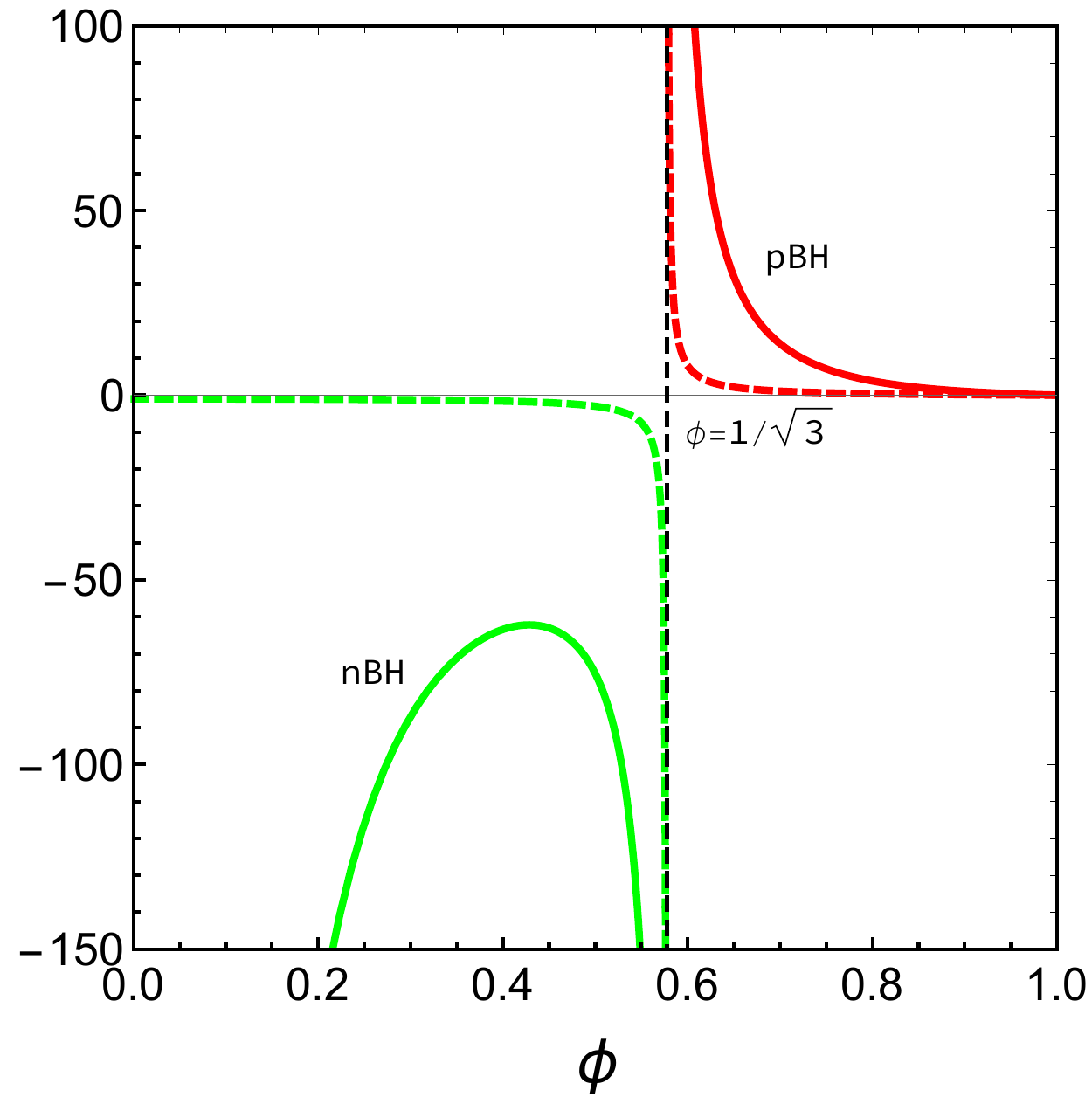}
	\end{tabular}
	\caption{Left: The isothermal curves on the $q-\phi$ plane with different temperatures. Right: the profiles of the heat capacity $C_q/q^2$ (solid lines) and the compressibility $\kappa_{T_\text{H}}q$ (dashed lines) with respect to the electrostatic potential $\phi$, where the green (red) lines are the quantities for nBH (pBH) branch.}\label{fig:q-phi and Cq-kappaT GB}
\end{figure*}

Moreover, it is possible to construct a constraint surface $F(q, \phi, T_\text{H})=0$ similar to the equation of state $F(P, V, T) = 0$ in the conventional thermodynamics. Using Eqs.~\eqref{charge} and \eqref{potential}, the Hawking temperature in Eq.~\eqref{Hawk temp} can be expressed in terms of  $\phi$ and $q$ in the form 
\begin{eqnarray}
	T_\text{H}=\frac{(1-\phi^2)\phi}{4\pi q}.
\end{eqnarray}
From the above expression, we can write
\begin{eqnarray}
	q=\frac{(1-\phi^2)\phi}{4\pi T_\text{H}},\label{eosGB}
\end{eqnarray}
which is indeed the equation of state of this system. The isothermal curves on the $q-\phi$ plane are plotted in the left panel in Fig.~\ref{fig:q-phi and Cq-kappaT GB}. The heat capacity can be expressed as 
\begin{eqnarray}
	C_q=-\frac{2\pi q^2}{\phi^2}\left[\frac{(1-\phi^2)}{(1-3\phi^2)}\right].\label{Cq GB}
\end{eqnarray} 
Note that the sign change of $C_q$ as well as its divergence occur at the value 
\begin{eqnarray}
	\phi_0=\frac{1}{\sqrt{3}},
\end{eqnarray}	
which actually corresponds to $r_0$. Each isothermal curve is, therefore, divided into the branches of pBH ($\phi>1/\sqrt{3}$) and nBH ($\phi<1/\sqrt{3}$).

Furthermore, the thermodynamical properties of RN-AF as discussed above will make sense if a black hole can be in locally thermal equilibrium against microscopic fluctuations. The conditions for local stability are studied in terms of both the heat capacity $C_q$ and compressibility $\kappa_{T_\text{H}}$ as
\begin{eqnarray}
	C_q>0,
	\hspace{1cm}
	\kappa_{T_\text{H}}>0.\label{local stab conds}
\end{eqnarray}
According to the \textit{Le Chatelier principle} which states that if a extensive variable fluctuate from its equilibrium value, its conjugate intensive variable will change in such a way that extensive variable is restored to its equilibrium \cite{landau1980, callen1985, Sekerka2015}. In this way, a thermal fluctuation will increase (decrease) the entropy so that the increasing (decreasing) temperature resists the heat flowing between the system and its environment, resulting to the reach into equilibrium. Therefore, the entropy monotonically increases with temperature. This means that the heat capacity is positive, $C_q>0$.

Let us consider another local stability. In the conventional fluid system, the work done on (by) a system decreases (increases) its volume. After that the pressure should increase (decrease) in order to enforce the system to restore its equilibrium state. This behavior also leads us to define the the isothermal compressibility, $\kappa_T=-\frac{1}{V}\big(\frac{\partial V}{\partial P}\big)_T$. The minus sign is introduced in order to identify the mechanically stable (unstable) system as a system with positive (negative) compressibility.

We have mentioned that $q$ and $\phi$ of the charged black holes play the similar roles to pressure $P$ and volume $V$ of the vdW fluid, respectively, in the aspect of phase structure. This might be interpreted that the mechanical stability in conventional system is actually the mechanical stability for charged black holes. Hence, the isothermal compressibility for RN-AF can be defined as
\begin{eqnarray}
	\kappa_{T_\text{H}}
	=-\frac{1}{\phi}\left(\frac{\partial\phi}{\partial q}\right)_{T_\text{H}}
	=-q\left[\frac{(1-\phi^2)}{(1-3\phi^2)}\right].\label{kappa GB}
\end{eqnarray}
Obviously, the vanishing and divergent points of $\kappa_{T_\text{H}}$ and $C_q$ are the same (see the terms in the square brackets in Eqs.~\eqref{Cq GB} and \eqref{kappa GB}). The signs of $C_q$ and $\kappa_{T_\text{H}}$ are also identical for any value of $\phi$. Therefore, the black hole with positive (negative) heat capacity has positive (negative) compressibility. In other words, the pBH (nBH) is both thermally and mechanically stable (unstable). The profiles of $C_q$ and $\kappa_{T_\text{H}}$ are illustrated in the right panel in Fig.~\ref{fig:q-phi and Cq-kappaT GB}. In addition, the discontinuous point of the compressibility $\kappa_{T_\text{H}}$ [\textit{i.e.}, $\phi=1/\sqrt{3}$] is indeed the place where the second-order phase transition occurs in the isothermal process, since the second derivative of the free energy $\mathcal{G}$ is proportional to $\kappa_{T_\text{H}}$ as follows: $\Big(\frac{\partial^2\mathcal{G}}{\partial q^2}\Big)_{T_\text{H}}=-\kappa_{T_\text{H}}\phi$.

It has been found that there does not exist a critical phenomenon in the RN-AF using the GB statistics. It is because there is no solution for $q$, $\phi$ and $T_\text{H}$ satisfying the criticality condition, $\big(\frac{\partial q}{\partial \phi} \big)_{T_\text{H}}=\big(\frac{\partial^2 q}{\partial \phi^2}\big)_{T_\text{H}}=0$. In the next section, the nonextensivity in the black hole entropy is introduced via the R\'enyi statistics. The critical phenomena and other interesting behaviors as well as the difference between the black holes with the GB and R\'enyi statistics will be studied.

%%%%%%%%%%%%%%%%%%%%%%%%%%%%%%%%%%%%%%%%%%%%%%%%%%%%%%%%%%%%%%%%%%%%%%%%%%%%%%%%%%%%%%%%%%%%%%%%%%%%%%%%%%%%%%%%%%%%%%%%%%%%%%%%%%%%%%%%%%%%%%%%%%%%%%%%%%%%%%%%%%%%%%%%%%%%%%%%%%%%%%%%%%%%%%%%%%%%%%%%%%%%%%%%%%%%%%%%%%%%%%%%%%%%%%%%%%%%%%%%%%%%%%%%%%%%%%%%%%%%%%%%%%%%%%%%%%%%%%%%%%%%%%%%%%%%%%%

\section{The emergent phase and swallowtail behavior of RN-AF in R\'enyi statistics}\label{RN-Renyi}

For an alternative R\'enyi statistics, the R\'enyi entropy $S_\text{R}$ of the black holes can be expressed in terms of the Bekenstein-Hawking entropy $S_{\text{BH}}$ described by the Tsallis entropy as discussed in Refs \cite{Czinner2016, Czinner2017, Promsiri2020, Promsiri:2021hhv, Alonso-Serrano:2020hpb, Tannukij2020, Nakarachinda2021, Samart:2020klx},
\begin{eqnarray}
	S_\text{R}=\frac{1}{\lambda}\ln(1+\lambda S_{\text{BH}}),
\end{eqnarray}
where $\lambda$ is the nonextensivity parameter. It is also found that the above entropic function obeys the additive composition rule so that the zeroth law is compatible \cite{Biro}. As mentioned in Refs.~\cite{Promsiri2020, Promsiri:2021hhv}, this approach has an energy (length) scale defined via the nonextensivity parameter $\lambda$ by
\begin{eqnarray}
	L_\lambda=\frac{1}{\sqrt{\pi \lambda}}.
\end{eqnarray}
The first law of black hole R\'enyi thermodynamics is given by 
\begin{eqnarray}
	dM=T_{\text{R}}dS_\text{R}+\phi dq.
\end{eqnarray} 
Note that the mass parameter still plays the same role of the enthalpy. Only the heat term for R\'enyi statistics $T_{\text{R}}dS_\text{R}$ is modified from that for GB statistics. The R\'enyi temperature $T_\text{R}$ and electrostatic potential $\phi$ corresponding to the R\'enyi entropy, respectively, are
\begin{eqnarray}
	T_\text{R}
	&=&\left( \frac{\partial M}{\partial S_\text{R}} \right)_q=\left( \frac{r_+ - r_-}{4\pi r_+^2} \right)\left(1+\frac{r_+^2}{L^2_{\lambda}}\right),\label{Tr} \\
	\phi
	&=& \left( \frac{\partial M}{\partial q} \right)_{S_\text{R}}=\frac{q}{r_+}.\label{phi in q r+}
\end{eqnarray}
The heat capacity and the Gibbs free energy are, respectively, computed as  
\begin{eqnarray}
	C_\text{R} 
	&=&T_\text{R}\left(\frac{\partial S_\text{R}}{\partial T_\text{R}}\right)_q 
	= \frac{2\pi r^2_+(r_+-r_-)}{3r_- - r_+ +\frac{r^2_+}{L^2_\lambda}(r_++r_-)}, \label{heat_capa}\\
	\mathcal{G}_\text{R} 
	&=&M - T_\text{R}S_\text{R}%\nonumber\\
	=\frac{r_++r_-}{2}-\frac{(r_+-r_-)(r_+^2+L_\lambda^2)}{4r_+^2}\ln\left(1+\frac{r_+^2}{L_\lambda^2}\right).
\end{eqnarray}
One can check that, in the limit $L_\lambda\to\infty$ (or $\lambda\to0$), $T_\text{R}, C_\text{R}$ and $\mathcal{G}_\text{R}$ are reduced to $T_\text{H}, C_q$ and $\mathcal{G}$ for the GB statistics case, respectively.

As mentioned previously, the RN-AF via R\'enyi statistics has a critical behavior. In other words, there exist the nontrivial solutions for $q$ and $\phi$ of the conditions determining the critical point,
\begin{eqnarray}
	\left(\frac{\partial q}{\partial \phi}\right)_{T_\text{R}}
	=\bigg(\frac{\partial^2 q}{\partial \phi^2}\bigg)_{T_\text{R}}=0.\label{critical cond}
\end{eqnarray}
Using Eqs.~\eqref{charge} \eqref{Tr} and \eqref{phi in q r+}, the R\'enyi temperature is written as
\begin{eqnarray}
	T_\text{R}=\frac{(1-\phi^2)\Big(\phi+\frac{q^2}{\phi L_\lambda^2}\Big)}{4\pi q}.\label{TR in q phi}
\end{eqnarray}
From the above equation of state, the conditions \eqref{critical cond} can be solved for the charge, potential and temperature at critical point which are, respectively, obtained in the following:
\begin{eqnarray}
	q_c^2&=&L_\lambda^2\big(7-4\sqrt{3}\,\big),\\
	\phi_c^2&=&\frac{2 - \sqrt{3}}{\sqrt{3}},\\
	T_c&=&T_\text{R}\big|_{q=q_c, \phi=\phi_c}=\frac{2}{3\pi L_\lambda}\sqrt{2\sqrt{3}-3}.
\end{eqnarray}

To investigate the phase structure of RN-AF via R\'enyi statistics, our analysis can be simplified by rescaling the parameters in the system as follows: 
\begin{eqnarray}
	q \rightarrow \bar{q}=\frac{q}{L_\lambda},\hspace{1cm}
	\phi \rightarrow \bar{\phi}=\phi,\hspace{1cm}
	r_\pm \rightarrow \bar{r}_\pm=\frac{r_\pm}{L_\lambda}.\label{rescale}
\end{eqnarray}
\begin{figure*}[!ht]
	\begin{tabular}{c c}
		\includegraphics[scale=0.493]{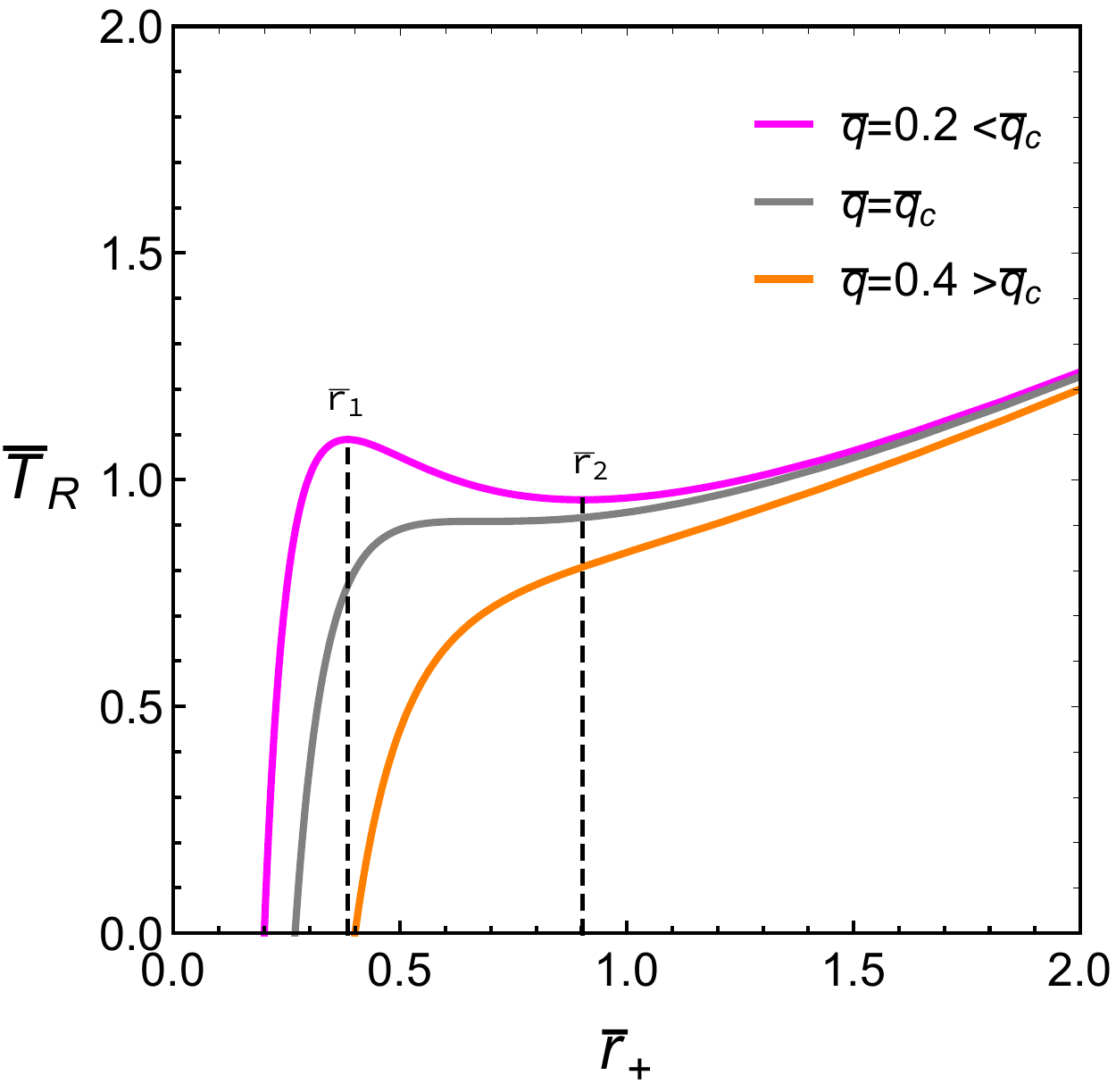} \hspace{1cm} 
		\includegraphics[scale=0.5]{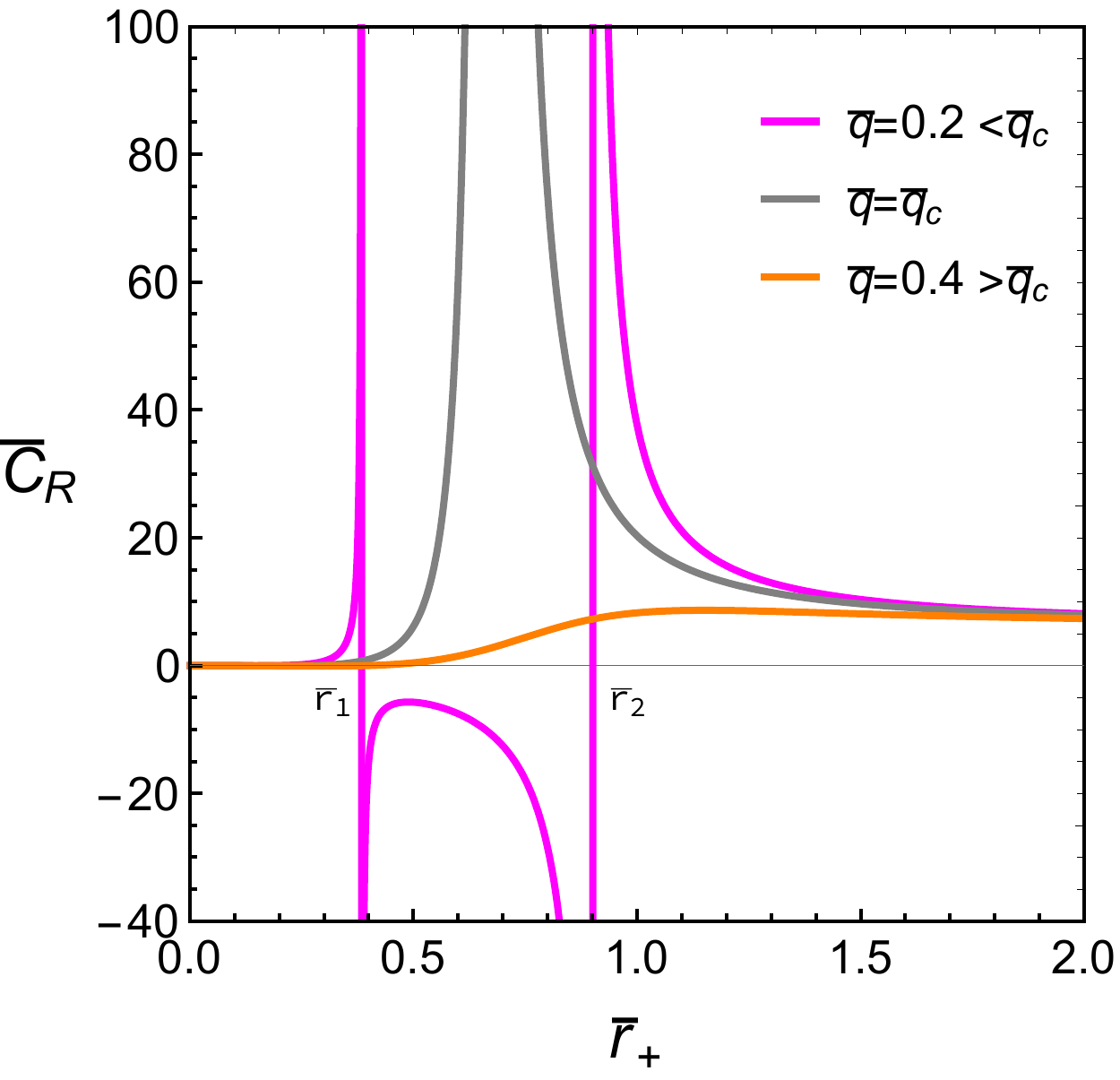}
	\end{tabular}
	\caption{The rescaled R\'enyi temperature $\bar{T}_\text{R}$ (left) and heat capacity $\bar{C}_\text{R}$ (right) of the RN-AF via R\'enyi statistics with respect to the rescaled outer horizon radius $\bar{r}_+$ for various values of the rescaled charge $\bar{q}$. The radii $\bar{r}_1$ and $\bar{r}_2$ in these plots are the values only for the case $\bar{q}=0.2$.}\label{fig:T Cq in Renyi vary q}
\end{figure*}
The R\'enyi temperature, heat capacity, compressibility and Gibbs free energy can be rescaled as
\begin{eqnarray}
	T_\text{R} &\rightarrow& \bar{T}_\text{R}=2\pi L_\lambda T_\text{R},\hspace{1cm}
	C_\text{R} \rightarrow \bar{C}_\text{R}=\frac{1}{L_\lambda^2}C_\text{R},\nonumber\\
	\kappa_{T_\text{R}} &\rightarrow& \bar{\kappa}_{T_\text{R}}=L_\lambda\kappa_{T_\text{R}},\hspace{1.2cm}
	\mathcal{G}_\text{R} \rightarrow \bar{\mathcal{G}}_\text{R}=\frac{1}{L_\lambda}\mathcal{G}_\text{R},\label{rescale S,T,C,G}
\end{eqnarray}
respectively. Note that, for the R\'enyi case, we do not consider the dimensionless quantities by treating the charge $q$ as a scaling parameter, as done in the GB case, since the system now has a fundamental length scale $L_\lambda$ due to the existence of the nonextensivity. Therefore it is more appropriate to measure the thermodynamic quantities with respect to this length scale $L_\lambda$ instead of $q$. The quantities at the critical point now become
\begin{eqnarray}
	\bar{q}_c&=&\sqrt{7-4\sqrt{3}} \approx 0.268, \\
	\bar{\phi}_c&=&\sqrt{\frac{2 - \sqrt{3}}{\sqrt{3}}} \approx 0.393,\\
	\bar{T}_c&=&\frac{4}{3}\sqrt{2\sqrt{3}-3} \approx 0.908.
\end{eqnarray}

Let us consider the local stability in the aspect of the heat capacity first. The behaviors of the rescaled temperature and heat capacity are illustrated in Fig.~\ref{fig:T Cq in Renyi vary q}. In these plots, we present the quantities with three different values of $\bar{q}$ which are $\bar{q}=0.2<\bar{q}_c$ (the magenta curves), $\bar{q}=\bar{q}_c$ (the gray curves), and $\bar{q}=0.4>\bar{q}_c$ (the orange curves). By considering its first and second derivatives with respect to $\bar{r}_+$, it is seen that, for $\bar{q}<\bar{q}_c$, there exist two local extrema in $\bar{T}_\text{R}$, where there is no finite maximum in this case, corresponding to the divergence of $\bar{C}_\text{R}$ occuring at $\bar{r}_+=\frac{1}{\sqrt{2}}\Big(1-\bar{q}^2\mp\sqrt{1-14\bar{q}^2+\bar{q}^4}\Big)^{1/2}\equiv\bar{r}_{1,2}$. The temperature reaches the local maximum and local minimum at $\bar{r}_1$ and $\bar{r}_2$, respectively. The three possible configurations exist due to the appearance of the nonextensivity. Two of them are the phases with positive heat capacity denoted as pBH1 (for the small black hole branch, $\bar{r}_e < \bar{r}_+ < \bar{r}_1$) and pBH2 (for the large black hole branch, $\bar{r}_+ > \bar{r}_2$). Another phase is the intermediate black hole branch, $\bar{r}_1 < \bar{r}_+ < \bar{r}_2$ with negative heat capacity denoted as nBH. Moreover, there are two second-order phase transitions which are the nBH-pBH1, and nBH-pBH2 transitions occurring at $\bar{r}_1$ and $\bar{r}_2$, respectively. Interestingly, the pBH2 phase does not appear in the standard GB statistics. Therefore, it could be an emergent phase within the R\'enyi statistics. One has seen that the nBH phase, which has negative slope in $\bar{T}_\text{R}$ or negative value in $\bar{C}_\text{R}$, disappears when $\bar{q}\geq\bar{q}_c$, since the radii $\bar{r}_1$ and $\bar{r}_2$ are identical when $\bar{q}=\bar{q}_c$ and then undefined when $\bar{q}>\bar{q}_c$. In other words, the pBH1 and pBH2 phases are degenerate for $\bar{q}>\bar{q}_c$. Hence, the phase structure is clearly dependent on the values of $\bar{q}$. This result is a key difference between the properties of the RN-AF described by the GB and R\'enyi statistics.

\begin{figure*}[!ht]
	\begin{tabular}{c c}
		\includegraphics[scale=0.4]{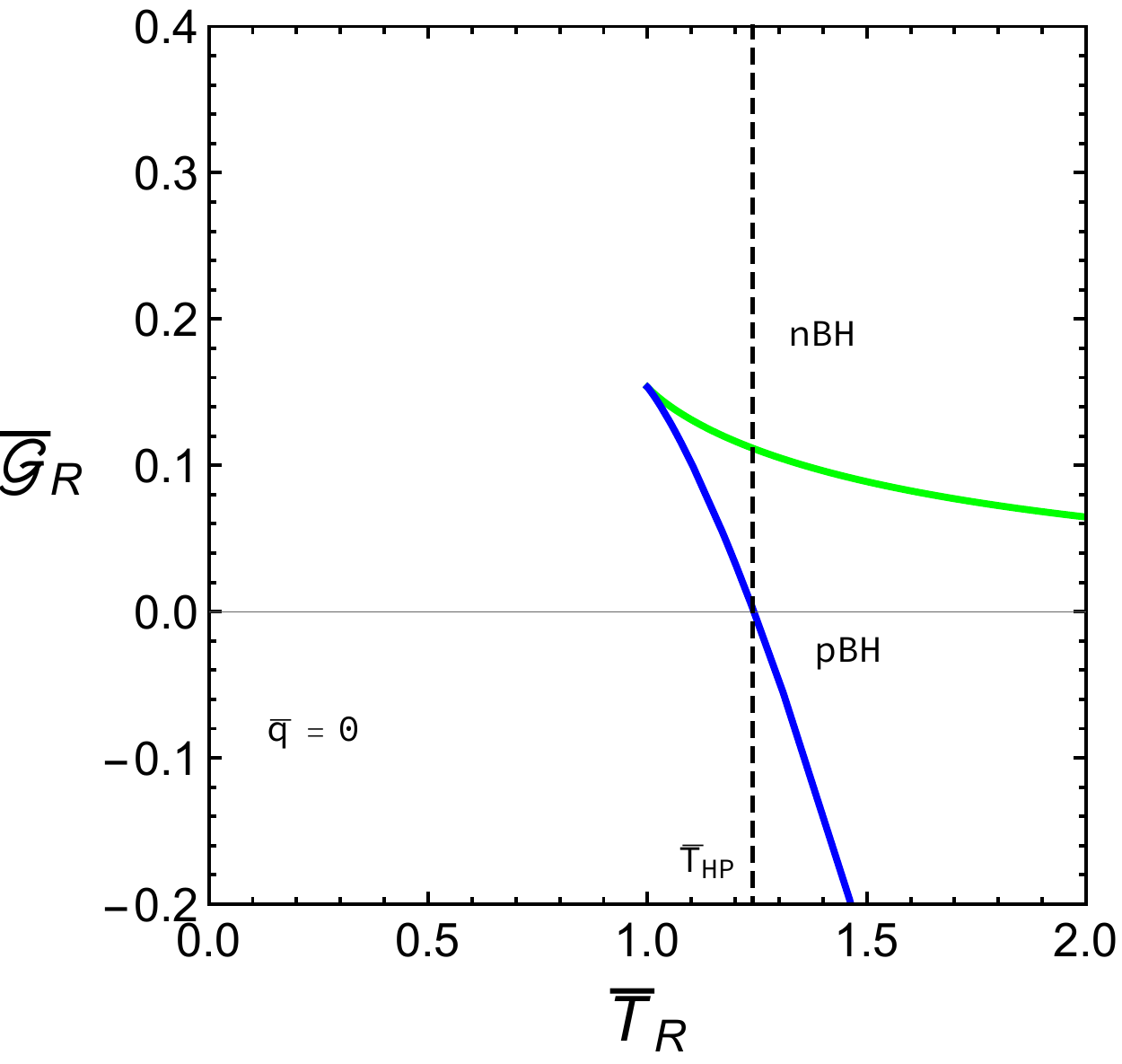}\hspace{1.cm}
		\includegraphics[scale=0.4]{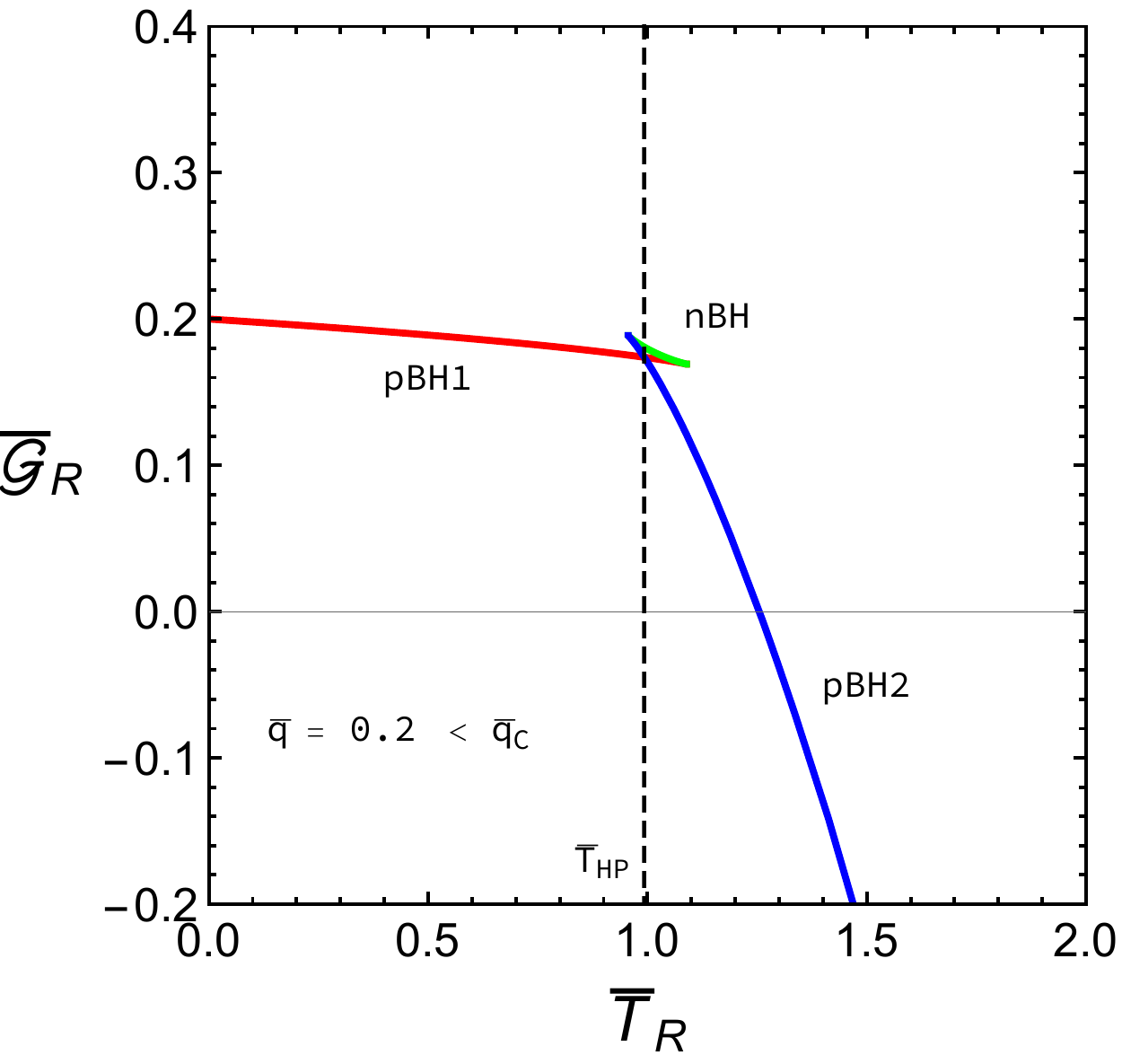}\\
		\includegraphics[scale=0.4]{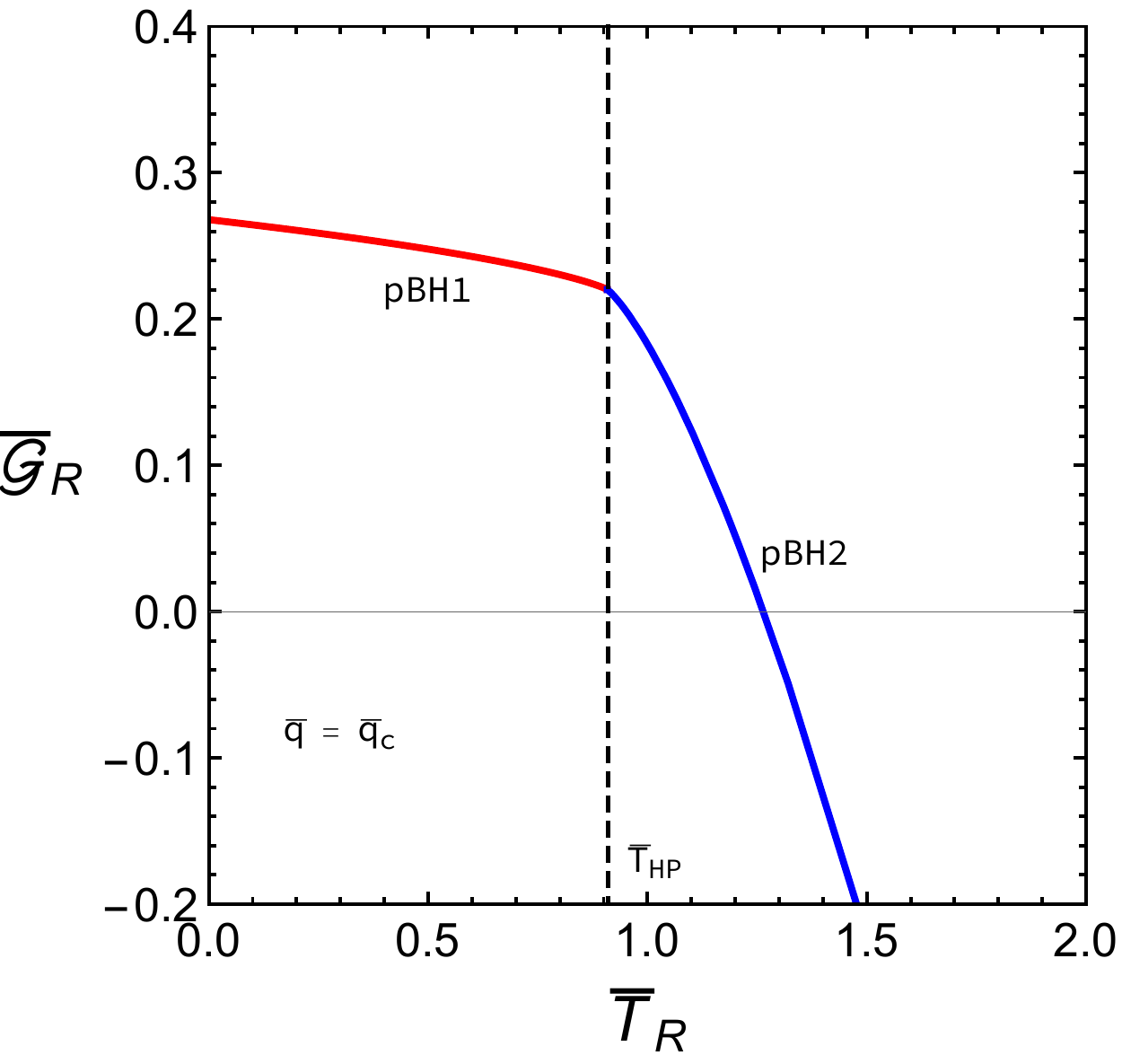}\hspace{1.cm}
		\includegraphics[scale=0.4]{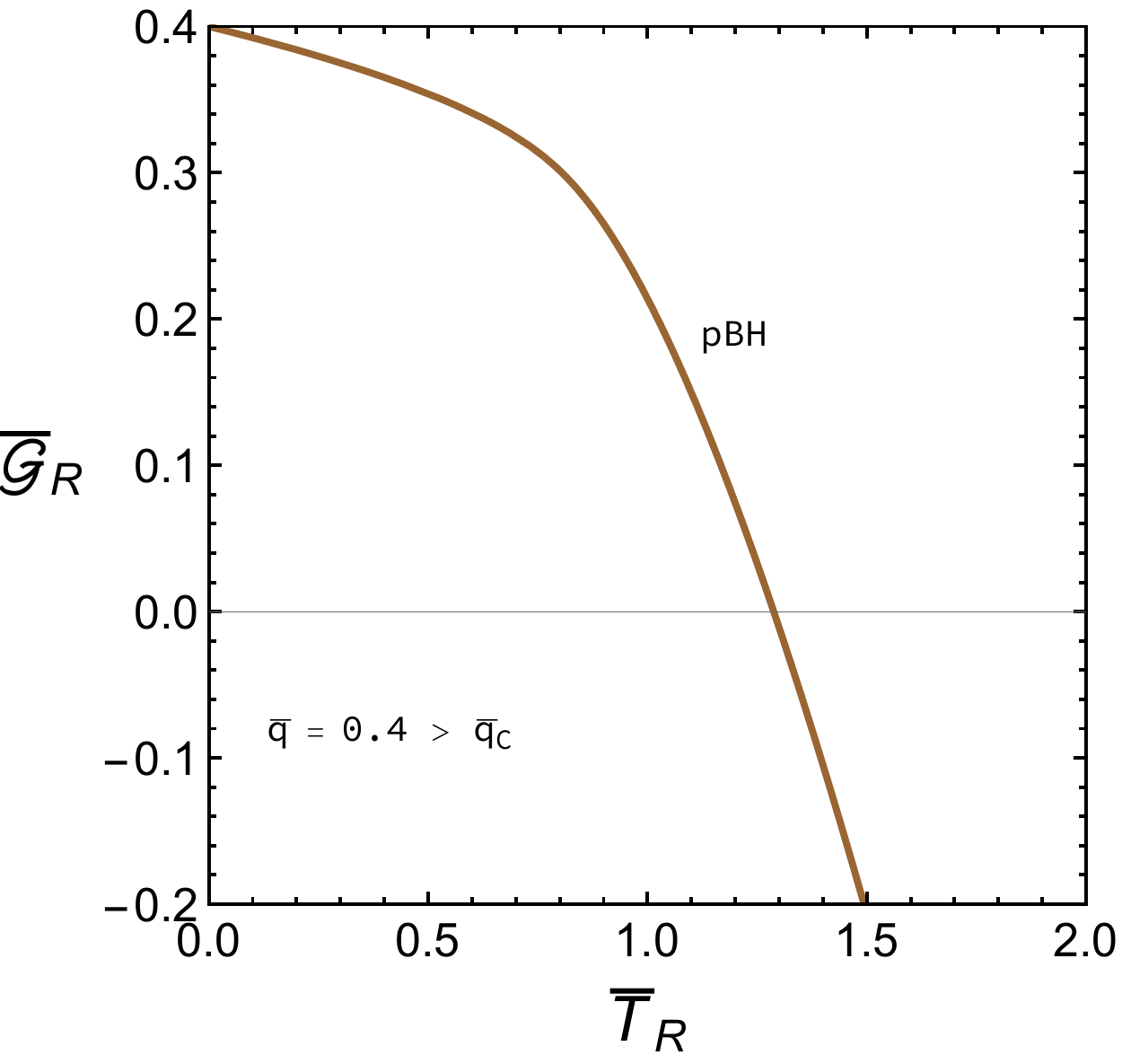}
	\end{tabular}
	\caption{The profiles of the free energy $\bar{\mathcal{G}}_\text{R}$ with respect to the temperature $\bar{T}_\text{R}$ for various fixed values of charge $\bar{q}$. Note that each of these with fixed value of charge corresponds to an isobaric process.}\label{fig:G in Renyi vary q}
\end{figure*}

\begin{figure*}[!ht]
	\begin{tabular}{c c}
		\includegraphics[scale=0.5]{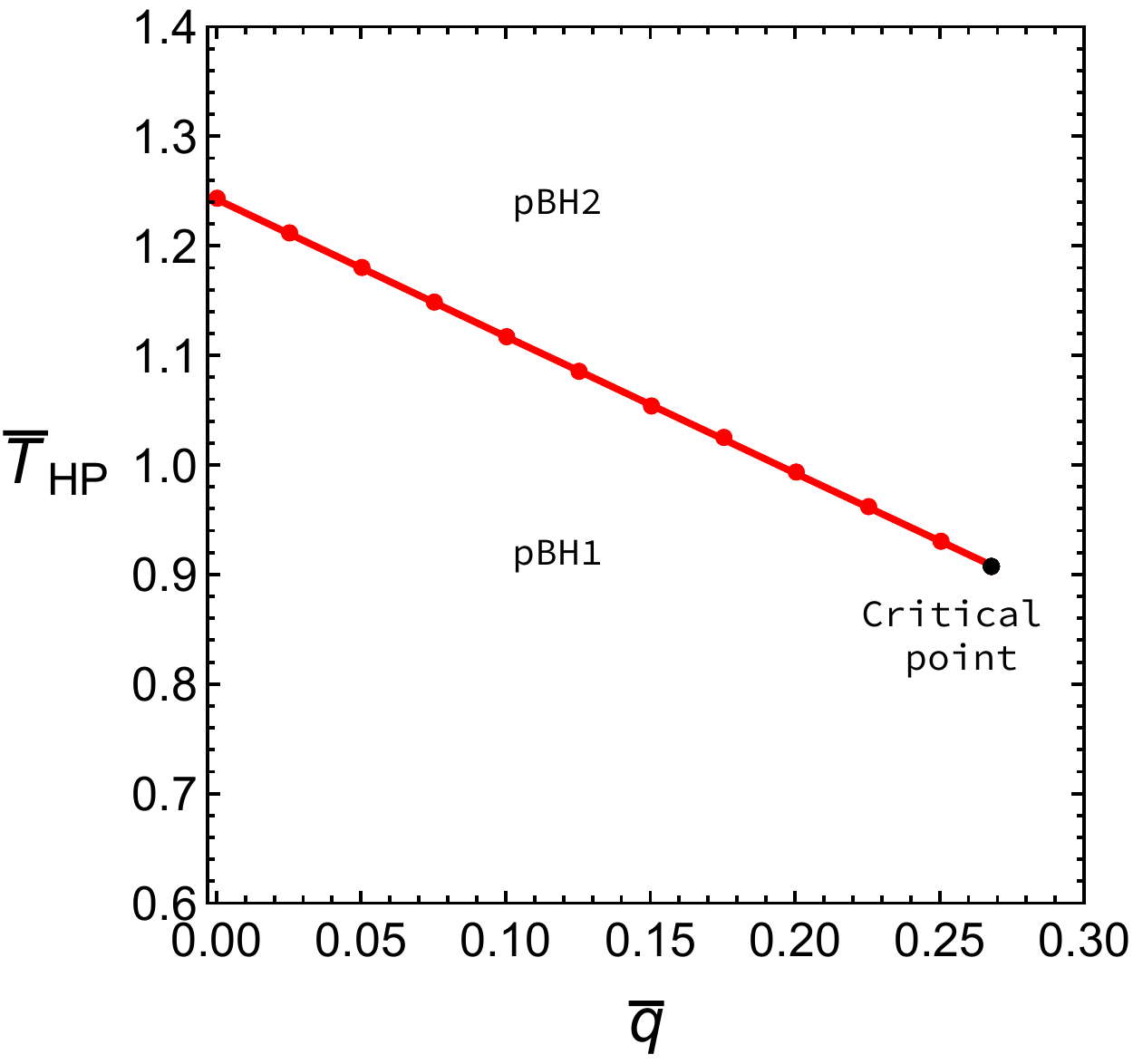}
	\end{tabular}
	\caption{The profile of the Hawking-Page temperature $\bar{T}_\text{HP}$ with respect to the charge $\bar{q}$.}\label{fig:TR-HP vs q}
\end{figure*}

For the global stability, the Gibbs free energy profiles with different $\bar{q}$ are illustrated in Fig.~\ref{fig:G in Renyi vary q}. Due to the appearance of the emergent pBH2 phase, there exists another cusp corresponding to $\bar{r}_2$ for the case $\bar{q}<\bar{q}_c$. The upper right panel in Fig.~\ref{fig:G in Renyi vary q} is actually shown the swallowtail behavior which is obviously different from the black hole via GB statistics. Moreover, the first-order phase transition from pBH1 to pBH2 interestingly occurs for the R\'enyi case with $\bar{q}\leq \bar{q}_c$. This transition is actually called the Hawking-Page phase transition (the temperature at this transition is, then, called the Hawking-Page temperature $\bar{T}_\text{HP}$) for the charged case. On the other hand, the uncharged black hole, \textit{i.e.}, the Schwarzschild black hole, also has the Hawking-Page transition which is the transition from the thermal radiation phase with zero free energy to the pBH phase as shown in the upper left panel in Fig.~\ref{fig:G in Renyi vary q}. It is noted that, for $0<\bar{q}\leq \bar{q}_c$, $\bar{T}_\text{HP}$ decreases as the charge increasing. In addition, the Hawking-Page temperature is maximized and minimized for the black hole with $\bar{q}=0$ and $\bar{q}=\bar{q}_c$, respectively, as shown in Fig.~\ref{fig:TR-HP vs q}.

We now move to analyze the mechanical stability by comparing to the thermal one. The rescaled heat capacity in Eq.~\eqref{rescale S,T,C,G} can be expressed in terms of $\bar{q}$ and $\bar{\phi}$ as
\begin{eqnarray}
	\bar{C}_\text{R}
	=2\pi\bar{q}^2\left[\frac{(1-\bar{\phi}^2)}{\bar{q}^2(1+\bar{\phi}^2)-\bar{\phi}^2(1-3\bar{\phi}^2)}\right].\label{CR in q phi} 
\end{eqnarray}
It is seen that the sign of this heat capacity depends directly on the sign of the denominator because its numerator is always positive for the whole range $0<\bar{\phi}<1$. For fixing $\bar{q}$, the denominator vanishes ($\bar{C}_\text{R}\rightarrow \infty$) at
\begin{eqnarray}
	\bar{\phi}_{\pm}=\sqrt{\frac{1}{6}\left(1-\bar{q}^2\pm\sqrt{1-14\bar{q}^2+\bar{q}^4}\,\right)}.\label{phi}
\end{eqnarray}
Straightforwardly, when $\bar{q}=\bar{q}_c=2-\sqrt{3}$, we found that the two solutions \eqref{phi} are degenerate, $\bar{\phi}_+=\bar{\phi}_-=\bar{\phi}_c\approx 0.393$. For the large charge case $\bar{q} > \bar{q}_c$, therefore, the range of negative value of the denominator in Eq.~\eqref{CR in q phi}  disappears, and hence $\bar{C}_\text{R}$ is always positive. For the small charge case $\bar{q}<\bar{q}_c$, there are three branches of charge black hole which are $0<\bar{\phi}<\bar{\phi}_-$ (positive $\bar{C}_\text{R}$), $\bar{\phi}_-<\bar{\phi}<\bar{\phi}_+$ (negative $\bar{C}_\text{R}$), and $\bar{\phi}_+<\bar{\phi}<1$ (positive $\bar{C}_\text{R}$). This behavior is equivalent to the phase structure as illustrated in the right panel in Fig.~\ref{fig:T Cq in Renyi vary q}. We also conclude that, for $\bar{q}<\bar{q}_c$, the regions $0<\bar{\phi}<\bar{\phi}_-$, $\bar{\phi}_-<\bar{\phi}<\bar{\phi}_+$ and $\bar{\phi}_+<\bar{\phi}<1$ are the regions for the phases of pBH2, nBH and pBH1, respectively.

\begin{figure*}[!ht]
	\begin{tabular}{c}
		\includegraphics[scale=0.36]{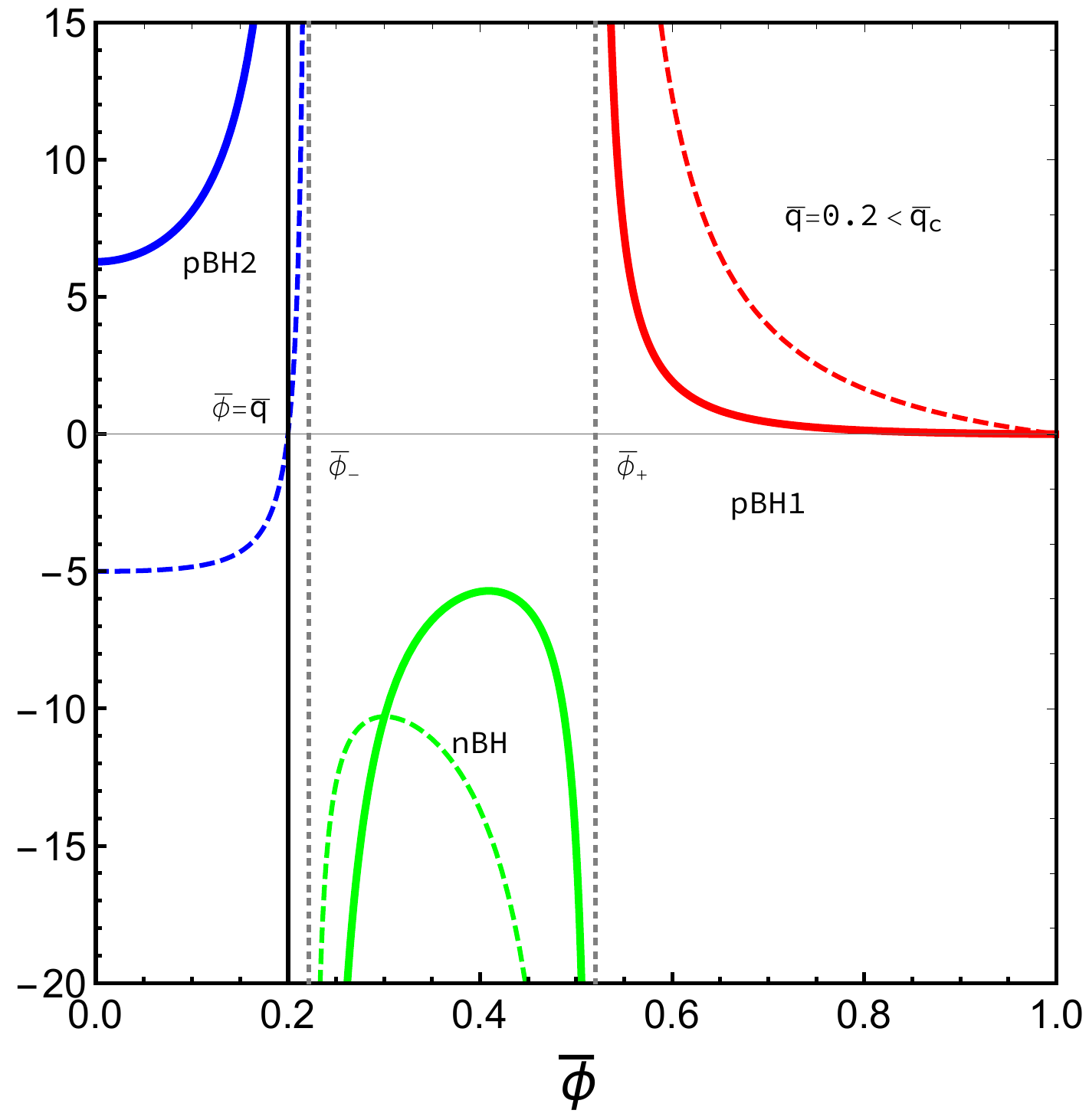}\hspace{.3cm}
		\includegraphics[scale=0.41]{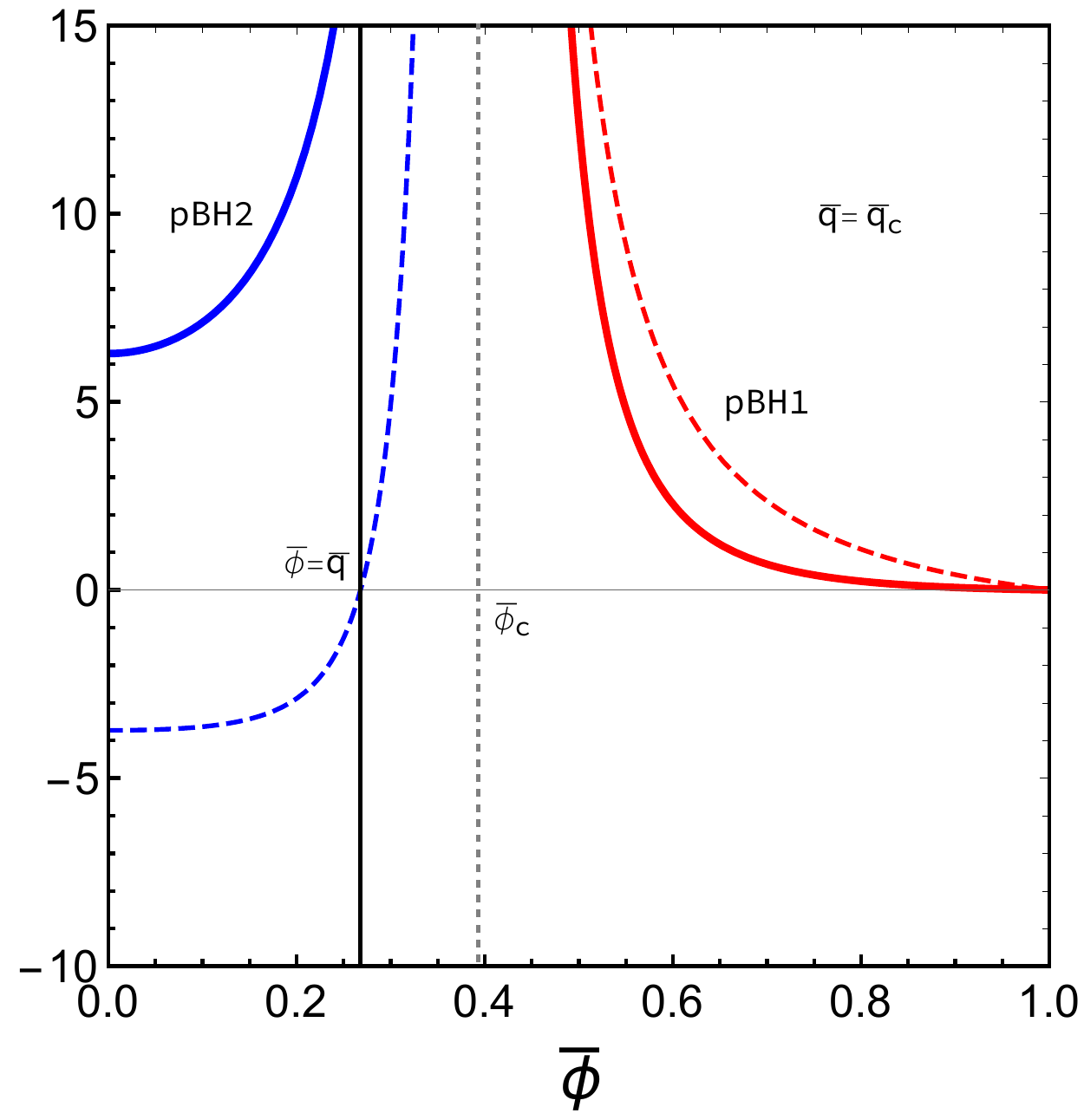}\hspace{.3cm}
		\includegraphics[scale=0.4]{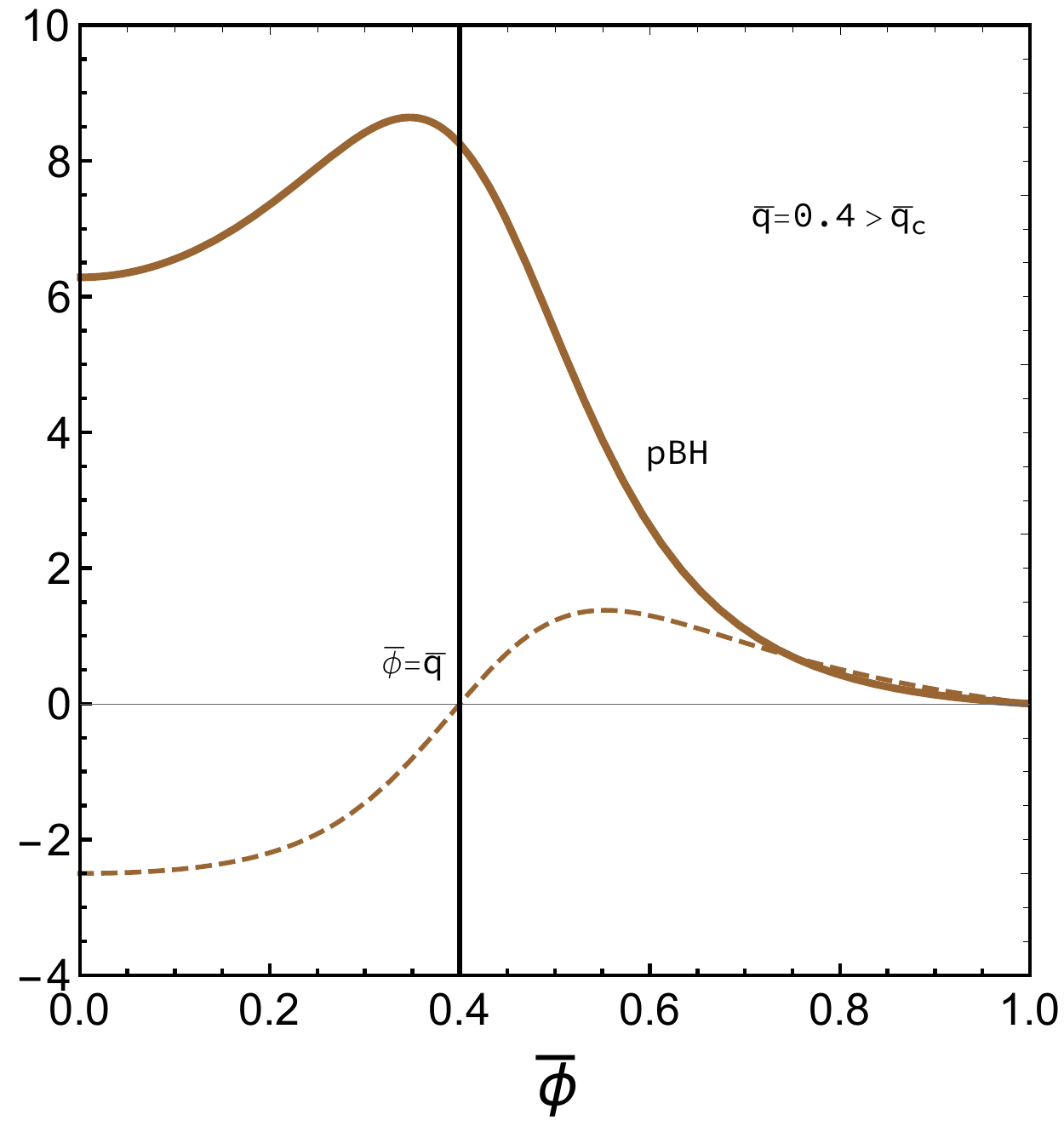}
	\end{tabular}
	\caption{The profiles of the heat capacity $\bar{C}_\text{R}$ (solid lines) and the compressibility $\bar{\kappa}_{T_\text{R}}$ (dashed lines) with respect to the electrostatic potential $\bar{\phi}$ for various values of $\bar{q}$. For the left and middle panels, the blue, green and red lines are the quantities for pBH2, nBH and pBH1, respectively. Note that the values of $\bar{\phi}_\pm$ is for the case $\bar{q}=0.2$. For the right panel, the quantities are shown only for the pBH phase.}\label{fig: C kappaT Renyi}
\end{figure*}

Using the rescaled version of Eq.~\eqref{TR in q phi}, one can compute the rescaled compressibility for the RN-AF in the R\'enyi case as follows:
\begin{eqnarray}
	\bar{\kappa}_{T_\text{R}}
	&=&-\frac{L_\lambda}{\phi}\left(\frac{\partial \phi}{\partial q}\right)_{T_\text{R}}%,\nonumber\\
	=\frac{(\bar{\phi}^2-\bar{q}^2)}{\bar{q}}\left[\frac{(1-\bar{\phi}^2)}{\bar{q}^2(1+\bar{\phi}^2)-\bar{\phi}^2(1-3\bar{\phi}^2)}\right].\label{kappa TR}
\end{eqnarray}
%Similar to the GB case, the compressibility diverges at the same place as the heat capacity does, since they have the same denominators, \textit{i.e.}, $\bar{q}^2(1+\bar{\phi}^2)-\bar{\phi}^2(1-3\bar{\phi}^2)$. However, the signs of the compressibility in the R\'enyi case also depends on a term of its numerator, $\bar{\phi}^2-\bar{q}^2$. When fixing charge, $\bar{\kappa}_{T_\text{R}}$ has the same (opposite) sign to $\bar{C}_\text{R}$ if the potential is larger (small) than the charge. Hence, the black hole with positive heat capacity does not necessarily have positive compressibility. In other words, the black hole with small enough outer horizon radius (large enough $\bar{\phi}$) can be both thermally and mechanically stable. Let us split our consideration into three cases which are $\bar{q}<\bar{q}_c$, $\bar{q}=\bar{q}_c$ and $\bar{q}>\bar{q}_c$. For $\bar{q}<\bar{q}_c$, the range of being both thermally and mechanically stable is $\bar{q}<\bar{\phi}<1$. For $\bar{q}=\bar{q}_c$, \fixme{the nBH phase will disappear and both $\bar{C}_\text{R}$ and $\bar{\kappa}_{T_\text{R}}$} are divergent at $\bar{\phi}_c$. 
The compressibility diverges at the same place as the heat capacity does, since they have the same denominators, \textit{i.e.}, $\bar{q}^2(1+\bar{\phi}^2)-\bar{\phi}^2(1-3\bar{\phi}^2)$. For fixing charge, the behaviour of the above compressibility is then analyzed as three different cases which are $\bar{q}<\bar{q}_c$, $\bar{q}=\bar{q}_c$ and $\bar{q}>\bar{q}_c$ as illustrated in the left, middle and right panels in Fig. \ref{fig: C kappaT Renyi}, respectively. The general features of the compressibility can be listed as follows: (i) by comparing Eqs.~\eqref{CR in q phi} and \eqref{kappa TR}, it is noticed that the sign of the compressibility is opposite to that of the heat capacity when $\bar{\phi}<\bar{q}$, and (ii) the compressibility vanishes when $\bar{\phi}=\bar{q}$. For the small charge $\bar{q}<\bar{q}_c$ case, it is found that, by using the Taylor series expansion around $\bar{q}=0$, the potential $\bar{\phi}_-$ in Eq.~\eqref{phi} is approximated as
\begin{eqnarray}
	\bar{\phi}_-\approx \bar{q}+2\bar{q}^3+12\bar{q}^5+\mathcal{O}(\bar{q}^6).
\end{eqnarray}
This yields that $\bar{q}$ is always less than $\bar{\phi}_-$. Hence, $\bar{\kappa}_{T_\text{R}}$ of pBH1 $(\bar{\phi}_+<\bar{\phi}<1)$ and nBH $(\bar{\phi}_-<\bar{\phi}<\bar{\phi}_+)$ phases have the same sign as $C_\text{R}$. In other words, the pBH1 and nBH phases have the positive and negative compressibility, respectively. It is very important to note that the emergent pBH2 phase is locally stable for both $\bar{C}_\text{R}>0$ and $\bar{\kappa}_{T_\text{R}}>0$ within the range $\bar{q}>\bar{\phi}>\bar{\phi}_-$ as shown in the left panel in Fig. \ref{fig: C kappaT Renyi}. For the other two cases $\bar{q}\geq\bar{q}_c$, the black hole is mechanically stable and unstable when $\bar{\phi}>\bar{q}$ and $\bar{\phi}<\bar{q}$, respectively, as shown in the middle and right panel in Fig. \ref{fig: C kappaT Renyi}.

We have investigated the thermodynamical properties of the black hole with the effect of nonextensivity. The important differences between the system described by the GB and R\'enyi statistics are pointed out. Especially, the black hole can be thermodynamically stable using the R\'enyi statistics, while it is not possible using the GB statistics. As will be discussed in the next section, it is possible to construct the Maxwell equal area law for the charged black holes similar to vdW fluid.

%%%%%%%%%%%%%%%%%%%%%%%%%%%%%%%%%%%%%%%%%%%%%%%%%%%%%%%%%%%%%%%%%%%%%%%%%%%%%%%%%%%%%%%%%%%%%%%%%%%%%%%%%%%%%%%%%%%%%%%%%%%%%%%%%%%%%%%%%%%%%%%%%%%%%%%%%%%%%%%%%%%%%%%%%%%%%%%%%%%%%%%%%%%%%%%%%%%%%%%%%%%%%%%%%%%%%%%%%%%%%%%%%%%%%%%%%%%%%%%%%%%%%%%%%%%%%%%%%%%%%%%%%%%%%%%%%%%%%%%%%%%%%%%%%%%%%%%

\section{Maxwell equal area law}\label{Maxwell}

\begin{figure*}[!ht]
	\begin{tabular}{c c}
		\includegraphics[scale=0.50]{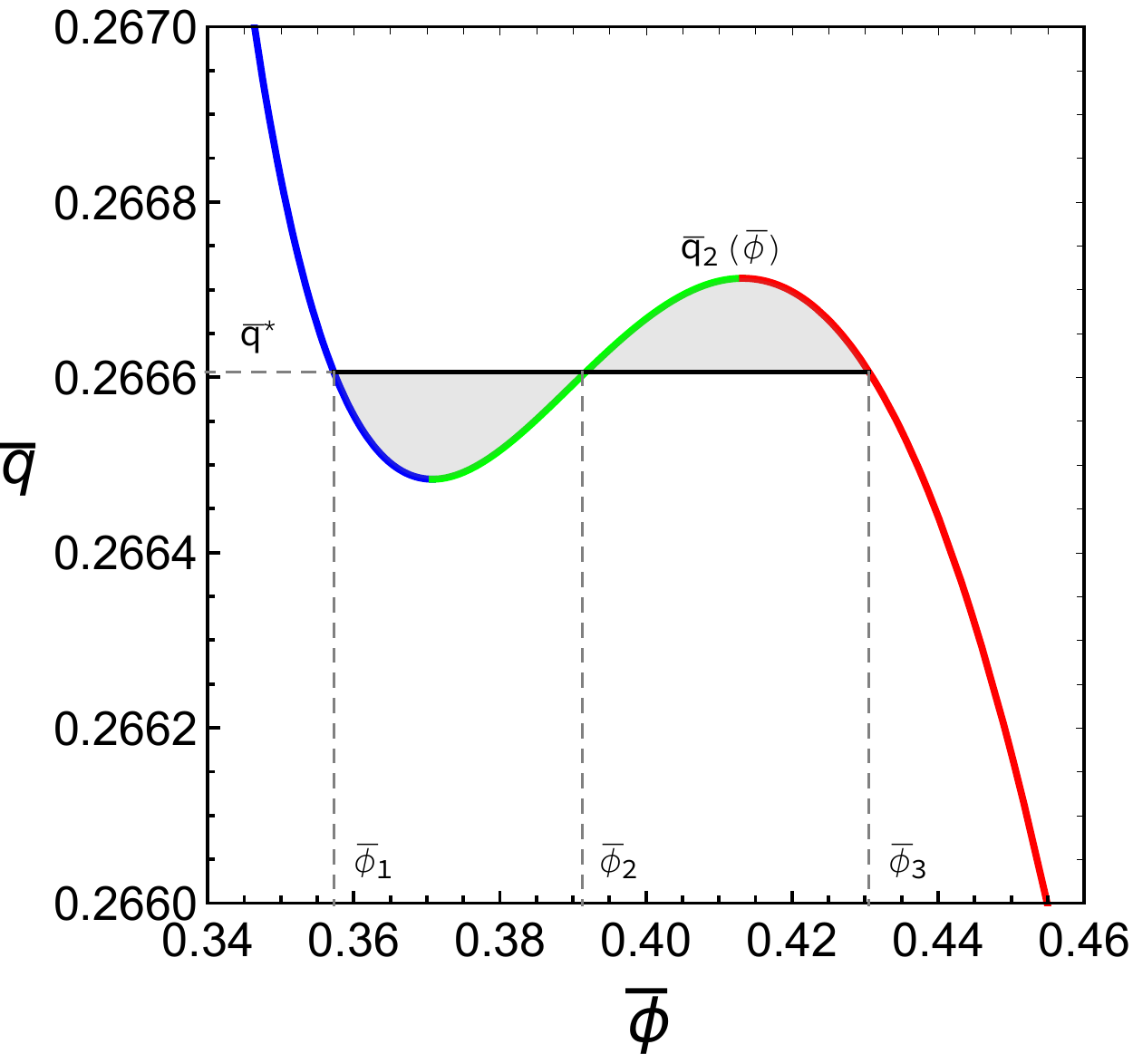}\hspace{1cm}
		\includegraphics[scale=0.47]{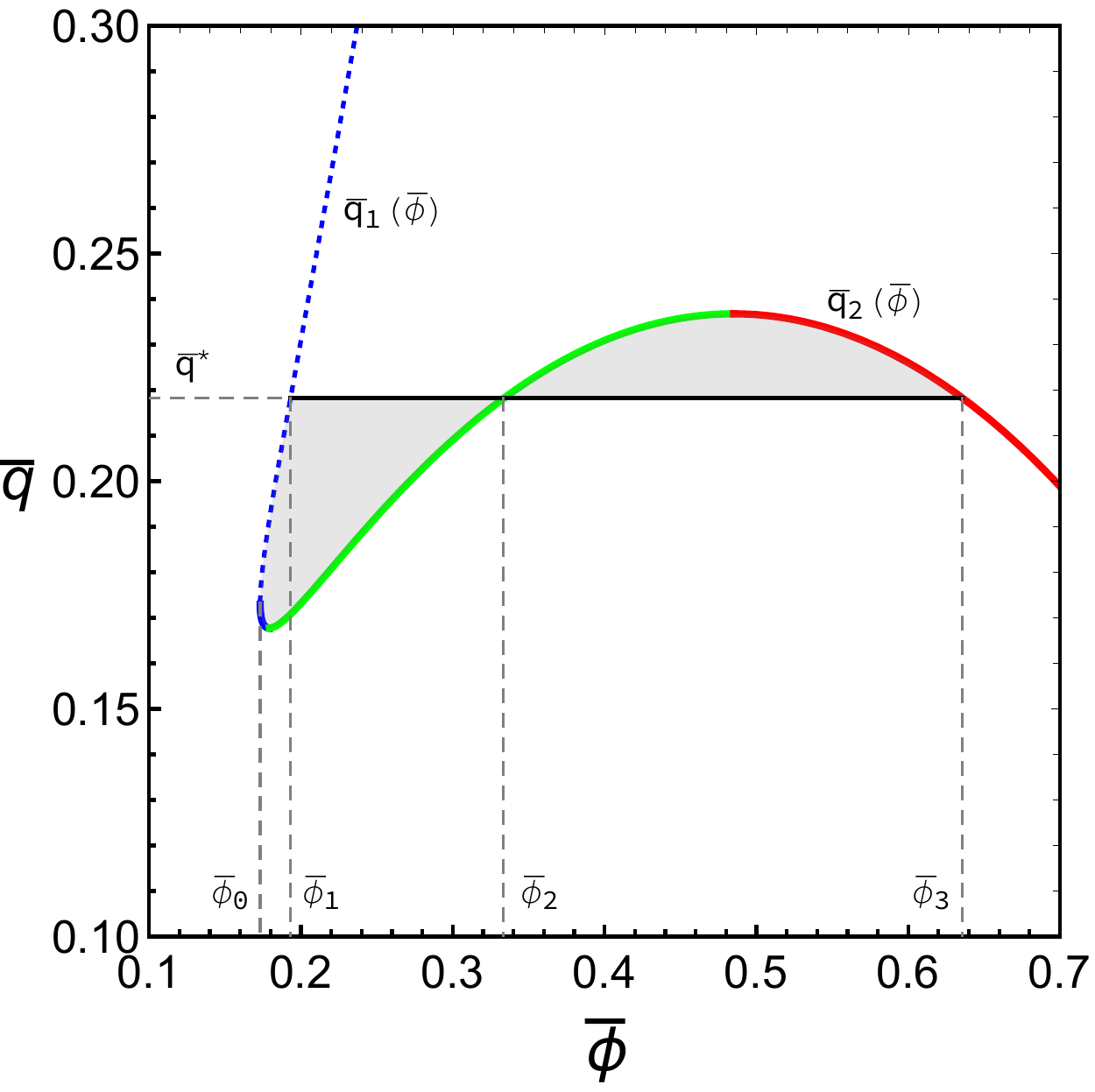}
	\end{tabular}
	\caption{Equal area law in the $\bar{q}-\bar{\phi}$ plane where the isothermal curves of $\bar{q}_1$ and $\bar{q}_2$ branches are represented as the dashed and solid curves, respectively. The blue, green and red lines denote the pBH2, nBH and pBH1 phases, respectively. Left: the case of near the critical point with fixed $\bar{T}_\text{R}=0.91$. Right: the case of far from the critical point with fixed $\bar{T}_\text{R}=0.97$. }\label{fig: Maxwell}
\end{figure*}

In this section, we are interested in finding the Maxwell equal area law on the $\bar{q}-\bar{\phi}$ plane (similar to the $P-V$ plane for the vdW fluid). By solving the rescaled version of Eq.~\eqref{TR in q phi} for $\bar{q}$, an equation of state $\bar{q}=\bar{q}(\bar{\phi}, \bar{T}_\text{R})$ is obtained in the form
\begin{eqnarray}
	\bar{q}=\frac{\bar{\phi}\bar{T}_\text{R}}{1-\bar{\phi}^2}\left(1\pm\sqrt{1-\left(\frac{1-\bar{\phi}^2}{\bar{T}_\text{R}}\right)^2}\,\right). \label{eof}
\end{eqnarray}
Let us define these two branches of $\bar{q}$ as follows:
\begin{eqnarray}
	\bar{q}_1&=&\frac{\bar{\phi}\bar{T}_\text{R}}{1-\bar{\phi}^2}\left(1+\sqrt{1-\left(\frac{1-\bar{\phi}^2}{\bar{T}_\text{R}}\right)^2}\,\right),\label{q1}\\
	\bar{q}_2&=&\frac{\bar{\phi}\bar{T}_\text{R}}{1-\bar{\phi}^2}\left(1-\sqrt{1-\left(\frac{1-\bar{\phi}^2}{\bar{T}_\text{R}}\right)^2}\,\right).\label{q2}
\end{eqnarray}
As discussed in Ref. \cite{Zhou:2019xai}, we have split this consideration into two cases, \textit{i.e.} the cases of near and far from the critical point, where the charge satisfying the Maxwell equal area law is denoted as $\bar{q}^\ast$. In the near critical point case $(\bar{T}_\text{R} \gtrsim \bar{T}_c)$, the line of charge $\bar{q}^\ast$ intersects the isothermal curve of only the $\bar{q}_2$ branch as shown in the left panel of Fig.~\ref{fig: Maxwell}. On the other hand, the line of charge $\bar{q}^\ast$ intersects the curve of both $\bar{q}_1$ branch (dashed curve) and $\bar{q}_2$ branch (solid curve) for the far from critical point case $(\bar{T}_\text{R} \gg \bar{T}_c)$ as illustrated in the right panel of Fig.~\ref{fig: Maxwell}.
It is found that the slope of the $\bar{q}_1$ branch for the case of far from the critical point is positive corresponding to the negative compressibility. Hence, this first-order phase transition is the transition between the mechanically unstable phase and the stable phase. This feature is actually called the novel equal area law \cite{Zhou:2019xai}.

\subsection{Near the critical point}

To satisfy the Maxwell equal area law, the line of constant charge $\bar{q}^\ast$ intersects the isothermal curve of $\bar{q}_2(\phi)$ branch at $\bar{\phi}_1$, $\bar{\phi}_2$ and $\bar{\phi}_3$ as shown in the left panel in Fig.~\ref{fig: Maxwell}. For this case, the condition for finding $\bar{q}^\ast$ is expressed as 
\begin{eqnarray}
	\bar{q}^\ast(\bar{\phi}_3-\bar{\phi}_1)&=&\int_{\bar{\phi}_1}^{\bar{\phi}_3}\bar{q}_2d\bar{\phi}.
\end{eqnarray}
As a result, one obtains
\begin{eqnarray}	
	\bar{q}^\ast&=&\frac{\bar{T}_\text{R}}{2(\bar{\phi}_3-\bar{\phi}_1)}
	\left[\sqrt{1-\left(\frac{1-\bar{\phi}^2}{\bar{T}_\text{R}}\right)^2}-\ln\left\{1+\sqrt{1-\left(\frac{1-\bar{\phi}^2}{\bar{T}_\text{R}}\right)^2}\,\right\}\right]_{\bar{\phi}_1}^{\bar{\phi}_3}.\label{qs0}
\end{eqnarray}
In other conditions in which the line of charge $\bar{q}^\ast$ also intersects the isothermal curve of the $\bar{q}_2$ branch at $\bar{\phi}_1$ and $\bar{\phi}_3$ that is $\bar{q}_2(\bar{\phi}_1)=\bar{q}_2(\bar{\phi}_3)=\bar{q}^\ast$, we obtain
\begin{eqnarray}
	\bar{q}^\ast&=&\frac{\bar{\phi}_1\bar{T}_\text{R}}{1-\bar{\phi}_1^2}\left[1-\sqrt{1-\left(\frac{1-\bar{\phi}_1^2}{\bar{T}_\text{R}}\right)^2}\,\right],\label{qs1}\\
	\bar{q}^\ast&=&\frac{\bar{\phi}_3\bar{T}_\text{R}}{1-\bar{\phi}_3^2}\left[1-\sqrt{1-\left(\frac{1-\bar{\phi}_3^2}{\bar{T}_\text{R}}\right)^2}\,\right].\label{qs2}
\end{eqnarray} 
It is obvious that there are three unknown variables with three equations. We consider $[\text{Eq.}~\eqref{qs1}]=[\text{Eq.}~\eqref{qs2}]$ to eliminate $\bar{\phi}_1$, and then, determine $\bar{\phi}_3$ from $2[\text{Eq.}~\eqref{qs0}]=\text{Eq.}~\eqref{qs1}]+[\text{Eq.}~\eqref{qs2}]$. Unfortunately, it is complicated to solve $\bar{\phi}_3$ analytically, since we have to deal with the terms of logarithmic function. Instead, we used the numerical calculation for each value of $\bar{T}_\text{R}$. The results for the case of near the critical point are shown in Table.~\ref{tab: near cp}
\begin{table}[!htbp]
\centering
\caption{The values of $\bar{q}^\ast$, $\bar{\phi}_1$ and $\bar{\phi}_3$ at various values of $\bar{T}_\text{R}$ for the case of  near critical point.}
\vspace{0.5cm}
\begin{tabular}{| c | c | c | c | }
\hline
\hspace{.5cm}$\bar{T}_\text{R}$\hspace{.5cm} & \hspace{1.cm}$\bar{q}^\ast$\hspace{1.cm} & \hspace{1.cm}$\bar{\phi}_1$\hspace{1.cm} & \hspace{1.cm}$\bar{\phi}_3$\hspace{1.cm} \\
\hline
0.91 & 0.266606 & 0.357255 & 0.430539 \\
0.92 & 0.258548 & 0.300521 & 0.494178 \\
0.93 & 0.250493 & 0.268988 & 0.532580 \\
0.94 & 0.242440 & 0.245060 & 0.563344 \\
\hline
\end{tabular}
\label{tab: near cp}
\end{table}

\subsection{Far from the critical point}

It is found that the minimum value of $\bar{\phi}$ (denoted as $\bar{\phi}_0$) is actually the value that the branches in Eqs.~\eqref{q1} and \eqref{q2} merge together. Therefore, it is given by
\begin{eqnarray}
%	1 - \left( \frac{1-\phi_0^2}{T_\text{R}} \right)^2=0,
%	\hspace{1cm}\to\hspace{1cm}
	\bar{\phi}_0=\sqrt{1-\bar{T}_\text{R}}.\label{phi0}
\end{eqnarray}
Obviously, the temperature $\bar{T}_\text{R}$ cannot be greater than one for the isothermal process. Therefore, the Maxwell construction of RN-AF on the $\bar{q}-\bar{\phi}$ plane with the R\'enyi model will make sense in the region $\bar{T}_\text{R}\leq 1$. Note that the bound in temperature does not hold for the system evolving in other types of process. As seen in the right panel in Fig.~\ref{fig: Maxwell}, the area satisfying the Maxwell equal area law for this case can be expressed as
\begin{eqnarray}
	\bar{q}^\ast(\bar{\phi}_3-\bar{\phi}_1)
	&=&-\int_{\bar{\phi}_0}^{\bar{\phi}_1}\bar{q}_1d\bar{\phi}+\int_{\bar{\phi}_0}^{\bar{\phi}_3}q_2d\bar{\phi},
\end{eqnarray}
where $\bar{\phi}_1$ is the value of $\bar{\phi}$ at which the line of $\bar{q}^\ast$ intersects the curve of  $\bar{q}_1(\bar{\phi})$ branch, and $\bar{\phi}_2$ and $\bar{\phi}_3$ are the values of $\bar{\phi}$ at which the line of $\bar{q}^\ast$ intersect the curve of $\bar{q}_2(\bar{\phi})$ branch. Eventually, one obtains
\begin{eqnarray}
	\bar{q}^\ast
	&=&\frac{\bar{T}_\text{R}}{2(\bar{\phi}_3-\bar{\phi}_1)}
	\Bigg[2\ln\left(\frac{1-\bar{\phi}_1^2}{\bar{T}_\text{R}}\right)+\sqrt{1-\left(\frac{1-\bar{\phi}_1^2}{\bar{T}_\text{R}}\right)^2}-\ln\Bigg\{1+\sqrt{1-\left(\frac{1-\bar{\phi}_1^2}{\bar{T}_\text{R}}\right)^2}\,\Bigg\}\nonumber\\
	&&\hspace{2.5cm}+\sqrt{1-\left(\frac{1-\bar{\phi}_3^2}{\bar{T}_\text{R}}\right)^2}-\ln\Bigg\{1+\sqrt{1-\left(\frac{1-\bar{\phi}_3^2}{\bar{T}_\text{R}}\right)^2}\,\Bigg\}\Bigg].\label{qs20}
\end{eqnarray}
The conditions from the line of $\bar{q}^\ast$ actually intersecting the isothermal curves of the $\bar{q}_1$ and $\bar{q}_2$ branches at $\bar{\phi}_1$ and $\bar{\phi}_3$, are, respectively written as
\begin{eqnarray}
	\bar{q}^\ast&=&\frac{\bar{\phi}_1\bar{T}_\text{R}}{1-\bar{\phi}_1^2}\left[1+\sqrt{1-\left(\frac{1-\bar{\phi}_1^2}{\bar{T}_\text{R}}\right)^2}\,\right],\label{qs21}\\
	\bar{q}^\ast&=&\frac{\bar{\phi}_3\bar{T}_\text{R}}{1-\bar{\phi}_3^2}\left[1-\sqrt{1-\left(\frac{1-\bar{\phi}_3^2}{\bar{T}_\text{R}}\right)^2}\,\right].\label{qs22}
\end{eqnarray}
Similar to the previous case, one can numerically determine the values of $\bar{\phi}_1$, $\bar{\phi}_3$ and $\bar{q}^\ast$ by using Eqs.~\eqref{qs20}$-$\eqref{qs22}. Eventually, the results are expressed in Table.~\ref{tab: far cp}.
\begin{table}[!htbp]
\centering
\caption{The values of $\bar{q}^\ast$, $\bar{\phi}_1$ and $\bar{\phi}_3$ at various values of $\bar{T}_\text{R}$ in the case of far from the critical point.}
\begin{tabular}{| c | c | c | c | }
\hline
\hspace{.5cm}$\bar{T}_\text{R}$\hspace{.5cm} & \hspace{1.cm}$\bar{q}^\ast$\hspace{1.cm} & \hspace{1.cm}$\bar{\phi}_1$\hspace{1.cm} & \hspace{1.cm}$\bar{\phi}_3$\hspace{1.cm} \\
\hline
0.95 & 0.234390 & 0.225272 & 0.589933 \\
0.96 & 0.226342 & 0.208192 & 0.613779 \\
0.97 & 0.218297 & 0.193062 & 0.635641 \\
0.98 & 0.210254 & 0.179423 & 0.655978 \\
0.99 & 0.202214 & 0.166972 & 0.675095 \\
1.00 & 0.194176 & 0.155495 & 0.693204 \\
\hline
\end{tabular}
\label{tab: far cp}
\end{table}

\begin{figure*}[!ht]
	\begin{tabular}{c c}
		\includegraphics[scale=0.5]{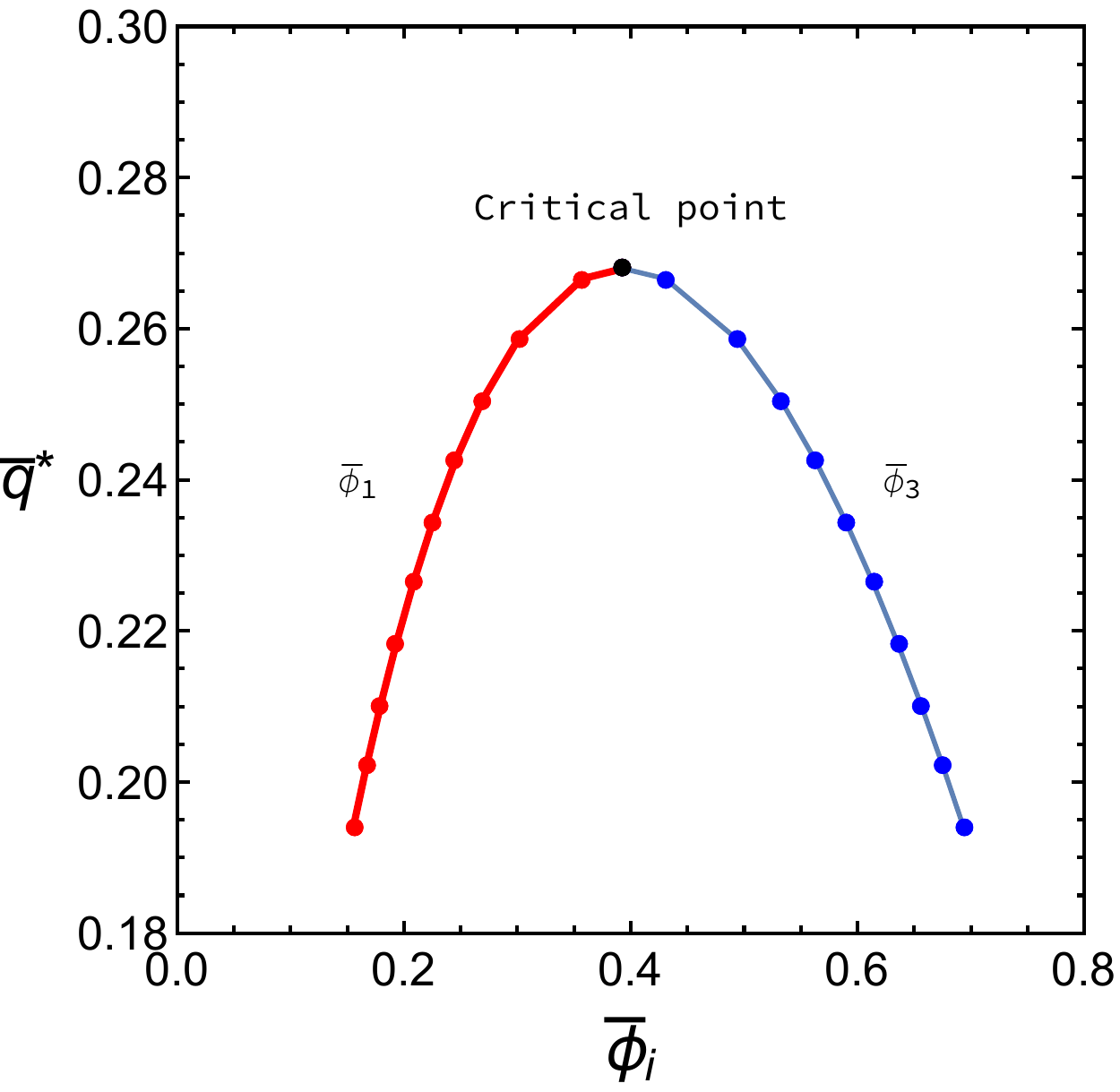}\hspace{1cm}
		\includegraphics[scale=0.48]{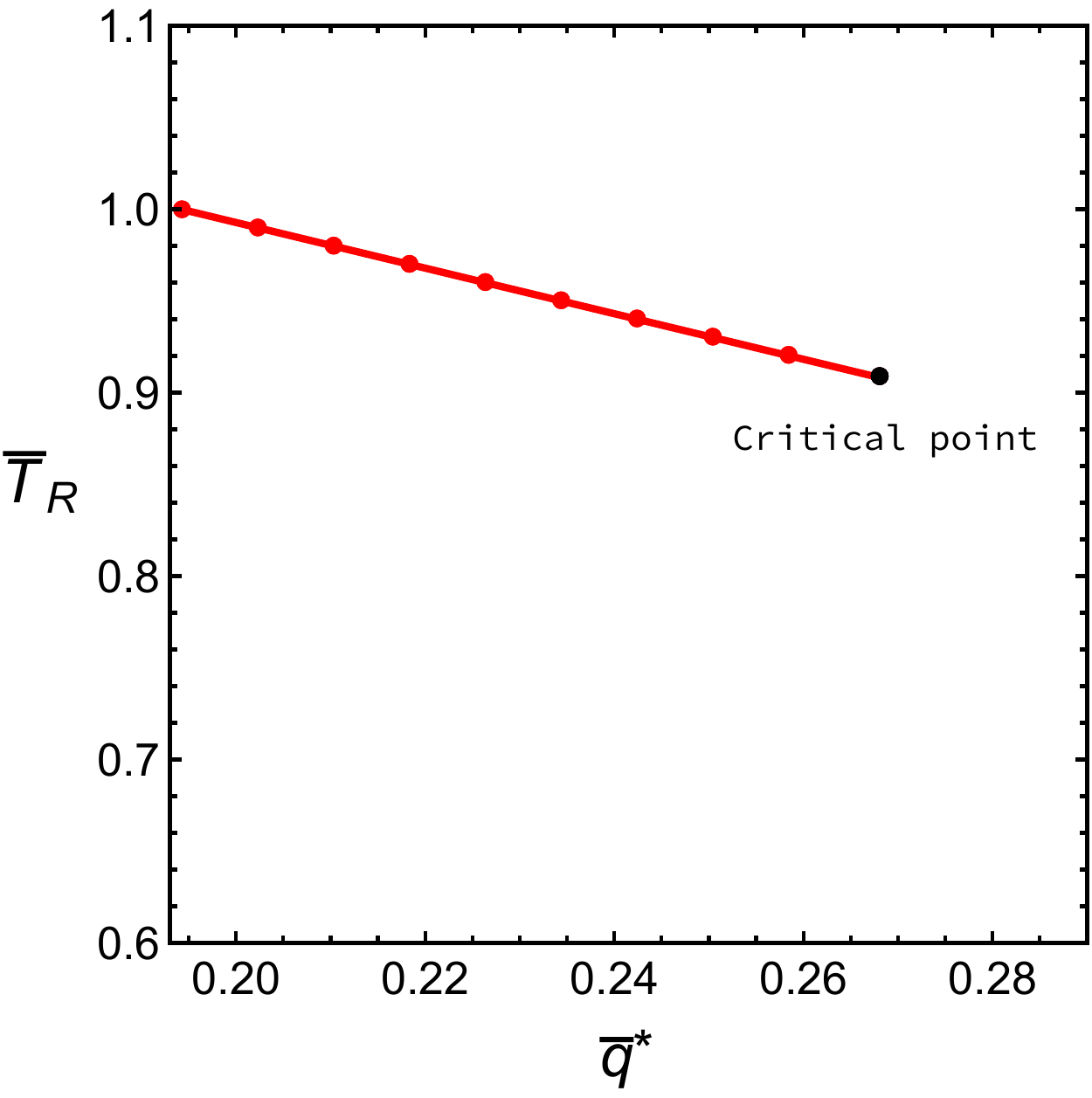}
	\end{tabular}
	\caption{Left: the behavior of $\bar{q}^\ast$ with respect to $\bar{\phi}_1$ (as a red line) and $\bar{\phi}_3$ (as a blue line) also called the Maxwell curve. Right: the R\'enyi temperature profile with respect to $q^\ast$.}\label{fig: qs}
\end{figure*}

Plotting in the $\bar{q}^\ast-\bar{\phi}$ plane, the points of $(\bar{\phi}_1, \bar{q}^\ast)$ and $(\bar{\phi}_3, \bar{q}^\ast)$ can be obtained from Tables \ref{tab: near cp} and \ref{tab: far cp}, where the line connecting between them can form a \textit{Maxwell curve} as shown in the left panel in Fig. \ref{fig: qs}. The lowest points are associated with the upper bound in temperature $\bar{T}_\text{R}=1$. At the critical point, the charge $\bar{q}^\ast$ approaches to the critical charge $\bar{q}_c$ while the potentials $\bar{\phi}_1$ and $\bar{\phi}_3$ merges together at the critical potential $\bar{\phi}_c$. It is also found that the charge $\bar{q}^\ast$ increases as the temperature decreases, and eventually ends at the critical values as shown in the right panel in Fig. \ref{fig: qs}.

\begin{figure*}[!ht]
	\begin{tabular}{c}
		\includegraphics[scale=0.15]{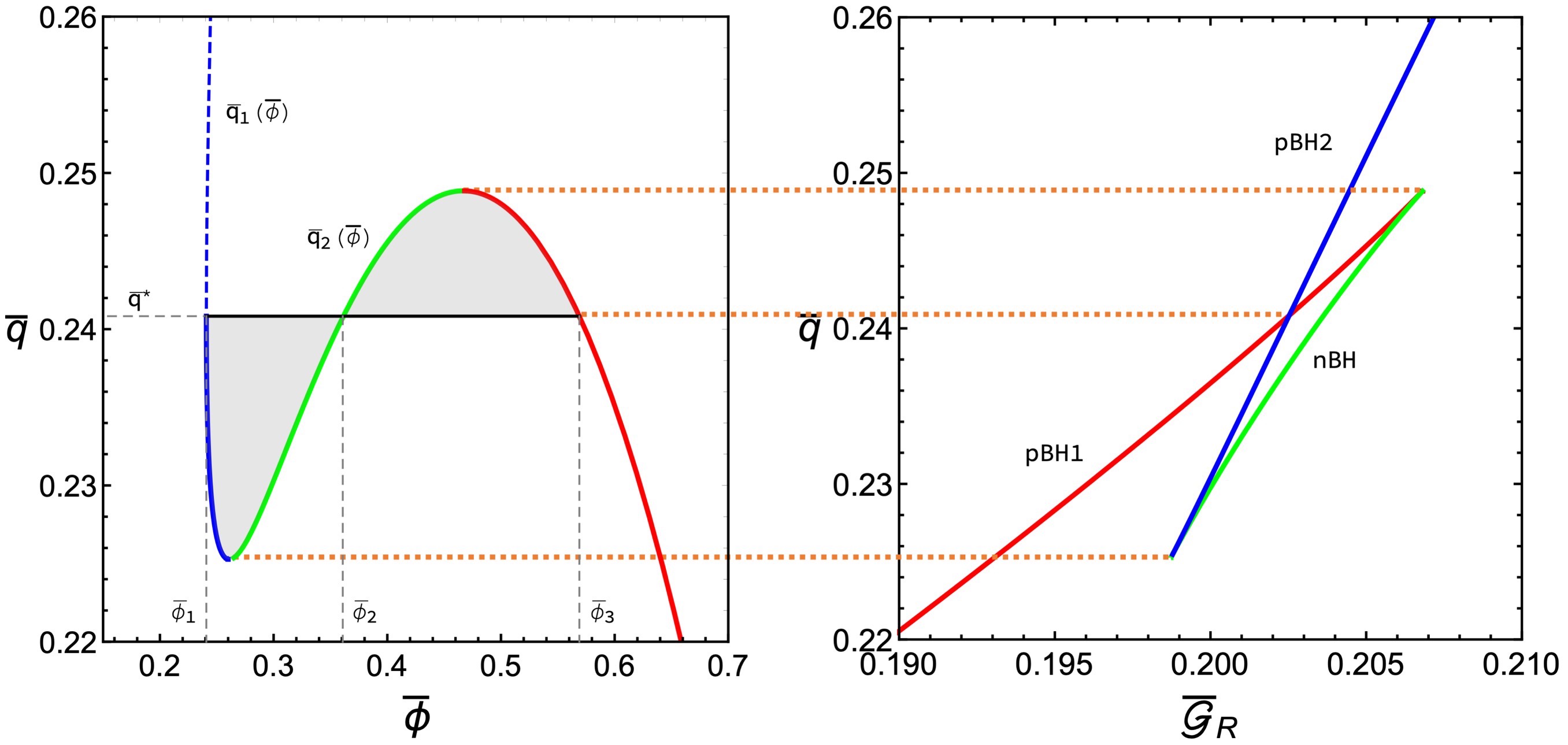}
	\end{tabular}
	\caption{The isothermal curves on the $\bar{q}-\bar{\phi}$ and $\bar{q}-\bar{\mathcal{G}}_\text{R}$ diagrams with $\bar{T}_\text{R}\approx0.942$.}\label{fig: Maxwell Tmax}
\end{figure*}

As mentioned that the first-order phase transition is not necessarily the transition locally stable phases for the charged black holes. In other words, the pBH2 phase with $\bar{q}^\ast$ and $\bar{\phi}_1$ at the transition point has negative compressibility if the temperature is not small enough. By numerical analysis, we have investigated the upper bound in temperature in which the first-order phase transition is the transition between mechanically stable pBH2 phase and pBH1 phase. As a result, this maximum temperature is numerically evaluated as
\begin{eqnarray}
    \bar{T}_\text{R}\approx0.942.\label{T max Maxwell}
\end{eqnarray} 
As seen in the Fig. \ref{fig: Maxwell Tmax}, the value $\bar{\phi}_1$ is indeed the point that the branches $\bar{q}_1$ and $\bar{q}_2$ merges, \textit{i.e.}, the point $\bar{\phi}_0$. Moreover, the figure also illustrates that the pBH1 and pBH2 phases with the charge $\bar{q}^\ast$ have the same Gibbs free energy (see right end point of the middle orange dotted line). The upper and lower orange dotted lines show that the local extrema on the $\bar{q}-\bar{\phi}$ plane correspond to the cusp in $\bar{q}-\bar{\mathcal{G}}_\text{R}$ plane in which the second-order phase transitions occur. 

\begin{figure*}[!ht]
	\begin{tabular}{c c}
		\includegraphics[scale=0.58]{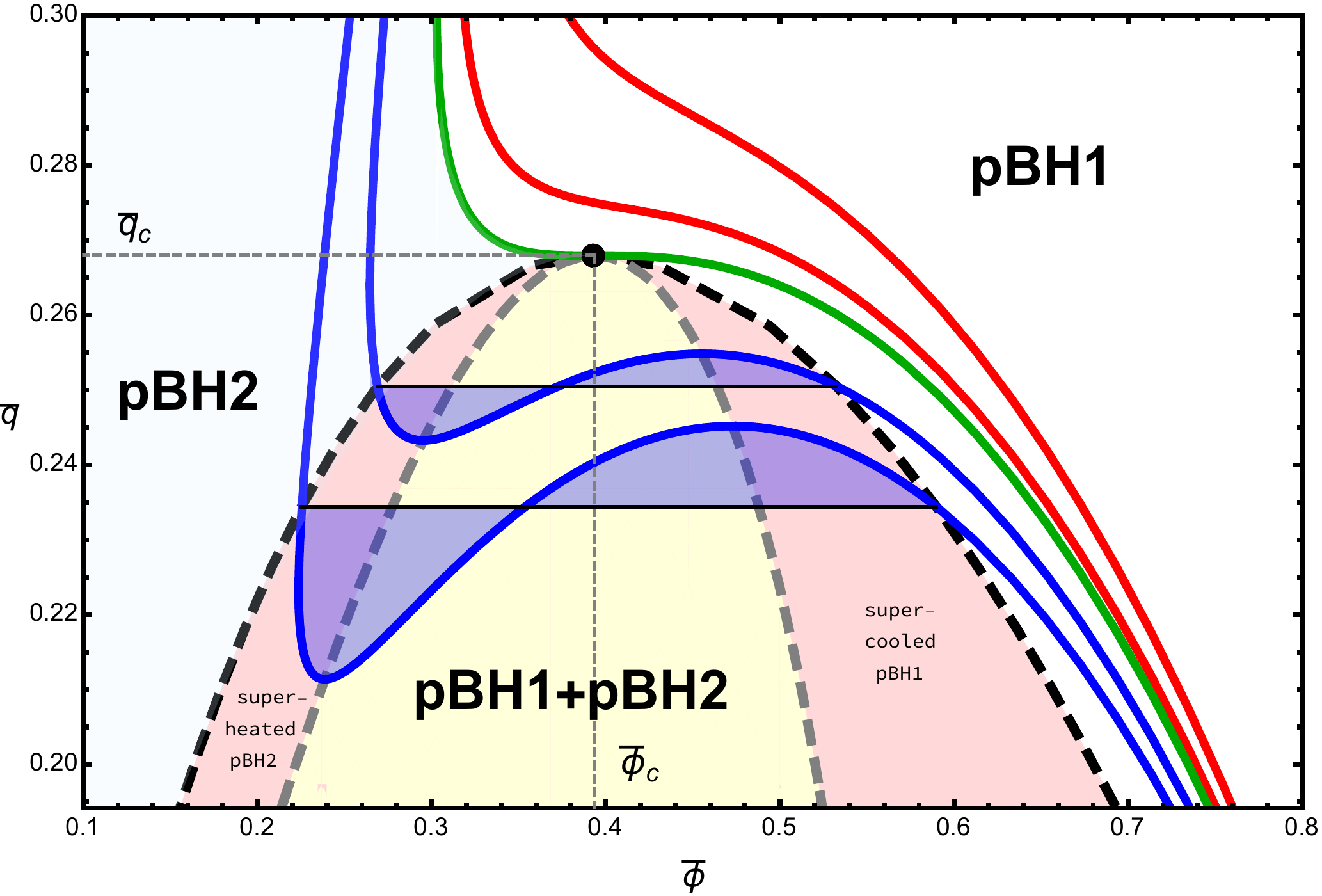}
	\end{tabular}
	\caption{The phase structure of the RN-AF with R\'enyi statistics on the $\bar{q}-\bar{\phi}$ plane is shown here, where the temperature of isotherm lines increases from top to bottom.}\label{fig: diagram-full}
\end{figure*}

We show the phase structure of RN-AF with R\'enyi statistics on the $\bar{q}-\bar{\phi}$ plane analogous to that of the vdW fluid on the $P-V$ plane in Fig. \ref{fig: diagram-full}. The red, green and blue curves are the isothermal curves for $\bar{T}_\text{R}<\bar{T}_c$, $\bar{T}_\text{R}=\bar{T}_c$ and $\bar{T}_\text{R}>\bar{T}_c$, respectively. The dashed black curve and dashed gray curve are the Maxwell curve or coexistence curve and the spinodal curve, respectively. Each point along the spinodal curve has infinite heat capacity or the point where the second-order phase transition between nBH and pBH1 (or pBH2) occurs. The blue and white regions stand for the pure pBH2 and pBH1 phases, respectively. The left and right red regions are the regions for the metastable phases called the superheated pBH2 and supercooled pBH1, respectively. In the yellow region, the phase is mixed between pBH1 and pBH2. It is seen that the black hole can be in all phases when $\bar{T}_\text{R}>\bar{T}_c$. This feature is similar to the vdW fluid with the temperature being less than its critical value. 

The interesting properties in the aspect of the thermodynamical stability of the charged black holes are discussed. The first-order phase transition between pBH1 and pBH2 obeying the Maxwell equal area law has been investigated. The phase structure for this black hole is also shown. In the next section, the thermodynamical implication of the black hole is discussed via the Ruppeiner geometry of the thermodynamical phase space.

%%%%%%%%%%%%%%%%%%%%%%%%%%%%%%%%%%%%%%%%%%%%%%%%%%%%%%%%%%%%%%%%%%%%%%%%%%%%%%%%%%%%%%%%%%%%%%%%%%%%%%%%%%%%%%%%%%%%%%%%%%%%%%%%%%%%%%%%%%%%%%%%%%%%%%%%%%%%%%%%%%%%%%%%%%%%%%%%%%%%%%%%%%%%%%%%%%%%%%%%%%%%%%%%%%%%%%%%%%%%%%%%%%%%%%%%%%%%%%%%%%%%%%%%%%%%%%%%%%%%%%%%%%%%%%%%%%%%%%%%%%%%%%%%%%%%%%%

\section{Black hole phase structure from Ruppeiner geometry}\label{TherGeo}

The thermodynamic geometry provides some understandings about the microscopic behavior from the macroscopic quantities of thermodynamical systems. Therefore, this approach is suitable for investigating the microscopic aspects of black holes because their complete information are hidden from an observer outside its event horizon. In the context of the Ruppeiner geometry, a line element $\Delta l^2$  between two neighbouring fluctuation states can be written in the form \cite{Ruppeiner1979}
\begin{eqnarray}
	\Delta l^2 &=& \frac{1}{k_\text{B}}g_{ij}\Delta x^i \Delta x^j,  \\
	g_{ij} &=& -\frac{\partial^2 S(x)}{\partial x^i\partial x^j},\label{Rup metric}
\end{eqnarray}
where $S(x^i)$ is the entropy function of the phase space coordinates $x^i$ chosen to be the thermodynamic variables, the metric tensor $g_{ij}$ is defined as the Ruppeiner metric, and $k_\text{B}$ is the Boltzmann constant. In general relativity, it is well known that the gravitational interaction is encoded in the form of the curvature of spacetime. In a similar way, the scalar curvature of the Ruppeiner metric, as shown in Eq.~\eqref{Rup metric} could encode the microscopic interactions of a thermal system. It has been convinced that the sign of the scalar curvature $R$ indicates the interaction between two microstructures of the system, namely, $R<0$ and $R>0$ represent attractive and repulsive interaction, respectively, while $R=0$ represents the case of interaction-free \cite{Ruppeiner2010}. Moreover, the divergence of the scalar curvature occurs at the second-order phase transition.

As suggested in Refs. \cite{Promsiri2020, Promsiri:2021hhv} that the microstates of a strongly self-gravitating system could be correlated in nonextensive description, this nontrivial correlation may be encoded in the nonextensivity parameter $\lambda$. In this section, we are interested in comparing between the interactions among microstates of the RN-AF with $\lambda > 0$ and those with $\lambda=0$ through considering the values of scalar curvature $R$ in some thermodynamical phase spaces.

\subsection{RN-AF in GB statistics}

\begin{figure*}[!ht]
	\begin{tabular}{c c}
		\includegraphics[scale=0.5]{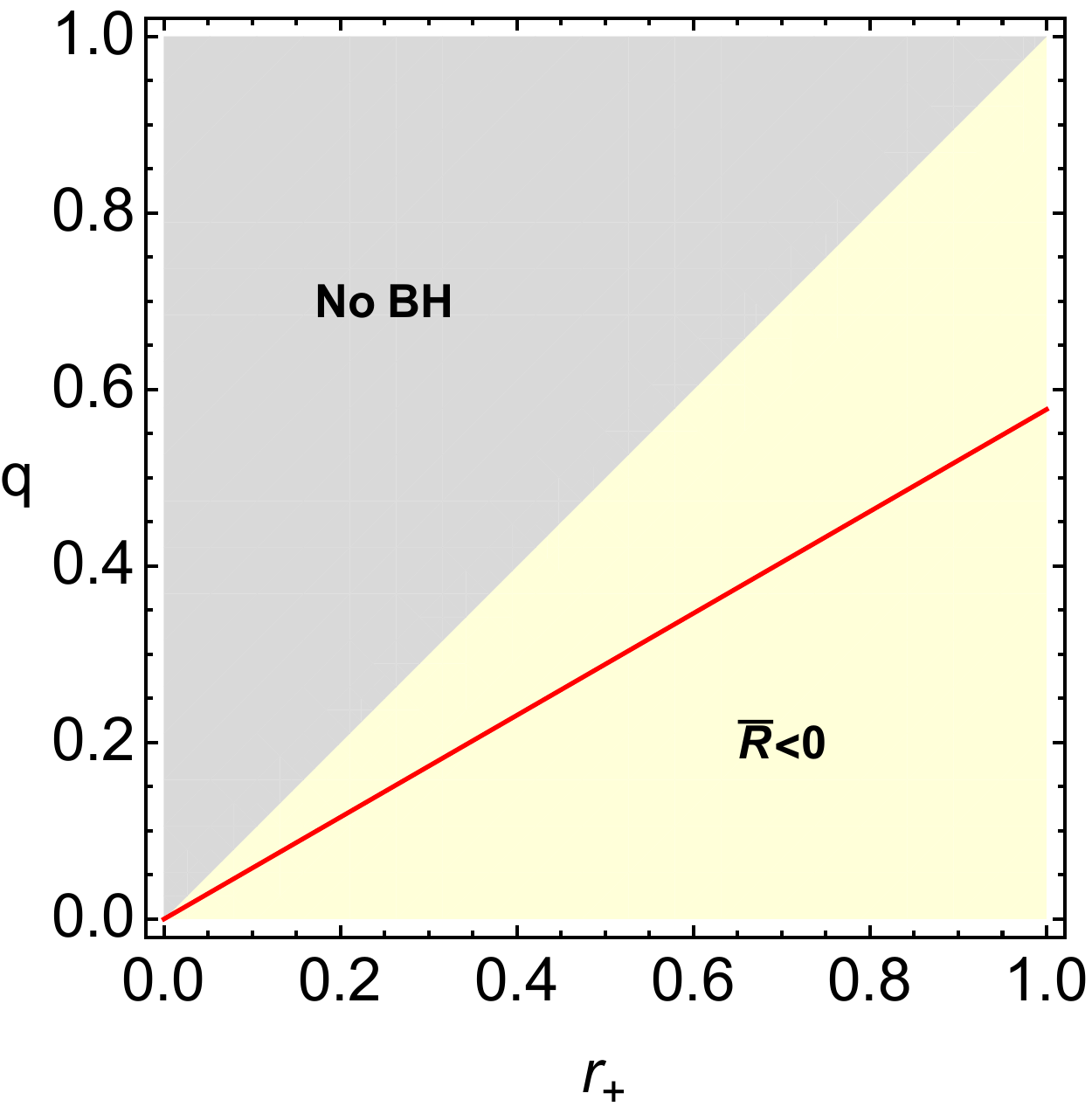}\hspace{1cm}
		\includegraphics[scale=0.5]{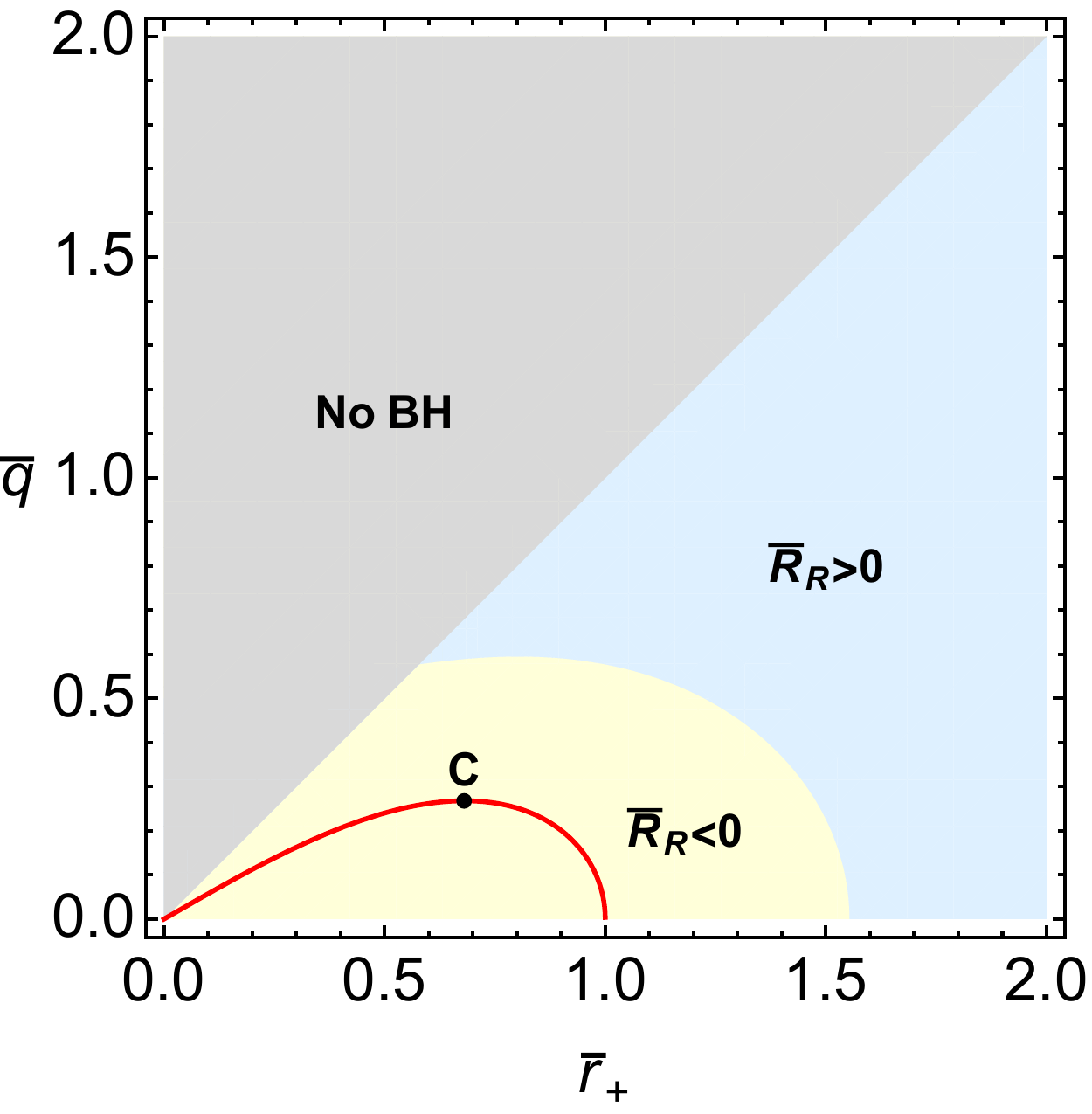}
	\end{tabular}
	\caption{The region plots for the signs of the Ricci scalar of the Ruppeiner geometry of black hole $R$ within the GB statistics (left) and the R\'enyi statistics (right) are shown, where the regions of negative and positive values of $R$ are in yellow and blue, respectively. Note that the black hole solution does not exist in the gray region. The red lines are the lines corresponding to $R\to-\infty$, and the point C represents the critical point.}\label{fig: Ricci region}
\end{figure*}

Firstly, we review the thermodynamic geometry of RN-AF in the standard GB statistics studied in \cite{Shen:2005nu, Wang:2019cax}. Then we proceed to consider the effects of nonextensivity parameter on the scalar curvature and phase transition of RN-AF subsequently. The Ruppeiner metric for the RN-AF can be defined as the second derivative of the Bekenstein-Hawking entropy in Eq.~\eqref{area} with respect to the phase space coordinates $x^i=(u, \phi)$ as
\begin{eqnarray}
	g_{ij}=-\frac{\partial^2 }{\partial x^i\partial x^j}S_\text{BH}(x). \label{Rup}
\end{eqnarray}
The internal energy $u$ has been defined in Eq.~\eqref{energy}, which can be written in the form
\begin{eqnarray}
	u = \frac{r_+}{2}\left( 1 - \phi^2  \right). \label{uRNflat}
\end{eqnarray}
Here, the coordinates $u$ and $\phi$ are assumed to be independent of each other. Note that these phase space coordinates are chosen to correspond with $S=S(u, \phi)$ (see Eq.~\eqref{du in S phi}). Using Eq.~\eqref{uRNflat}, the entropy formula \eqref{area} can be written in term of the coordinates $x^i$ as in the following:
\begin{eqnarray}
	S_\text{BH}(u, \phi)=\frac{4\pi u^2}{(1-\phi^2)^2}.
\end{eqnarray}
As a result, the metric tensor in $u-\phi$ phase space is obtained as follows:
\begin{eqnarray}
	g_{ij}(u, \phi)
	&=& - 
	\begin{pmatrix}
	\partial_u\partial_u S_\text{BH} & \partial_\phi \partial_u S_\text{BH} \\
	\partial_u\partial_\phi S_\text{BH} & \partial_\phi \partial_\phi S_\text{BH} 
	\end{pmatrix}	
	= -\frac{8\pi}{(1-\phi^2)^2}
	\begin{pmatrix}
	1 & \frac{4 u \phi}{1-\phi^2} \\
	\frac{4 u \phi}{1-\phi^2} & \frac{2 u^2 ( 1 + 5\phi^2 )}{(1-\phi^2)^2}
	\end{pmatrix}.
\end{eqnarray}
%the line element in $(u,\phi)$-phase space is obtained as follows: %becomes
%\begin{eqnarray}
%	ds^2(u, \phi)=-\frac{8\pi}{(1-\phi^2)^2}du^2 - \frac{64\pi u\phi}{(1-\phi^2)^3}dud\phi - \frac{16\pi u^2(1+5\phi^2)}{(1-\phi^2)^4}d\phi^2.
%\end{eqnarray}
Under the general coordinate transformation, $x^i \rightarrow x'^i$, the metric $g_{ij}$ transform as
\begin{eqnarray}
	g'_{ij} = \frac{\partial x^m}{\partial x'^i}\frac{\partial x^n}{\partial x'^j}g_{mn}, \label{transform}
\end{eqnarray}
which is just the metric tensor in the Riemannian geometry. Using Eqs.~\eqref{potential} and \eqref{uRNflat}, the metric tensor in the phase space coordinates $x'^i = (r_+,r_-)$ can be written in the form
%\begin{eqnarray}
%	ds^2(r_+, r_-)=-\frac{\pi (2r_+-5r_-)}{r_+-r_-}dr_+^2 - \frac{2\pi r_+}{r_+-r_-}dr_+dr_- - \frac{\pi r_+^2}{r_-(r_+-r_-)}dr_-^2.
%\end{eqnarray}
%where the Ruppeiner metric reads
\begin{eqnarray}
	g_{ij}(r_+, r_-) = -\frac{\pi}{(r_+-r_-)}
	\begin{pmatrix}
	2r_+ - 5r_- &&& r_+ \\
	r_+ &&& r^2_+/r_-
	\end{pmatrix}.
\end{eqnarray}
Eventually, the Ricci curvature scalar $R$ is in the form
\begin{eqnarray}
	R=-\frac{r_+-r_-}{\pi r_+(3r_--r_+)^2}.\label{R GB}
\end{eqnarray}
By comparing to the expression of heat capacity $C_q$ in Eq.~\eqref{C GB}, it is interestingly found that both Ricci scalar $R$ and heat capacity $C_q$ vanish, and diverge at the same positions which are $r_+=r_-$, and $r_+=r_0$, respectively. This implies that the thermodynamic geometry approach can indicate an occurrence of  the second-order phase transition through the divergence of $R$. In the left panel in Fig.~\ref{fig: Ricci region}, the region of negative value of $R$ for the RN-AF via the GB statistics, the regular one with the condition $M^2 > q^2$, in a parameter space of $r_+$ and $q$ is shown in yellow, whereas the gray upper-left region corresponds to the non-existence of charged black hole due to $M^2 < q^2$. It is interesting to note that the red line in the panel is the line of which $R \to -\infty$ corresponding to $|C_q| \to \infty$, representing the dividing line between the nBH and pBH phases. The Ricci scalar profile versus the outer event horizon $r_+$ and temperature $T_\text{H}$ are plotted in Fig.~\ref{fig: R vs rb, T, q GB}, whose the left panel shows that the nBH-pBH transition occurs at the point $r_0$, given by Eq.~\eqref{r0}, at which $R$ approaches negatively infinite.  Moreover, the right panel shows that $R$ also turns out to be negatively infinite at $T_\text{max}$, which corresponds to $r_0$. Obviously, the Ricci scalar vanishes at both extremal and very large black hole limits. Note that, at the same temperature, the Ricci scalar of pBH phase has more negative value than the nBH phase as shown in the right panel of Fig.~\ref{fig: R vs rb, T, q GB}. 
\begin{figure*}[!ht]
	\begin{tabular}{c c}
		\includegraphics[scale=0.5]{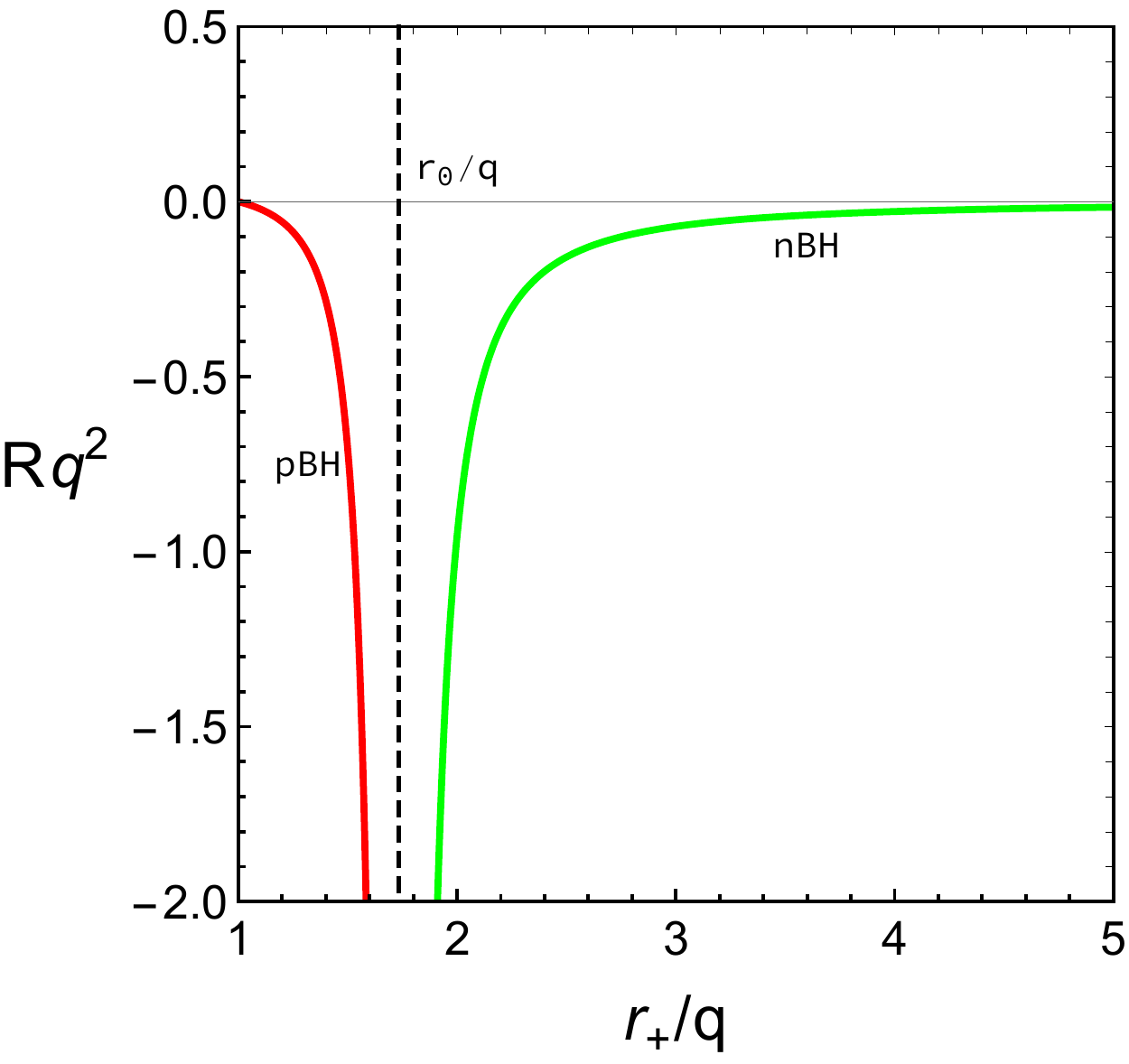}\hspace{1cm}
		\includegraphics[scale=0.515]{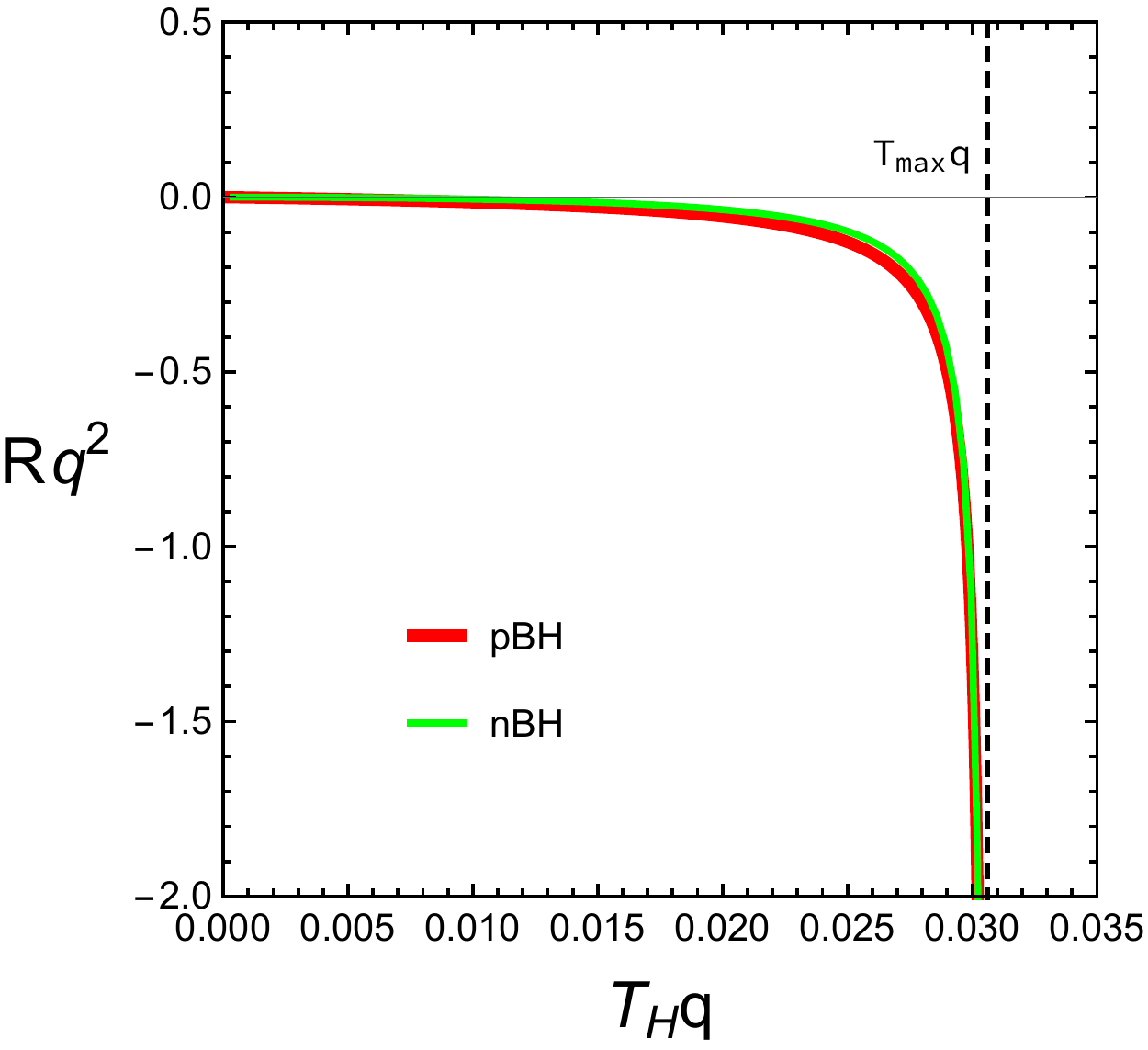}%\hspace{1cm}
	\end{tabular}
	\caption{The dimensionless Ricci scalar $Rq^2$ with respect to the dimensionless outer horizon radius $r_+/q$ (left) and Hawking temperature $T_\text{H}q$ (right). %Right: the profile of the rescaled Ricci scalar $R$ with respect to the rescaled charge $q$ with fixed rescaled $T_\text{H}=0.01$.
	}\label{fig: R vs rb, T, q GB}
\end{figure*}

According to Eq.~\eqref{R GB}, $R$ is always negative implying that the interactions between microscopic constituents of the RN-AF with the GB thermodynamics is attractive and becomes vanished at the extremal black holes. Hence, the extremal limit could be regarded as an analogue of the limit to an ideal gas. To illustrate this, let us discuss the competing roles between the mass $M$ and the charge $q$ terms in a charged black hole's horizon function $f(r)$ in Eq.~\eqref{f of r}. Even though we are working in the framework of general relativity, $M$ and $q$ could  be thought in some sense to be responsible to an attractive gravitational interaction and the repulsive electrostatic interaction, respectively.
Comparing these two parameters in the geometric unit, the condition $M^2 > q^2$ may imply that the attractive gravitational force is stronger than the repulsive electric force for this charged black hole system.  In the extremal black hole ($M^2 = q^2$), the gravitational and electrostatic interactions are balanced. In this way, the interactions between the constituents of the black hole may be always attractive due to the gravitational force domination, but the interactions becomes absent in the extremal case. Since the solution does not contain a naked singularity obeying the cosmic censorship conjecture \cite{Penrose2002}, the repulsive electrostatic force cannot be stronger than the attractive gravitational force. Therefore, the Ricci scalar curvature of the RN-AF in the GB statistics is less than or equal to zero as shown in Fig.~\ref{fig: R vs rb, T, q GB}.

%%%%%%%%%%%%%%%%%%%%%%%%%%%%%%%%%%%%%%%%%%%%%%%%%%%%%%%%%%%%%%%%%%%%%%%%%%%%%%%%%%%%%%%%%%%%%%%%%%%%%%%%%%%%%%%%%%%%%%%%%%%%%%%%%%%%%%%%%%%%%%%%%%%%%%%%%%%%%%%%%%%%%%%%%%%%%%%%%%%%%%%%%%%%%%%%%%%%%%%%%%%%%%%%%%%%%%%%%%%%%%%%%%%%%%%%%%%%%%%%%%%%%%%%%%%%%%%%%%%%%%%%%%%%%%%%%%%%%%%%%%%%%%%%%%%%%%%

\subsection{RN-AF in R\'enyi statistics}

In the section \ref{RN-Renyi}, we found that the thermodynamics of RN-AF with the R\'enyi statistics has different thermal phase structure compared to the one with the GB statistics, namely, the emergent pBH2 phase has occured in this case. It is interesting to investigate some aspects of the connections between phase structure and interaction among microscopic constituents of black hole in the R\'enyi case via the thermodynamic geometry framework. In a similar way as working with the GB statistics, we start by writing the R\'enyi entropy in terms of $u$ and $\phi$ as follows:
\begin{eqnarray}
	S_\text{R} = \pi L^2_\lambda \ln \left[ 1+\frac{4u^2}{L^2_\lambda (1-\phi^2)^2} \right].
\end{eqnarray}
With the approach in a similar way as Eq.~\eqref{Rup}, the metric tensor in $u-\phi$ phase space can be computed as
\begin{eqnarray}
	g_{ij}(u, \phi) 
%	&=& - \begin{pmatrix}
%	\partial_u\partial_u S_\text{R} & \partial_\phi \partial_u S_\text{R} \\
%	\partial_u\partial_\phi S_\text{R} & \partial_\phi \partial_\phi S_\text{R} 
%	\end{pmatrix} \nonumber \\
	&=& \frac{-8\pi}{\left(1-\phi^2\right)^3\left[ 1-\frac{4u^2}{L^2_\lambda\left(1-\phi^2\right)^2} \right]^2}
	\begin{pmatrix}
	1-\phi^2+\frac{4u^2}{L^2_\lambda (1-\phi^2)} 
	& 4 u \phi \\
	4 u \phi
	& \frac{2 u^2}{1-\phi^2}\left[ 1 + 5\phi^2 + \frac{4u^2(1 + \phi^2)}{L^2_\lambda\left(1-\phi^2\right)^2} \right]
\end{pmatrix}. \nonumber \\
\end{eqnarray} 
In $r_+-r_-$ phase space, it becomes
\begin{eqnarray}
	g_{ij}(r_+, r_-) 
	=-\frac{\pi r_+}{(r_+ - r_-)\left( 1 + \frac{r^2_+}{L^2_\lambda} \right)}
	\begin{pmatrix}
	\frac{\left[2r_+ - 5r_- - \frac{r^2_+}{L^2_\lambda} (2r_+ + r_-)\right]}{r_+\left(1 + \frac{r^2_+}{L^2_\lambda}\right)} 
	&&& 1 \\
	1
	&&& r_+/r_-
	\end{pmatrix}.
\end{eqnarray}
The Ricci scalar corresponding to the R\'enyi entropy is given by
\begin{eqnarray}
	R_\text{R} 
	=-\frac{(r_+-r_-)}{\pi r_+	\left[ 3r_--r_++\frac{r^2_+}{L^2_\lambda}(r_++r_-) \right]^2}\left[ 1-\frac{2r_+}{L^2_\lambda}(2r_--r_+) - \frac{r^3_+}{L^4_\lambda}(r_++2r_-) \right]. \label{RRenyi}
\end{eqnarray}
By rescaling the parameters as introduced in \eqref{rescale}, the curvature scalar in Eq.~\eqref{RRenyi} is transformed as $R_\text{R} \rightarrow \bar{R}_\text{R}=\pi L^2_\lambda R_\text{R}$, and hence we obtain
\begin{eqnarray}
	\bar{R}_\text{R} = -\frac{(\bar{q}^2 - \bar{r}^2_+)\big[\bar{r}^4_+ -2\bar{r}^2_+ (1-\bar{q}^2) +4\bar{q}^2 - 1\big]}{\left[ 3\bar{q}^2 - \bar{r}^2_+ + \bar{r}^2_+ (\bar{q}^2 + \bar{r}^2_+)   \right]^2}.\label{RbarR}
\end{eqnarray}

\begin{figure*}[!ht]
	\begin{tabular}{c c}
		\includegraphics[scale=0.37]{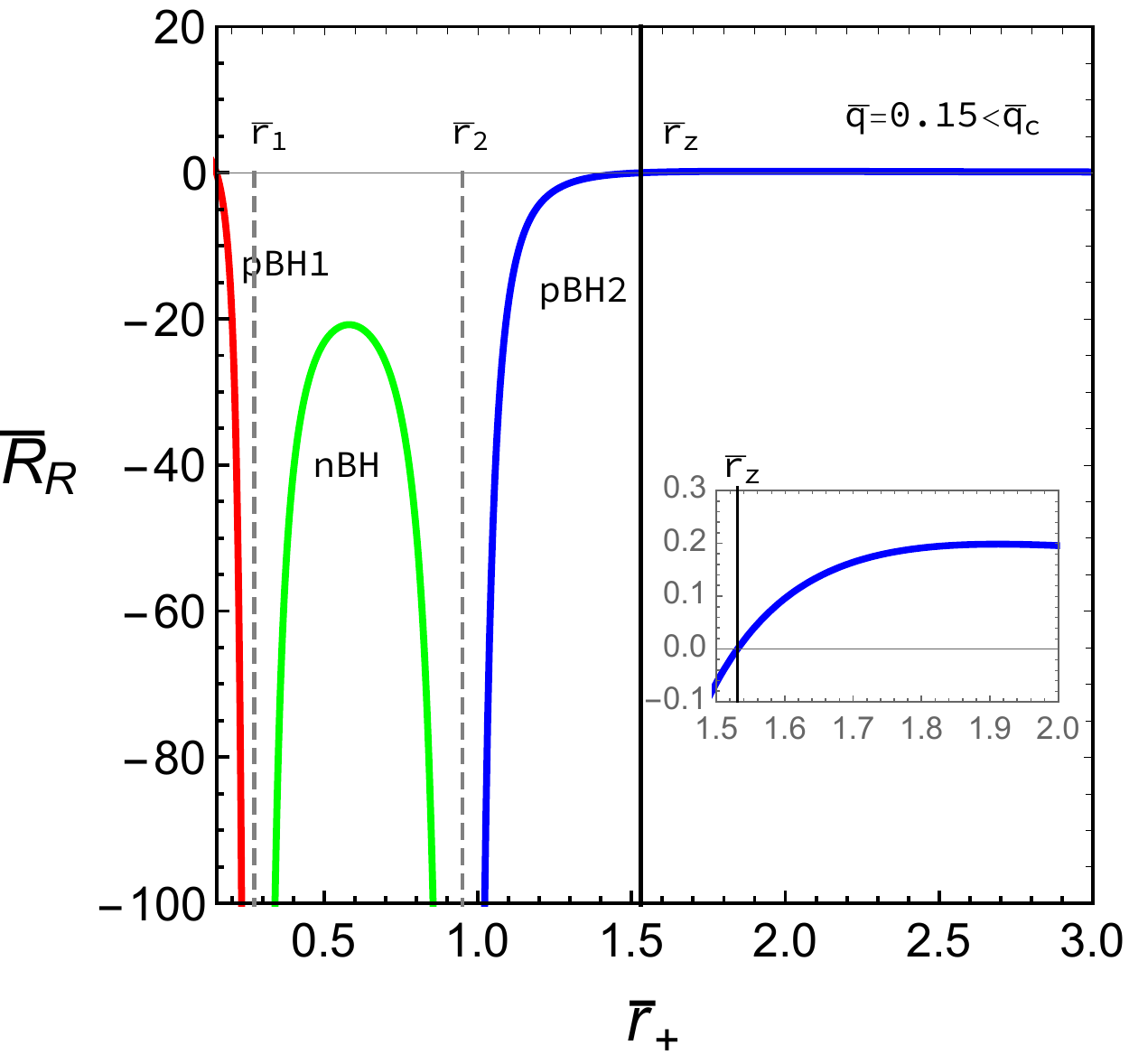}\hspace{1cm}
		\includegraphics[scale=0.37]{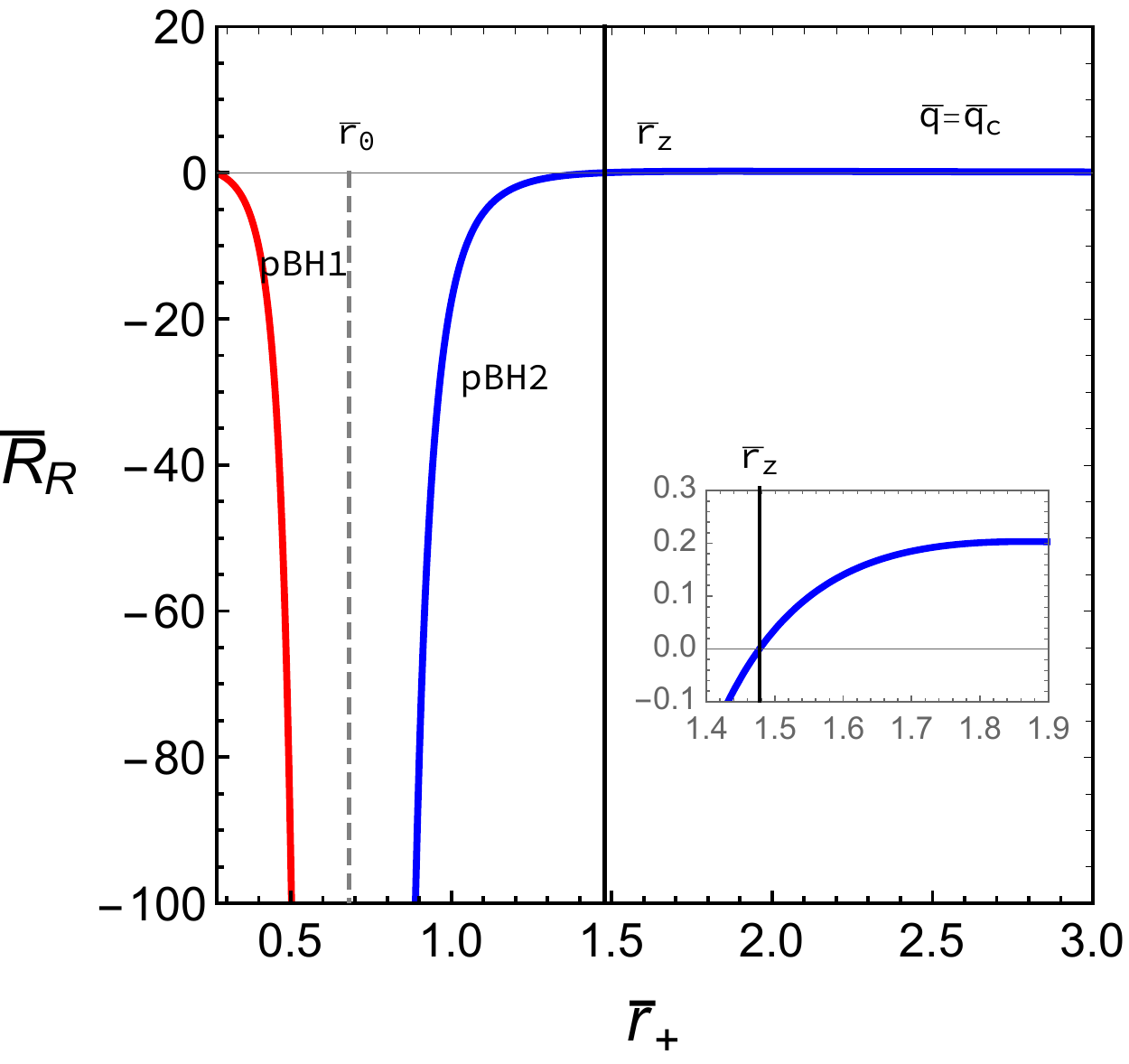}\hspace{1cm}
		\includegraphics[scale=0.37]{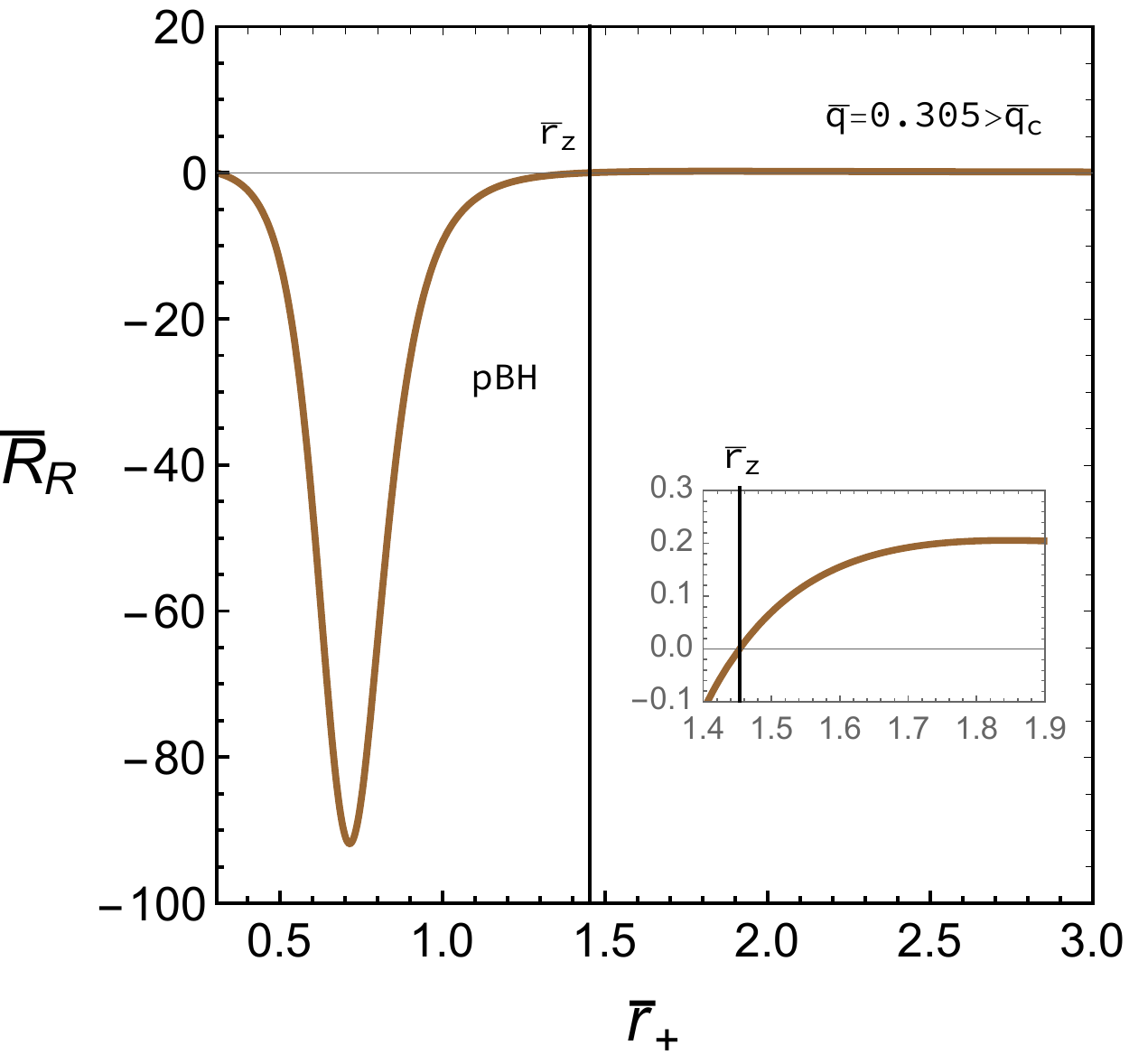}
	\end{tabular}
	\caption{The Ricci scalar $\bar{R}_\text{R}$ versus the outer horizon radius $\bar{r}_+$ of the RN-AF from the R\'enyi statistics for fixed charges $\bar{q} < \bar{q}_c$ (left), $\bar{q} = \bar{q}_c$ (middle) and $\bar{q} > \bar{q}_c$ (right), respectively.}\label{fig:Rvsrb-Renyi}
\end{figure*}

Using Eq.~\eqref{RbarR} to illustrate the phase structure in $(\bar{r}_+,\bar{q})$ plane as shown in the right panel in Fig.~\ref{fig: Ricci region}, even though the black hole solutions are allowed due to the condition $M^2>q^2$ in a similar fashion as in the GB thermodynamics, the charged black holes can exist not only with $\bar{R}_\text{R} < 0$ (yellow region) but also with  $\bar{R}_\text{R} > 0$ (blue region) in the alternative Re\'nyi thermodynamics. The emergence of the additional blue region in the right panel in Fig.~\ref{fig: Ricci region} leads us to wonder what is the nature of the critical phase transitions between these several phases, which include the new emergent black hole phase with $\bar{R}_\text{R}>0$. To explore some relations between critical phase transition and interactions among black hole molecules, we will investigate $\bar{R}_\text{R}$ as a function of $\bar{r}_+$ and $\bar{T}_\text{R}$. In Fig.~\ref{fig:Rvsrb-Renyi}, we plot $\bar{R}_\text{R}$ versus $\bar{r}_+$ for $\bar{q} < \bar{q}_c$, $\bar{q} = \bar{q}_c$ and $\bar{q} > \bar{q}_c$, respectively. For small charge $\bar{q} < \bar{q}_c$, the $\bar{R}_\text{R}$ curve can be separated into the curves for three branches of black holes, namely, pBH1 (red curve), nBH (green curve) and pBH2 (blue curve) as shown in the left panel in Fig.~\ref{fig:Rvsrb-Renyi}. Note that $\bar{R}_\text{R} \to -\infty$ at the horizon radius of the size $\bar{r}_1$ and $\bar{r}_2$ corresponds to $|\bar{C}_\text{R}| \to \infty$. The negativity of the Ricci scalar $\bar{R}_\text{R}$ for both pBH1 and nBH phases implies attractive interactions between black hole microstructures, in common with the consideration using the GB statistics. Interestingly, $\bar{R}_\text{R}$ of the emergent pBH2 phase has negative and positive values when $\bar{r}_2 < \bar{r}_+ < \bar{r}_z$ and $ \bar{r}_+ > \bar{r}_z$, respectively. This means that the interactions between microstructures of the black hole turn out to be repulsive when the horizon size is sufficiently larger than the nonextensivity length scale, namely $\bar{r}^2_+ \geq \bar{r}^2_z = 1 - \bar{q}^2 + \sqrt{\bar{q}^4-6\bar{q}^2+2}$. At the critical point ($\bar{q} = \bar{q}_c$), horizon radius of the sizes $\bar{r}_1$ and $\bar{r}_2$ are degenerate with the value of $\bar{r}_0$, such that the nBH phase disappears.  As previously discussed, apart from its sign, the value of $|\bar{R}_R|$ could be associated with the strength of interactions or correlations among black hole constituents. In other words, the larger the value of $|\bar{R}_R|$ is, the stronger the interactions between black hole molecules are.  As well known that the correlation length $\xi$ of the system diverges near the critical point, the infinity of $|\bar{R}_R|$ at this point as shown in the middle panel in Fig.~\ref{fig:Rvsrb-Renyi} is therefore not so weird. For $\bar{q} > \bar{q}_c$, there exists only a single phase with positive heat capacity where $\bar{R}_R$ is negative and positive in the ranges $\bar{q} < \bar{r}_+  < \bar{r}_z$ and $\bar{r}_+ > \bar{r}_z$, respectively (the right panel in Fig.~\ref{fig:Rvsrb-Renyi}). 

\begin{figure*}[!ht]
	\begin{tabular}{c c}
		\includegraphics[scale=0.37]{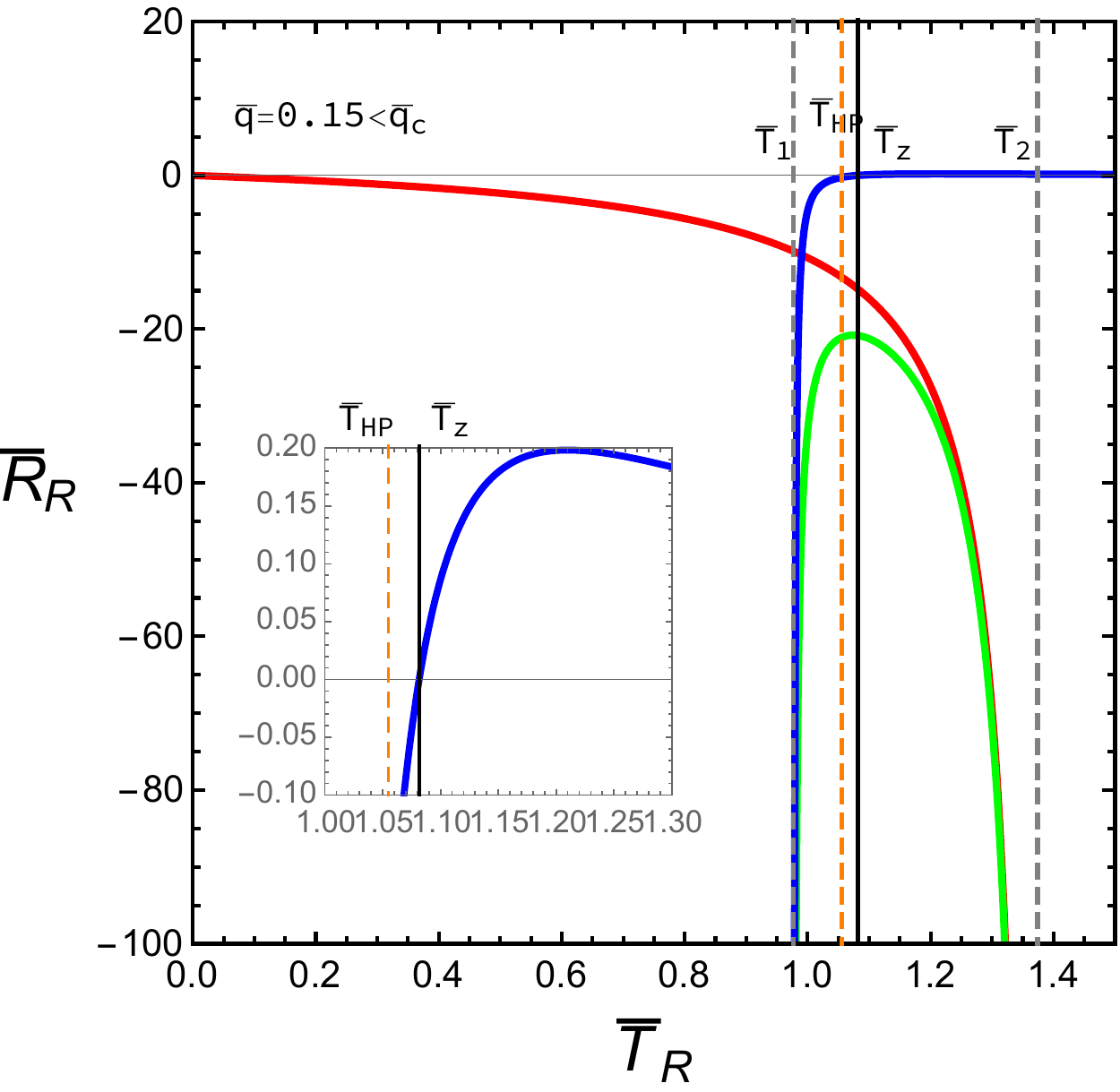}\hspace{1cm}
		\includegraphics[scale=0.37]{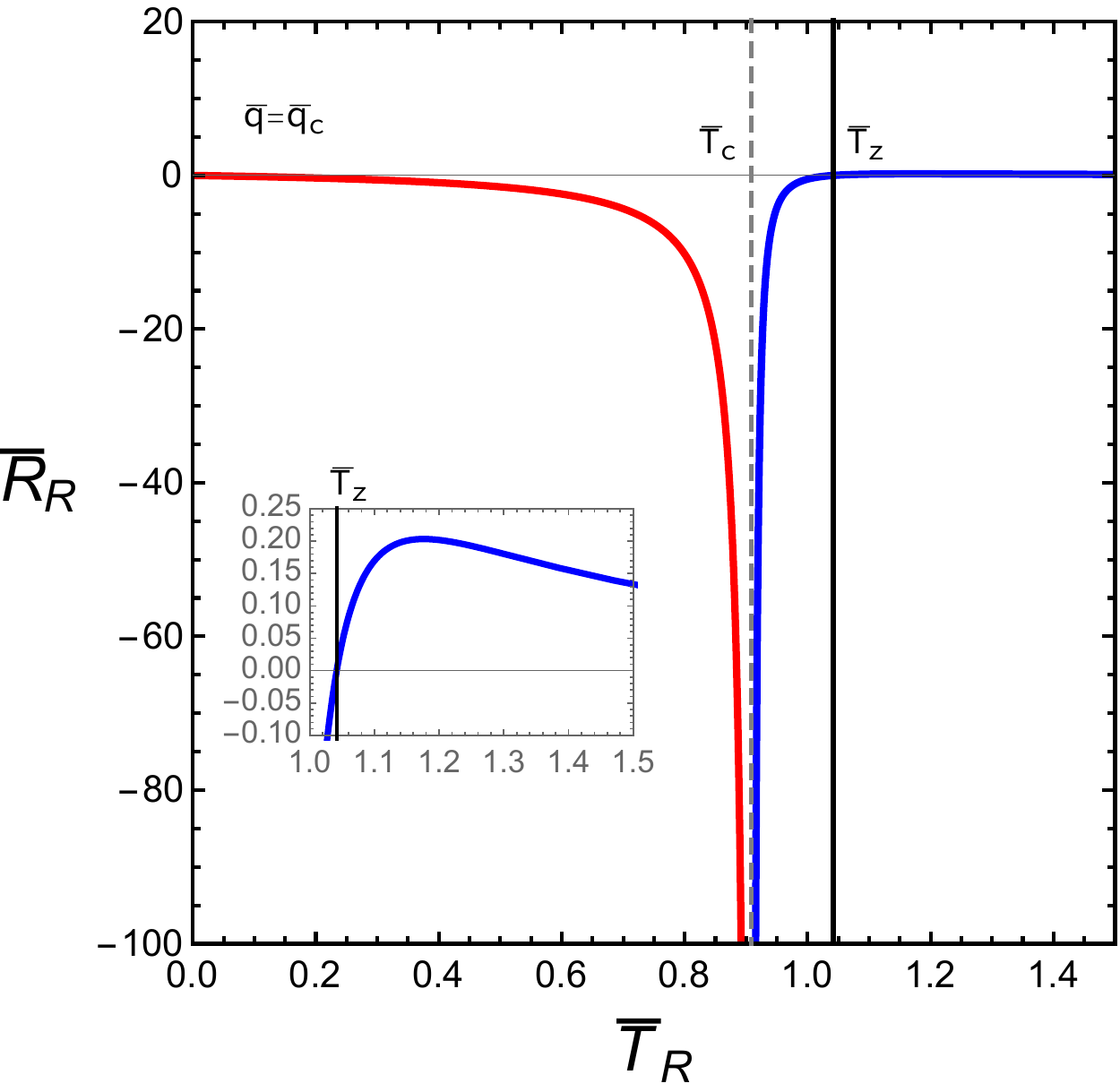}\hspace{1cm}
		\includegraphics[scale=0.37]{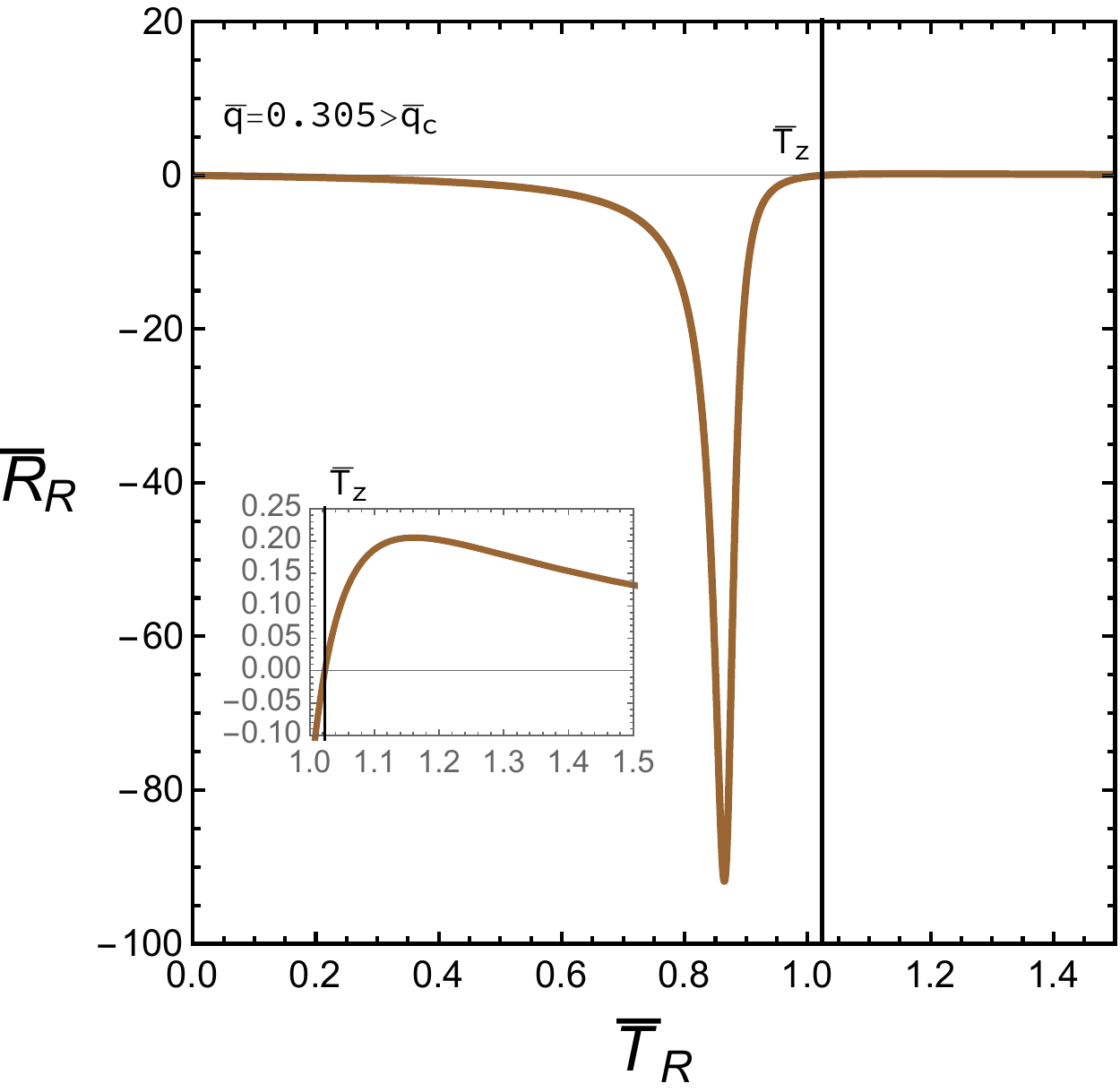}
	\end{tabular}
	\caption{The Ricci scalar $\bar{R}_\text{R}$ versus the R\'enyi temperature $\bar{T}_\text{R}$ of the RN-AF with the R\'enyi statistics for fixed charges $\bar{q} < \bar{q}_c$ (left), $\bar{q} = \bar{q}_c$ (middle) and $\bar{q} > \bar{q}_c$ (right), respectively. }\label{fig:RvsT-Renyi}
\end{figure*}

The phase structure can be explored by considering $\bar{R}_R$ as a function of $\bar{T}_\text{R}$ which corresponds to the plot of Gibbs free energy in Fig.~\ref{fig:G in Renyi vary q}. For small charge $\bar{q} < \bar{q}_c$, there are three branches of charged black hole which are pBH1, nBH and pBH2 phases represented by red, green and blue curves, respectively, as shown in the left panel in Fig.~\ref{fig:RvsT-Renyi}. At the extremal limit corresponding to $\bar{T}_\text{R}=0$, we have found that $\bar{R}_\text{R}=0$ in the same way as in the RN-AF with GB statistics. However, it is different from the case of the Reissner-Nordstr\"om black hole in asymptotically anti-de Sitter (RN-AdS) with GB statistics where the Ricci scalar negatively diverges at the extremal black hole \cite{Shen:2005nu}. Since the heat capacity and the Ricci scalar diverge at the same value of horizon radius, \textit{i.e.} $\bar{r}_1$ and $\bar{r}_2$, the nBH-pBH1 and nBH-pBH2 phase transitions occur at the temperatures denoted by $\bar{T}_2$ and $\bar{T}_1(<\bar{T}_2)$, respectively.  One can split the temperature  into three ranges. As shown in the left panel in the figure, the pBH1 phase exists at low temperature ($\bar{T}_\text{R}<\bar{T}_1$) as well as at a certain temperature in the range $\bar{T}_1<\bar{T}_\text{R}<\bar{T}_2$, in which the nBH and pBH2 phases simultaneously also appear corresponding to the swallowtail behavior in the free energy diagram in the upper left panel in Fig.~\ref{fig:G in Renyi vary q}.  It is also seen that there exists the crossing between the Ricci scalar for the pBH1 (red) and pBH2 (blue) phases. The temperature of this crossing point is not the Hawking-Page temperature $T_\text{HP}$ (see the orange dashed line). Unfortunately, the Ricci scalar for the Ruppeiner geometry does not contain the information of the first-order phase transition. When the temperature increases to $\bar{T}_\text{R}>\bar{T}_2$, the pBH1 and nBH phases disappear, while pBH2 phase persists for this range of high temperature. This feature can be also found in the upper left panel in Fig.~\ref{fig:G in Renyi vary q}. Moreover, the pBH1 and nBH phases have $\bar{R}_\text{R} < 0$ at any temperature. Therefore, the interactions between two black hole microstructures are attractive. For pBH2 phase, we find that $\bar{R}_\text{R} < 0$ when $\bar{T}_1 < \bar{T}_\text{R} < \bar{T}_z$. Interestingly, at a sufficiently high temperature $\bar{T}_\text{R}>\bar{T}_z$, the pBH2 phase begins to has $\bar{R}_\text{R}>0$ and hence the interactions between two black hole microstructures become repulsive as found in the RN-AdS with the GB statistics \cite{Wei:2019uqg, Wei:2019yvs}.  

Moreover, the temperatures $\bar{T}_1$ and $\bar{T}_2$ get closer as the electric charge $\bar{q}(<\bar{q_c})$ increases. The temperatures then merge together at $\bar{T}_1=\bar{T}_2=\bar{T}_z$ when the charge reaches its critical value. This merger corresponds to the absence of the nBH phase. At the critical point $\bar{q} = \bar{q}_c$, the crossing of $\bar{R}_\text{R}$ for pBH1 and pBH2 phases disappears resulting that $\bar{R}_\text{R}$ negatively diverges as shown in the middle panel in Fig.~\ref{fig:RvsT-Renyi}. For $\bar{q} > \bar{q}_c$, there is only a single pBH phase of black hole represented as the brown curve in the right panel in Fig.~\ref{fig:RvsT-Renyi}. Obviously, the divergence in $\bar{R}_\text{R}$ becomes absent for a large amount of charge.

It is very important to note that the pBH2 phase emerging within the framework of the alternative R\'enyi statistics allows $\bar{R}_\text{R}$ to be positive value, which is in the same way as the large black hole phase of the RN-AdS from the GB statistics.  This implies that the interactions between black hole molecules are repulsive.

%\newpage
%\begin{figure*}[!ht]
%	\begin{tabular}{c c}
%		\includegraphics[scale=0.3]{Rvsrh_a.pdf}\quad
%		\includegraphics[scale=0.3]{Rvsrh_b.pdf}\quad
%		\includegraphics[scale=0.3]{Rvsrh_c.pdf}\quad
%	\end{tabular}
%	\caption{The Ruppeiner curvature $R$ versus the outer horizon radius $r_+$ of the RN-AF with the R\'enyi statistics for a fixed charge $q < q_c$ (left), $q = q_c$ (middle) and $q > q_c$ (right) respectively.}\label{fig:2}
%\end{figure*}
%
%
%
%\begin{figure*}[!ht]
%	\begin{tabular}{c c}
%		\includegraphics[scale=0.3]{RvsTRen.pdf}\quad
%		\includegraphics[scale=0.3]{RvsTRenc.pdf}\quad
%		\includegraphics[scale=0.3]{RvsTRenq033.pdf}
%	\end{tabular}
%	\caption{The Ruppeiner curvature $R$ versus the Hawking temperature $T_\text{R}$ of the RN-AF with the R\'enyi statistics for a fixed charge $q < q_c$ (left), $q = q_c$ (middle) and $q > q_c$ (right) respectively..}\label{fig:2}
%\end{figure*}

%%%%%%%%%%%%%%%%%%%%%%%%%%%%%%%%%%%%%%%%%%%%%%%%%%%%%%%%%%%%%%%%%%%%%%%%%%%%%%%%%%%%%%%%%%%%%%%%%%%%%%%%%%%%%%%%%%%%%%%%%%%%%%%%%%%%%%%%%%%%%%%%%%%%%%%%%%%%%%%%%%%%%%%%%%%%%%%%%%%%%%%%%%%%%%%%%%%%%%%%%%%%%%%%%%%%%%%%%%%%%%%%%%%%%%%%%%%%%%%%%%%%%%%%%%%%%%%%%%%%%%%%%%%%%%%%%%%%%%%%%%%%%%%%%%%%%%%
%\newpage
\section{Critical Behavior of the RN-AF in R\'enyi statistics}\label{criticality}

It has been found that thermodynamic behaviors of matter near a critical point of phase transition in different systems tend to have the property of \textit{universality}.  Remarkably, this behavior does not depend on the types of particles and of interactions between them.  The \textit{scaling behavior} of physical quantities in the vicinity of critical phase transition can be described by a set of \textit{critical exponents}. For example, the scaling behavior of a vdW fluid can be written as
\begin{eqnarray}
    P-P_c &\sim & (V-V_c)^\delta , \label{P_V_crit} \\
    V_g-V_l &\sim & (T-T_c)^\beta , \label{V_T_crit} \\
    C_V &\sim & (T-T_c)^{-\alpha} , \\
    \kappa_\text{T} &\sim & (T-T_c)^{-\gamma},
\end{eqnarray}
where the critical exponents are $\delta =3$, $\beta = 1/2$, $\alpha =0$ and $\gamma =1$. $V_g$ and $V_l$ are the volumes for the gas and liquid phases of the vdW fluid.  For different systems which are governed by the same set of critical exponents, we can say that they are in the same \textit{universality class}. Interestingly, it has been found that the critical exponents of RN-AdS in the canonical ensemble are identical with the vdW fluid, and hence these systems are in the same universality class of phase transition \cite{Niu:2011tb,Shen:2005nu, Wang:2019cax}. As discussed in section \ref{RN-Renyi}, the phase transition of the RN-AF  with nonextensive description and the vdW liquid-gas system are very similar. There are the swallowtail behaviors of free energy as shown in Fig.~\ref{fig:G in Renyi vary q}. There also exists the finite coexistence line which terminates at the critical point as indicated in Fig.~\ref{fig:TR-HP vs q} and the transitions between two phases, \textit{i.e.}, pBH1-pBH2 transition for the RN-AF with R\'enyi statistics and liquid-gas transition for the vdW fluid, across the coexistence line are the first-order type of phase transition. Notice that, in both cases, these first-order phase transitions disappear when the systems approach to the critical point at $T=T_c$. Moreover, by considering the Maxwell construction in section \ref{Maxwell}, we have found that the pBH1 and pBH2 phases coexist on the isothermal curve at the charge $\bar{q}^*$ as shown in Fig.~\ref{fig: Maxwell}.  In other words, this value of charge can render the vapor pressure analogue, where the pBH1-pBH2 coexistence can be associated with the dynamic equilibrium between the liquid and gas phases at constant temperature on the $P-V$ plane in the vdW fluid.    

In this section, we are interested in categorizing the phase transition of the RN-AF in R\'enyi description by considering the scaling behavior of physical quantities. We identify the quantities $q$ and $\phi$ of the RN-AF as $P$ and $V$ of the vdW fluid, respectively. Considering the system near the critical point, Eq.~\eqref{q2} as an equation of state can be written in the form of power series expansion in terms of a set of the variables $\tau$ and $\omega$ as
\begin{eqnarray} 
	\bar{q}&=&a_{mn}\tau^m \omega^n \nonumber \\
	&=& a_{00}+a_{10}\tau +a_{01}\omega + a_{11}\tau \omega + a_{20}\tau^2 \nonumber \\
	&&+ a_{02}\omega^2+a_{21}\tau^2\omega + a_{12}\tau \omega^2 + a_{30}\tau^3 + a_{03}\omega^3 + \hdots, \label{power}
\end{eqnarray}
where $a_{mn}$ is the coefficient of the expansion, and the variables $\tau$ and $\omega$ are defined by
\begin{eqnarray}
	\tau = \frac{\bar{T}_\text{R}}{\bar{T}_c}-1,
	\quad
	\omega = \frac{\bar{\phi}}{\bar{\phi}_c}-1. \label{parameter}
\end{eqnarray}
The power series in Eq.~\eqref{power} can be numerically obtained by 
\begin{eqnarray}
	\bar{q}&=& \bar{q}_c - 0.732\tau + 1.732 \tau \omega +3.732 \tau^2 \nonumber \\
	&&
	-24.12\tau^2\omega -5.598 \tau \omega^2 -27.856 \tau^3 -0.366\omega^3 +\hdots , \label{expansion}
\end{eqnarray}
where $a_{00}=\bar{q}_c=0.268$ and $a_{01}=a_{02}=0$. Firstly, we consider the change of $\bar{q}$ with $\bar{\phi}$ being near the critical point when the temperature $\bar{T}_R$ is set to be $\bar{T}_c$, hence we have $\tau =0$. Interestingly, Eq.~\eqref{expansion} can be written in the form of scaling law as 
\begin{eqnarray}
	\bar{q}-\bar{q}_c\sim (\bar{\phi} - \bar{\phi}_c)^3.
\end{eqnarray}
Comparing this with Eq.~\eqref{P_V_crit}, the critical exponent $\delta$ can be read off as 
\begin{eqnarray}
\delta = 3.
\end{eqnarray}

In section \ref{Maxwell}, we have shown the first-order pBH1-pBH2 phase transition at $\bar{q}^*$, which can be associated with a jump between $\bar{\phi}_3$ and $\bar{\phi}_1$.  To investigate this behavior near the critical point of the second-order phase transition, we need to consider the Gibbs free energy. The change of the Gibbs free energy can be written as  
\begin{eqnarray}
d\mathcal{\bar{G}}_\text{R} = -\bar{S}_\text{R}d\bar{T}_\text{R}+\bar{\phi}d\bar{q}.
\end{eqnarray}
Since the pBH1 and pBH2 coexist, this implies that their Gibbs free energies are identical, namely, we  have $\Delta \mathcal{\bar{G}}_\text{R} =\mathcal{\bar{G}}_\text{pBH1}-\mathcal{\bar{G}}_\text{pBH2}=0$.  Therefore, the values of $\bar{\phi}$ on either side of the coexistence curve at fixed temperature can be found from two following conditions
\begin{eqnarray}
	\Delta \mathcal{\bar{G}}_\text{R} = \int_{\bar{\phi}_1}^{\bar{\phi}_3}\bar{\phi} d\bar{q} = 0, \label{coexis1}
\end{eqnarray}
and
\begin{eqnarray}
	\bar{q}^*=\bar{q}(\bar{\phi}_1) = \bar{q}(\bar{\phi}_3). \label{coexis2}
\end{eqnarray}
Differentiating Eq.~\eqref{power} with respect to $\omega$, one obtains
\begin{eqnarray}
	d\bar{q} = (a_{11}\tau +a_{21}\tau^2+2a_{12}\tau \omega +3a_{03}\omega^2)d\omega. \label{diffq}
\end{eqnarray}
Note that we have kept the dimensionless perturbation variables $\tau$ and $\omega$ in the above formula up to the third order expansion.  Let us define $\bar{\phi}_1=\bar{\phi}_c(1-\omega_1)$ and $\bar{\phi}_3=\bar{\phi}_c(1+\omega_3)$. Using Eqs.~\eqref{parameter} and \eqref{diffq}, the conditions \eqref{coexis1} and \eqref{coexis2} can be, respectively, expressed in terms of the expansion coefficients as follows:
\begin{eqnarray}
	%&& 
	a_{11}\tau (\omega_3 +\omega_1) + a_{21}\tau^2 (\omega_3 +\omega_1)
	%\nonumber \\
	%&& \ \ \ \ \ \ \ \ \ \  
	+ \frac{1}{2}(a_{11}+2a_{12})\tau (\omega^2_3 -\omega^2_1) + a_{03}(\omega^3_3 +\omega^3_1) &=& 0,\\
	%&& 
	a_{11}\tau (\omega_3 +\omega_1) + a_{21}\tau^2 (\omega_3 +\omega_1)  
	%\nonumber \\
	%&& \ \ \ \ \ \ \ \ \ \  
	+ a_{12} \tau (\omega^2_3 -\omega^2_1) + a_{03}(\omega^3_3 +\omega^3_1) &=& 0.
\end{eqnarray}
When $\omega_1 =\omega_3 = \omega$, these two conditions have a non-trivial solution which is given by
\begin{eqnarray}
	\omega^2 = -\frac{1}{a_{03}}(a_{11}\tau + a_{21}\tau^2) \approx -\frac{a_{11}}{a_{03}}\tau = 4.732\tau . \label{omega_tau}
\end{eqnarray}
Recalling Eq.~\eqref{parameter}, Eq.~\eqref{omega_tau} can be written in the form of a scaling law relating $\bar{\phi}$ with $\bar{T}_R$ near the critical point as
\begin{eqnarray}
	\bar{\phi} -\bar{\phi}_c \sim (\bar{T}_\text{R} - \bar{T}_c)^{\frac{1}{2}},
\end{eqnarray}
where, comparing Eq.~\eqref{V_T_crit}, we can read off the critical exponent 
\begin{eqnarray}
\beta = \frac{1}{2}.
\end{eqnarray}

The behavior of heat capacity at constant $\bar{\phi}$ near the critical point can be described by the exponent $\alpha$. Since $C_\phi$ does not display a singularity behavior at the critical point, the critical exponent therefore vanishes, \textit{i.e.} $\alpha =0$. Finally, we will calculate the critical exponent $\gamma$ of the compressibility at constant temperature,
\begin{eqnarray}
\bar{\kappa}_{T_\text{R}} = -\frac{1}{\bar{\phi}}\left( \frac{\partial \bar{\phi}}{\partial \bar{q}} \right)_{\bar{T}_\text{R}}. \label{compress}
\end{eqnarray}
By using Eq.~\eqref{power}, one can compute 
\begin{eqnarray}
\left( \frac{\partial \bar{q}}{\partial \omega} \right)_\tau = a_{11}\tau + a_{21}\tau^2 + 2a_{12}\tau \omega +3a_{03}\omega^2. \label{dqdo}
\end{eqnarray}
As the system approaches the critical point, we can use Eqs.~\eqref{compress} and \eqref{dqdo} to find that 
\begin{eqnarray}
\bar{\kappa}_{T_\text{R}} \sim (\bar{T}_\text{R} - \bar{T}_c)^{-1}, 
\end{eqnarray}
which gives us the critical exponent
\begin{eqnarray}
\gamma = 1.
\end{eqnarray}
Interestingly, our studies have demonstrated that the critical exponents of the RN-AF system from the R\'enyi statistics are identical to those of the vdW fluid and of the RN-AdS from the GB statistics. Manifestly, these are in the same universality class.

%%%%%%%%%%%%%%%%%%%%%%%%%%%%%%%%%%%%%%%%%%%%%%%%%%%%%%%%%%%%%%%%%%%%%%%%%%%%%%%%%%%%%%%%%%%%%%%%%%%%%%%%%%%%%%%%%%%%%%%%%%%%%%%%%%%%%%%%%%%%%%%%%%%%%%%%%%%%%%%%%%%%%%%%%%%%%%%%%%%%%%%%%%%%%%%%%%%%%%%%%%%%%%%%%%%%%%%%%%%%%%%%%%%%%%%%%%%%%%%%%%%%%%%%%%%%%%%%%%%%%%%%%%%%%%%%%%%%%%%%%%%%%%%%%%%%%%%

%\newpage

\section{Conclusion and Discussion}\label{conclusion}

%\fixme{[What we have done and key differences in aspect of stability]}

In the present work, we have studied the thermodynamical properties of the charged black hole in asymptotically flat spacetime using the nonextensive R\'enyi statistics. Considering the phase space in which the electric charge and electrostatic potential are treated as the thermodynamical pressure and volume, respectively, we find that there exist the emergent pBH2 phase and the critical behavior of the phase transition. These novel features due to the nonextensive nature of the R\'enyi entropy do not appear in the case of the GB statistics.  Moreover, the swallowtail behavior appears in the free energy diagrams. This behavior is reminiscent of the phase transition in the vdW fluid.  The local stability of the black hole can also be determined by the positiveness of compressibility, in addition to that of heat capacity (see Eq.~\eqref{local stab conds}). Interestingly, the black hole in the emergent pBH2 phase is locally stable when the conditions of both thermal and mechanical stabilities in the range $\bar{q}<\bar{\phi}<\bar{\phi}_-$ are satisfied, where $\bar{\phi}_{-}$ has been defined in Eq.~\eqref{phi}.  On the other hand, the nBH and pBH1 are always locally unstable and locally stable, respectively, as discussed in section \ref{RN-Renyi}. For global stability, the sufficiently large black hole in pBH2 phase is thermodynamically preferred than those in other phases, since its Gibbs free energy is relatively lower.  Remarkably, there is the Hawking-Page phase transition by jumping from the pBH1 to pBH2 phases caused by the presence of nonextensivity (see Fig.~\ref{fig:G in Renyi vary q}). The Hawking-Page temperature decreases as charge increases and then terminates at its critical value (see Fig. \ref{fig:TR-HP vs q}).

%\fixme{[Maxwell equal area law]}

For the isothermal process, the RN-AF from the R\'enyi statistics is possible to undergo the first-order phase transition by jumping between the pBH1 and pBH2 phases instead of passing the locally unstable nBH phase via the second-order phase transitions along the isotherm. This jump can be described by the Maxwell construction similar to those of the vdW fluid and the RN-AdS with the GB statistics. In other words, we can find the certain charge $\bar{q}^\ast$ at which the Gibbs free energies (corresponding to the areas between the isotherms and the horizontal line $\bar{q}=\bar{q}^\ast$ in the $\bar{q}-\bar{\phi}$ plane) of the pBH1 and pBH2 phases are equal for $\bar{T}_\text{R}>\bar{T}_c$. To study the Maxwell construction, we split our consideration into two cases which are near the critical point ($\bar{T}_\text{R}\gtrsim\bar{T}_c$) and far away from the critical point ($\bar{T}_\text{R}\gg\bar{T}_c$). 
For the former (latter) case, the aforementioned phase transition is the transition between the pBH1 phase and mechanically stable (unstable) pBH2 phase. As discussed in section~\ref{Maxwell}, the pBH2 phases in the $\bar{q}_2(\bar{\phi})$ branch are mechanically stable while those in the  $\bar{q}_1(\bar{\phi})$ branch are mechanically unstable, respectively. Interestingly, a cusp of isotherm at the origin $\bar{q}=\bar{\phi}=0$ appears when $\bar{T}_\text{R}\geq 1$. Note that this has been found in a similar way as in the RN-AdS case \cite{Chamblin19992nd, Zhou:2019xai}.    From the analysis with the Maxwell construction in this work, the first-order phase transition seems to occur at very small $\bar{q}$ and $\bar{\phi}$ with arbitrarily large $\bar{T}_\text{R}$. Nevertheless, this tends to be inconsistent with the result from the approach of comparing free energy, in which the phase transition between pBH1 and pBH2 phases in the RN-AF with small $\bar{q}$ and $\bar{\phi}$ from R\'enyi statistics can occur only at the temperature below $T_\text{HP}$ of the Schwarzschild black hole as seen in Fig. \ref{fig:TR-HP vs q}. To address this ambiguity, we consider the merging point between $\bar{q}_1(\bar{\phi})$ and $\bar{q}_2(\bar{\phi})$ branches. We have interestingly found that the temperature is constrained as $\bar{T}_\text{R}\leq1$ (see Eq.~\eqref{phi0}). In other words, the isotherms with $\bar{T}_\text{R}>1$ are undefined. It is also noted that the black hole with $\bar{T}_\text{R}=1$ corresponds to the Schwarzschild black hole ($\bar{q}=\bar{\phi}=0$). The maximum temperature of occurrence the phase transition between mechanically stable pBH1 and pBH2 phases has been evaluated in Eq.~\eqref{T max Maxwell}. The isotherms at this temperature are also illustrated in Fig. \ref{fig: Maxwell Tmax}. From this figure, it can be seen that the first-order phase transition in $\bar{q}-\bar{\phi}$ plane is directly related the swallowtail behavior in the free energy diagram. The phase structure of the black hole similar to those of the vdW fluid and of the RN-AdS from the GB statistics can be obtained as shown in Fig. \ref{fig: diagram-full}. 

%\fixme{[Ruppeiner geometry]}

Some aspects of black hole phase structure can be seen further through the approach of thermodynamic geometry.  On the one hand, the second-order phase transition in black hole system can be demonstrated through the divergence of the Ricci scalar curvature in the Ruppeiner geometry, which corresponds to the divergence of heat capacity. On the other hand, there is no information of the first-order phase transition in the Ricci scalar.  These results hold for the black hole thermodynamics from both the GB statistics and the R\'enyi one.  However, a different result from these two framework can be seen in the sign of the Ricci scalar curvature which may imply the types of the interaction of the black hole microstructures, namely, either attractive or repulsive.  It has been found that the interaction can be repulsive in some range of pBH2 phase in the R\'enyi statistics, whereas it would be only attractive in the GB case. 

We then further interpret that, apart from the charge, the nonextensivity of the black hole could be another contribution which lead to a repulsive interaction between the constituents in microscopic structure of black hole.  Remarkably, the occurrence of repulsive interactions other than the electric forces between charges of black hole has confirmed our previous suggestion that some non-trivial correlations between the microstates of a self-gravitating system could be emerged via the nonextensive nature of long-range interaction systems \cite{Promsiri2020, Promsiri:2021hhv}.    Interestingly, the correlations could be manifested themselves as the nonextensive nature in black hole thermodynamics, and vice versa. Recently, some peculiar phenomena arising from the nonextensivity nature of correlated systems have been studied in gravitational physics \cite{Komatsu:2016vof, PhysRevD.96.123504, 2018EPJC...78..829M, Tavayef:2018xwx, Jiulin:2004bg, Du:2006kg, 2008EL.....8459001C, PhysRevD.102.043017}.  See also \cite{Tsallis:2009zex} for a good review of $q$ statistics, together with its related thermodynamic aspects and its applications in a wide range of physics.  

%\fixme{[Criticallity]}

Although, the phase structures of the RN-AdS and the RN-AF with R\'enyi statistics in the fixed charge ensemble are very similar, their microscopic aspects are quite different. By considering the thermodynamic geometry, our study indicates that the value of the scalar curvature near the extremal limit of the RN-AdS negatively diverges while that of the RN-AF with R\'enyi statistics vanishes. This implies that the interaction among the microstructures of the RN-AdS is highly attractive while the RN-AF with R\'enyi description has no interaction between their microscopic constituents. However, we have also investigated the behavior of the RN-AF in the vicinity of the critical point of the second-order phase transition in the R\'enyi model. The results show that the critical exponents, \textit{i.e.}, $\delta, \beta, \alpha$ and $\gamma$, are identical with those in the vdW fluid and the RN-AdS from the GB statistics. Therefore, these three systems could be said to be in the same universality class. This is a further evidence that the behaviors of many different physical systems near the critical point are independent of the type of microscopic constituents and interaction between them.  
%\fixme{[Comments on Odinksov's work]}

One of the most intriguing discoveries is the connection between geometry and thermodynamics of black holes, \textit{i.e.}, the surface gravity and the surface area at event horizon are related to the Hawking temperature and the Bekenstein-Hawking entropy, respectively. These relations are seem to be spoiled in the R\'enyi description, and one may argue that the black hole thermodynamics with this approach is not relevant, as discussed in \cite{Nojiri:2021czz}. However, the violation of the entropy area law can occur in many quantum systems with the strong correlation or mutual information between microscopic degrees of freedom \cite{Wolf:2006zzb, PhysRevLett.100.070502, Vitagliano:2010db, PhysRevLett.111.210402}. Moreover, a loophole in black hole thermodynamics from the conventional approach can be seen from that the Hawking temperature has to be derived based on the GB statistics~\cite{Hawking1975,Hawking:1976ra}, while its entropy implies the existence of nonextensive nature. Hence, the entropic nature of a black hole can be debatable and a study on black hole thermodynamics from an alternative statistics with the presence of nonextensive nature cannot be said to make no sense. Based on one- or multi-parameter deformation statistics including the R\'enyi one, various publications study the systems of bosonic and fermionic fields \cite{Lavagno:1999ru, Lavagno:2002kv, Lavagno:2009ba, Dil:2017apl}. This might be a suggestion to derive the Hawking radiation of a black hole using an appropriate quantum field description.

\section*{Acknowledgement}
We are grateful to Supakchai Ponglertsakul for helpful discussion. This research project is supported by Thailand Science Research and Innovation (TSRI) Basic Research Fund: Fiscal year 2021 under project number FRB640008.  RN is supported by National Research Council of Thailand (NRCT) : NRCT5-RGJ63009-110. CP has been supported by the Petchra Pra Jom Klao Ph.D. Research Scholarship from King Mongkut's University of Technology Thonburi (KMUTT).

\bibliography{ref}
\nocite{*}

\end{document}